\documentclass[12pt,aps,prd,reprint,showpacs,showkeys,eqsecnum,nofootinbib,notitlepage,superscriptaddress,floatfix]{revtex4-1}

\usepackage[tbtags]{amsmath}
\usepackage{amssymb}
\usepackage{stmaryrd}	
\usepackage{natbib}
\usepackage{hyperref}
\usepackage{graphicx}
\usepackage{threeparttable}
\usepackage{url}
\usepackage{bm}		
\usepackage{color}
\usepackage[dvipsnames,svgnames,table]{xcolor}

\usepackage{mathtools}
\usepackage{wasysym}
\usepackage{scalerel}

\newcommand\bvert[1][\relax]{\makebox[\widthof{$#1\vert$}]{\scalerel*[1.2pt]{\brokenvert}{\raisebox{-.2pt}{\vstretch{1.04}{#1\vert}}}}}
\newcommand\lbvert[1][\relax]{\mathopen{\bvert[#1]}}
\newcommand\rbvert[1][\relax]{\mathclose{\bvert[#1]}}

\makeatletter
\newcommand\lrbvert{%
  \@ifstar{\lrbvertscaled}{\lrbvertfixed}%
}
\newcommand\lrbvertscaled[2][\relax]{%
  \mathopen{\ThisStyle{\mathmakebox[\widthof{$\SavedStyle\left\lvert\mathclap{#2}\right.$}]{\scalerel*[1.2pt]{\brokenvert}{\raisebox{-.2pt}{$\SavedStyle\vstretch{1.04}{\left\lvert#2\right\rvert}$}}}}}%
  #2%
  \mathclose{\ThisStyle{\mathmakebox[\widthof{$\SavedStyle\left.\mathclap{#2}\right\rvert$}]{\scalerel*[1.2pt]{\brokenvert}{\raisebox{-.2pt}{$\SavedStyle\vstretch{1.04}{\left\lvert#2\right\rvert}$}}}}}%
}
\newcommand\lrbvertfixed[2][\relax]{%
  \lbvert[#1]#2\rbvert[#1]%
}
\makeatother

\newcommand{\ddd}{\mathtt{d}}
\newcommand{\ii}{\mathtt{i}}
\newcommand{\mm}{\mathtt{m}}
\newcommand{\op}{\mathtt{op}}
\newcommand{\WW}{\mathtt{W}}

\newcommand{\Software}[1]{\textsf{#1}}

\newcommand{\allbf}[1]{\textbf{\mathversion{bold}{#1}}}

\newcommand{\Cplusplus}{\hbox{C\raise.2ex\hbox{\footnotesize ++}}}

\newcommand{\Z}{\phantom{0}}		
\newcommand{\dd}{\allbf{$\delta$}}
\newcommand{\ZZ}{\mathbf{0}}
\newcommand{\U}[1]{\underline{#1}}

\newcommand{\LorenzGaugeSeff}{\Software{Lorenz\-Gauge\-1D\-Effective\-Source}}

\newcommand{\pseudopar}{\\[1ex]}

\newcommand{\half}{\frac{1}{2}}
\newcommand{\dhalf}{\dfrac{1}{2}}
\newcommand{\thalf}{\tfrac{1}{2}}

\newcommand{\lwso}[2]{{\setbox0=\hbox{\phantom{#2}}\hbox to\wd0{{#1}\hss}}}
\newcommand{\cwso}[2]{{\setbox0=\hbox{\phantom{#2}}\hbox to\wd0{\hss{#1}\hss}}}
\newcommand{\rwso}[2]{{\setbox0=\hbox{\phantom{#2}}\hbox to\wd0{\hss{#1}}}}

\newcommand{\ScriptsizeAtop}[2]{{\genfrac{}{}{0pt}{2}{#1}{#2}}}
\newcommand{\ParenScriptsizeAtop}[2]{{\genfrac{(}{)}{0pt}{2}{#1}{#2}}}

\newcommand{\ParenScriptsizeTextAtop}[2]
  {{\raisebox{1ex}{\hbox{$\ParenScriptsizeAtop{\text{#1}}{\text{#2}}$}}}}

\newcommand{\adjusted}{\text{adjusted}}
\newcommand{\base}{{(\text{base})}}
\newcommand{\bndry}{\text{bndry}}
\newcommand{\constant}{\text{constant}}
\newcommand{\eff}{\text{eff}}
\newcommand{\fixed}{\text{(fixed)}}
\renewcommand{\hom}{\text{(hom)}}
\newcommand{\initial}{\text{(initial)}}
\newcommand{\instantaneous}{\text{(instantaneous)}}

\newcommand{\nops}{\text{(nops)}}
\newcommand{\numerical}{\text{num}}
\newcommand{\OrbitAveraged}{\text{(orbit~averaged)}}
\newcommand{\OccasionallyUpdated}{\ParenScriptsizeTextAtop{occasionally}{updated}}
\newcommand{\hi}{\text{(hi)}}

\newcommand{\lo}{\text{(lo)}}

\newcommand{\ortho}{\text{(ortho)}}

\newcommand{\physical}{\text{(physical)}}
\newcommand{\puncture}{\text{(puncture)}}
\newcommand{\residual}{\text{(residual)}}
\newcommand{\RHS}{{\mathsf{RHS}}}

\newcommand{\singular}{\text{(singular)}}
\newcommand{\sourced}{\text{(sourced)}}
\newcommand{\total}{\text{(total)}}
\newcommand{\unstable}{\text{(unstable)}}
\newcommand{\vacuum}{\text{(vacuum)}}

\DeclareMathOperator{\gto}{\sf{gto}}

\DeclareMathOperator*{\erf}{erf}
\DeclareMathOperator*{\erfc}{erfc}

\newcommand{\nosum}{\hbox{\hbox to 0em{\hspace{0.20em}$\Big/$}$\sum$}\strut}
\newcommand{\M}{\mathcal{M}}
\renewcommand{\hbar}{{\bar{h}}}
\renewcommand{\O}{\mathcal{O}}
\newcommand{\Sterm}{\mathcal{S}}
\newcommand{\nti}[1]{{(#1)}}		
\newcommand{\A}{\nti{A}}
\newcommand{\B}{\nti{B}}
\newcommand{\I}{\nti{I}}
\newcommand{\J}{\nti{J}}

\newcommand{\Iellm}{\nti{I \ell m}}
\newcommand{\YabIellm}{Y_{ab}^\Iellm}
\newcommand{\hbarI}{\hbar^\I}
\newcommand{\hbarIellm}{\hbar^\Iellm}
\newcommand{\boxop}{\Box}
\newcommand{\boundary}[1]{{\partial #1}}
\newcommand{\del}{\nabla}
\newcommand{\Diss}{\mathsf{D}}
\newcommand{\conjugate}[1]{\mathop{\text{conj}}\!\left[#1\right]}

\newcommand{\RealPart}[1]{\mathop{\text{Re}}\!\left[#1\right]}
\newcommand{\ImagPart}[1]{\mathop{\text{Im}}\!\left[#1\right]}
\newcommand{\seq}{\,{=}\,}		
\newcommand{\slt}{\,{<}\,}		
\newcommand{\sgt}{\,{>}\,}		
\newcommand{\sle}{\,{\le}\,}		
\newcommand{\sge}{\,{\ge}\,}		
\newcommand{\splus}{\,{+}\,}		
\newcommand{\sminus}{\,{-}\,}		
\newcommand{\sapprox}{\,{\approx}\,}	
\newcommand{\Sun}{\odot}

\newcommand{\ltsim}{\lesssim}
\newcommand{\gtsim}{\gtrsim}

\newcommand{\InnerProduct}[2]{\bigl\langle #1, #2 \bigr\rangle}
\newcommand{\BigInnerProduct}[2]{\Bigl\langle #1, #2 \Bigr\rangle}
\newcommand{\Norm}[1]{\bigl\| #1 \bigr\|}

\newcommand{\UnitVector}[1]{\widehat{#1}}

\newcommand{\PointwiseNorm}[1]{\lrbvert{\lrbvert{#1}}}

\newcommand{\FloorWRTSet}[2]{\lfloor #1 \rfloor_{#2}}

\newcommand{\bigAverage}[2]{\text{average}\bigl(#1, #2\bigr)}
\newcommand{\BigAverage}[2]{\text{average}\Bigl(#1, #2\Bigr)}

\newcommand{\widen}{\text{widen}}
\newcommand{\quantize}{\text{quantize}}

\newcommand{\uu}{\mathsf{u}}
\renewcommand{\ss}{\mathsf{s}}
\newcommand{\TT}{\mathsf{T}}
\newcommand{\XX}{\mathsf{X}}
\newcommand{\VV}{\mathsf{V}}

\newcommand{\Scri}{\mathcal{J}}

\newcommand{\rescaledG}{\widetilde{G}}

\newcommand{\EvolutionName}[1]{\textsf{#1}}

\newcommand{\CenterWithSizeOf}[2]{{\setbox0=\hbox{#2}\hbox to\wd0{\hss{#1}\hss}}}

\begin{document}
\title{\allbf{Time-domain evolution of Lorenz-gauge metric perturbations:
       taming the $\ell \,{=}\, m \,{=}\, 1$ gauge instability}}
\author{Jonathan Thornburg}
\email{dr.j.thornburg@gmail.com}
\affiliation{BKIS Orchards, Box 87, Thetis Island BC V0R 2Y0, Canada}
\affiliation{Department of Astronomy and Center for Spacetime Symmetries,
	     Indiana University, Bloomington, Indiana 47405,
	     USA}
\affiliation{Max-Planck-Institut f\"{u}r Gravitationsphysik,
             Albert-Einstein-Institut,
             Am M\"{u}hlenberg 1, D-14476 Potsdam-Golm,
             Germany}


\begin{abstract}
Calculating the spacetime metric perturbation (MP) sourced by
a small ``particle'' of mass~$\mu M$ (with $0 < \mu \ll 1$)
moving in a Schwarzschild or Kerr ``background'' black hole spacetime
of mass~$M$ is a longstanding research area in general relativity.
This calculation also has an important astrophysical motivation
as a major step in calculating the gravitational waves emitted
by an extreme-mass-ratio inspiral (EMRI) system.
Here I consider the specific problem of the time-domain calculation
of the $\O(\mu)$~Lorenz-gauge MP $h_{ab}$ sourced by the particle.
Decomposing the Schwarzschild-background MP into $e^{im\phi}$ modes,
Dolan and Barack [\textit{Phys.\ Rev.\ D}\textbf{87}, 084066 (2013)]
found that the $m=1$ time-domain Lorenz-gauge MP generically contains
an \emph{unstable gauge mode} which grows linearly with time.
Here I demonstrate a method for computing a Lorenz-gauge time-domain
evolution which is mostly free of this gauge mode.
This method computes an ``orthogonalized'' MP $h_{ab}^\ortho$
as a linear combination of the sourced MP and a homogeneous MP $h_{ab}^\hom$
(evolved in parallel with the sourced MP).
The linear combination is updated ``occasionally''
to make $h_{ab}^\ortho$ orthogonal to $h_{ab}^\hom$
with respect to a chosen inner product on MPs.
I show that, for the test case of a particle in a circular orbit
in a Schwarzschild background, the resulting $h_{ab}^\ortho$
satisfies the $\O(\mu)$~Einstein equations
and the $\O(\mu)$~Lorenz gauge conditions,
remains bounded as $t \to \infty$, and
at late (finite) times
contains only a small component of the unstable gauge mode.
These results hold both for the case where the particle is modelled
as a point particle with MP jump conditions across the particle,
and for the case where
the particle is modelled by a Barack-Goldburn-Vega-Detweiler
``effective source''.
My numerical code for evolving the MP, computing $h_{ab}^\ortho$,
and verifying the above results is included as online supplemental
material with this paper, and will be deposited in the open-source
Black Hole Perturbation Toolkit.
\end{abstract}


\maketitle


\begin{quote}
\textit{This paper is dedicated to the memory of
my late friend and colleague Professor Steven \hbox{Detweiler}.}
\end{quote}


\section{Introduction}
\label{sect-intro}

Consider a small body (``particle'') of mass $\mu M$
(with $0 < \mu \ll 1$) moving freely in an asymptotically flat
and (exterior to any event horizon) globally hyperbolic
background spacetime (e.g., Schwarzschild or Kerr spacetime) of mass~$M$.
Self-consistently calculating the resulting spacetime and the motion
of the small body is a long-standing research question in general relativity.

There is also an astrophysical motivation for this calculation:
If a neutron star or stellar-mass black hole of mass~${\sim}\, 1$--$100 M_\Sun$
orbits a massive black hole of mass~${\sim}\, 10^5$--$10^7 M_\Sun$,
\footnote{
	 $M_\Sun$ denotes the solar mass.
	 }
{} the resulting ``extreme--mass-ratio inspiral'' (EMRI) system will be
a strong astrophysical gravitational-wave (GW) source, likely detectable by
the planned Laser Interferometer Space Antenna (LISA) space-based
gravitational-wave detector.
LISA is expected to observe many such systems, some of them at quite
high signal/noise ratios.
The data analysis for, and indeed the detection of, such systems
will generally require matched-filtering the detector data stream
against large numbers of precomputed
GW templates~(\textcite{LISA-Consortium-waveform-modelling-white-paper-2023}).
These templates, in turn, will mainly be generated by
``fast waveform models'' (e.g., \textcite{Katz-etal-2021:FEW-package})
calibrated against smaller numbers of waveforms calculated \textit{ab initio}.
The problem of computing these latter waveforms provides the
astrophysical motivation for this work.

Here I consider the case where the particle's orbit is too
relativistic for post-Newtonian methods
(e.g.,~\textcite{Futamase-Itoh-2007:PN-review,
Schaeffer-Jaranowski-2024-living-review,
Blanchet-2024-living-review})
to be reliably accurate.
Since the radiation-reaction orbital-evolution timescale is long
($\sim \mu^{-1} M$) while the required resolution near the particle
is high ($\ltsim \mu M$), a direct ``numerical relativity''
integration of the Einstein equations would be very expensive
(and possibly insufficiently accurate) for this problem.

Instead, I use black hole perturbation theory, treating the particle
as an $\O(\mu)$~perturbation on the background spacetime.
For this work I consider only linear (1st-order) perturbation effects:
I assume the particle moves on a pre-specified trajectory in the
background spacetime, and consider the problem of calculating the
resulting $\O(\mu)$~metric perturbation via a time-domain numerical
evolution of the Lorenz-gauge perturbed Einstein equations sourced
by the particle, in the manner of \textcite{Barack-Lousto-2005}
(hereinafter BL05).

As discussed by BL05, using the Lorenz gauge for such a calculation
has the notable advantages that the Lorenz-gauge metric perturbation
from a point particle is well-behaved both close to and far from the
the particle, and that the Lorenz-gauge perturbed Einstein equations
are a Z4 system (and hence are hyperbolic at leading order).

Time-domain calculations of this type are a useful complement to
the more common frequency-domain calculations, and inter-comparing
these different types of calculations can help improve our confidence
in the correctness and accuracy of both.
Two important advantages of time-domain calculations are that they're
relatively easy to extend to highly eccentric particle orbits, and
that (with somewhat more effort) they can be extended from a
Schwarzschild background to a Kerr background,
\footnote{
	 Frequency-domain calculations on a Kerr background
	 are difficult for various reasons, notably the
	 non-separability of the Lorenz-gauge
	 metric perturbation equations (see, e.g.,
	 \protect\textcite{Whiting-Price-2005,Teukolsky-2014:Kerr-review}),
	 although there has been major progress in recent years
	 (\protect\textcite{Dolan-Kavanagh-Wardell-2022,
	  Dolan-etal-2024,Wardell-Kavanagh-Dolan-2025}).
	 }
{}
although I will not further consider either of these extensions here.

\textcite{Dolan-Barack-2013} (hereinafter DB13) considered
the test case of the Lorenz-gauge metric perturbation
on a Schwarzschild background,
with the metric perturbation decomposed into $e^{im\phi}$~modes
(so as to prototype a future extension to a Kerr background).
They found that time-domain $m=1$ Cauchy evolutions exhibit an
\emph{unstable gauge mode}, i.e., a metric perturbation which grows
unboundedly with time while still satisfying the Lorenz gauge condition
at any finite time.
\footnote{
	 They also found unstable gauge modes in $m=0$
	 evolutions, but they were able to stabilize these
	 evolutions using a generalization of the Lorenz
	 gauge condition.
	 }
{}
At any finite time, this gauge mode satisfies all appropriate physical
boundary conditions at the horizon and at spatial infinity.
At a fixed spatial position, this gauge mode grows linearly with time
and so is unbounded as $t \to \infty$.

Decomposing the metric perturbation into Barack-Lousto-Sago $(\ell,m)$
tensor-spherical-harmonic modes (BL05, with the modification described
by \textcite[note~37]{Barack-Sago-2007}, hereinafter BS07),
DB13 found that the unstable gauge mode is an $\ell \seq m \seq 1$
even-parity mode.  DB13 also found an explicit analytical solution
(their equation~(131)) with properties matching their numerical results.
It appears highly likely that a similar mode would also be generically
present in time-domain Lorenz-gauge metric-perturbation evolutions
on a Kerr background.

This unstable gauge mode poses a major problem for time-domain evolutions
of Lorenz-gauge metric perturbations.  Several approaches have been proposed
to resolve this problem, but thus far all require either analytical
knowledge of the unstable gauge mode (available only for a Schwarzschild
background), restrictions on the particle orbit, or leaving the time
domain.  (A frequency-domain evolution naturally enforces boundedness
of the metric perturbation as $t \to \infty$, thus eliminating the
unstable gauge mode.)

In their time-domain calculations of gravitational self-force on a
Schwarzschild background, BL05
and \textcite{Barack-Sago-2007,Barack-Sago-2009,Barack-Sago-2010,
Barack-Sago-2011}
avoided the gauge mode problem by solving for the $\ell \seq m \seq 1$
metric perturbation using the method of \textcite{Detweiler-Poisson-2004},
but this method only applies for the case of a Schwarzschild background.

For the case of a circular geodesic particle orbit
(where the metric perturbation's time dependence
at a fixed spatial position contains only a single frequency),
DB13 were able to remove the unstable gauge mode
by applying a frequency-domain filter to the numerically-computed
metric perturbation.
This method will very likely also work for circular equatorial
particle orbits on a Kerr background, but, unfortunately,
there does not appear to be any way to generalize this method
to more generic particle orbits.
DB13 also found that the frequency filter introduces considerable
high-frequency numerical noise.

For zero-energy zoom-whirl orbits in Schwarzschild,
\textcite{Barack-etal-2019} were able to cancel out the
unstable gauge mode in time-domain evolutions, but their method
requires knowing the unstable gauge mode analytically, and is thus
only applicable for a Schwarzschild background.

Here I present a new method for mostly removing the unstable gauge mode
in $m \seq 1$ time-domain Lorenz-gauge metric-perturbation evolutions.
This method is based on \emph{orthogonalizing} the metric perturbation
with respect to an auxiliary homogeneous metric-perturbation evolution
run in parallel with the main evolution.
At late times the homogeneous metric perturbation is dominated by
the unstable gauge mode, so orthogonalizing with respect to the
homogeneous metric perturbation is approximately the same as
orthogonalizing with respect to the unstable gauge mode.
The result is that the orthogonalized metric perturbation
is mostly free of (i.e., contains only a small component of)
the unstable gauge mode.

I present numerical tests of the orthogonalization scheme for the test case
of a particle in a circular geodesic orbit on a Schwarzschild background,
using the BL05 formalism for time-domain Cauchy evolution of the
metric perturbation in $1{+}1$~numerical dimensions.
I consider two different schemes for modelling the particle:
either as a point particle via jump conditions for the
Barack-Lousto-Sago fields and their derivatives across the particle
(following BL05),
or via a Barack-Golbourn-Vega-Detweiler effective source,
using Wardell's \LorenzGaugeSeff{} effective source code
(\textcite{Wardell-LorenzGauge1DEffectiveSource}).

To preview the main result of this paper,
the orthogonalization technique works well
for both point-particle and effective-source schemes.
More precisely, I show that for both schemes,
the orthogonalized metric perturbation
\begin{itemize}
\item	satisfies the $\O(\mu)$~Einstein equations
	and the $\O(\mu)$~Lorenz gauge conditions
	(up to finite-differencing accuracy),
\item	remains bounded as $t \to \infty$, and
\item	at any finite time, contains only a small component
	(bounded as $t \to \infty$) of the unstable gauge mode.
\end{itemize}

The remainder of this paper is organized as follows:
Section~\ref{sect-intro/notation} summarizes my notation.
Section~\ref{sect-theory-generic} describes
the $\O(\mu)$~perturbed (linearized) Einstein equations,
the point-particle and effective-source schemes for solving these,
the unstable gauge mode,
and the basic idea of the orthogonalization scheme,
all for a generic background spacetime.
Section~\ref{sect-Schw} specializes to a Schwarzschild background,
and describes the Barack-Lousto-Sago tensor-spherical-harmonic decomposition
of metric perturbations
and the resulting BL05 Cauchy evolution scheme for metric perturbations,
the point-particle and effective-source schemes,
the implementation of the orthogonalization scheme,
and some points about the computation of diagnostics.
Section~\ref{sect-numerical-tests} describes numerical tests
of my implementation of the BL05 evolution scheme
and of the basic orthogonalization scheme.
Section~\ref{sect-discussion+conclusions} presents conclusions
and directions for further research.
Appendix~\ref{app-point-particle-jump-conditions}
describes the computation of the jumps in the Barack-Lousto-Sago fields
across the particle when using a point-particle scheme.
Appendix~\ref{app-numerical} describes my numerical scheme.
Appendix~\ref{app-ortho-variants} presents tests of four variants
of the basic orthogonalization scheme:
using a shorter orthogonalization time spacing,
orbit-averaging~$\lambda$,
gradual turnon of the puncture and effective source,
and using a fixed (time-independent)~$\lambda$,
Appendix~\ref{app-convergence-tests} presents numerical tests of
the convergence of the orthogonalized metric perturbation
to a continuum limit, and the convergence of the corresponding
Lorenz gauge constraints and independently-computed Einstein tensor to zero.


\subsection{Notation}
\label{sect-intro/notation}

I generally follow the sign and notation conventions of \textcite{Wald-1984},
with $G = c = 1$ units and a $(-,+,+,+)$ metric signature.
I assume spacetime to be globally hyperbolic (exterior to any event horizon),
foliated by the 3-dimensional spacelike Cauchy hypersurfaces $\Sigma_t$,
and asymptotically flat with (ADM) mass $M$.

I use the Penrose abstract-index notation,
with lower-case Latin indices $abcd$ from the beginning of the alphabet
running over spacetime coordinates.
Upper-case Latin indices $IJK$ range over the integers
from~$1$ to~$10$ inclusive, indexing the Barack-Lousto-Sago
tensor-spherical-harmonic basis elements.
Non-tensor indices and labels are generally parenthesized,
as in $\hbar^\I_\ortho$.
A lower-case typewriter-font Latin $\ii$~indexes grid points,
and a lower-case typewriter-font Latin $\mm$~indexes
the individual points of finite-difference molecules,
as described in detail in appendix~\ref{app-numerical/notation}.
The indices $\ell$ and $m$ index tensor spherical harmonics,
as described in section~\ref{sect-Schw/ell-emm-decomposition}.

Spacetime coordinate indices $abcd$ are subject to
the Einstein summation convention.
Non-tensor indices (including the spherical-harmonic indices $\ell$ and $m$,
any parenthesized indices,
and any indices subscripting or superscripting the $\nosum$ symbol)
are not subject to the Einstein summation convention.
$g_{ab}$ is the background spacetime metric,
and is assumed to be Ricci-flat everywhere away from the particle.
$g_{ab}$ is used to raise and lower all indices.
$\partial_a = \partial \big/ \partial x^a$ is the usual coordinate
partial derivative operator,
$\del_a$ is the covariant derivative operator associated with $g_{ab}$,
and $\boxop = \del^a \del_a$ is the covariant D'Alembertian operator.
$h_{ab}$ is the metric perturbation,
$h = h_a{}^a$ is its trace,
and $\hbar_{ab} = h_{ab} - \half h g_{ab}$
is the trace-reversed metric perturbation.

$\InnerProduct{h_{ab}^\A}{h_{ab}^\B}$
is an inner product on metric perturbations $h_{ab}$
in the constant-time slice $\Sigma_t$,
$\Norm{h_{ab}}$ is the associated norm,
and $h_{ab}^\A \perp h_{ab}^\B$ means
$\InnerProduct{h_{ab}^\A}{h_{ab}^\B} = 0$.
For any given $(\ell,m)$,
$\InnerProduct{\hbarI_\A}{\hbarI_\B}$
is an inner product on the Barack-Lousto-Sago metric-perturbation
fields $\hbarI = \hbarIellm$ in the constant-time slice $\Sigma_t$,
$\Norm{\hbarI}$ is the associated norm,
and $\hbarI_\A \perp \hbarI_\B$ means
$\InnerProduct{\hbarI_\A}{\hbarI_\B} = 0$.
Indices repeated between the two arguments of the inner product,
as in $\InnerProduct{h_{ab}^\A}{h_{ab}^\B}$
or $\InnerProduct{\hbarI_\A}{\hbarI_\B}$
are not subject to the Einstein summation convention.
A caret accent denotes a unit vector defined with respect to the norm,
as in $\UnitVector{\hbarI} := \hbarI \big/ \Norm{\hbarI}$ ($\nosum^I$).
($X := E$ or $E =: X$ means that $X$ is defined to be (the expression) $E$.)

In the context of a specific coordinate system,
$\PointwiseNorm{T_{ab}}$ is a pointwise norm over the
coordinate components of a tensor or tensor-like expression $T_{ab}$
at an event (so that the value of $\PointwiseNorm{T_{ab}}$ at
an event is a single nonnegative real number).

$\varepsilon$ is the difference between $1.0$ and the next larger
floating-point number.  $\varepsilon \approx 1.1\,{\times}\,10^{-16}$
for IEEE-standard double-precision floating-point
arithmetic~(\textcite{Goldberg91}).

$\boundary{S}$ is the boundary of the set of events $S$.
$\conjugate{z}$ is the complex conjugate of the complex number~$z$.
Given a sequence of times $\{t_k \,\,|\,\, k=1, 2, 3, \dots \}$,
$\FloorWRTSet{t}{\{t_k\}}$ is the largest $t_k \le t$.
$\bigAverage{f}{[A,B]}$ is the average of the function $f$ over the
interval $[A,B]$, i.e., $\Bigl( \int_A^B f(t) \, dt \Bigr) \big/ (B{-}A)$.

A subscript $p$ denotes quantities evaluated at the particle position;
the particle is in a circular geodesic orbit at coordinate radius $r=r_p$,
with (coordinate) orbit period~$P$.

In the context of a specific coordinate system (for example, the usual
Schwarzschild $(t,r,\theta,\phi)$ coordinates for Schwarzschild spacetime),
I use $\times$ to denote the component-by-component multiplication
of (coordinate) tensor components, as in $S_{ab} \times G_{ab}$.
Indices appearing in such products are not subject to the
Einstein summation convention.

In appendix~\ref{app-point-particle-jump-conditions} only,
$x := r_*$;
for any variable or expression $q$,
$q' := \partial_x q$ and
$[q]_p$ is the jump in $q$ across the particle;
$N$ is the number of nontrivial Barack-Lousto-Sago
Schwarzschild-metric-perturbation fields $\hbarIellm$
($N \seq 6$ for the $\ell \seq m \seq 1$ case
which is the main focus here);
$\TT$, $\XX$, and $\VV$ are $N{\times}N$~matrices;
and for any given $(\ell,m)$,
$\uu$ and $\ss$ are (time-dependent) $N$-element column vectors
of field variables.


\section{Theory: Generic Background}
\label{sect-theory-generic}


\subsection{Perturbed (Linearized) Einstein Equations}
\label{sect-theory-generic/perturbed-Einstein-eqns}

Consider an asymptotically-flat Ricci-flat ``background'' spacetime of mass~$M$
(in practice, this will be either Kerr or Schwarzschild spacetime)
with metric $g_{ab}$ and associated covariant derivative operator $\del$,
perturbed by a small metric perturbation $h_{ab}$
(so that the physical metric is $g^\physical_{ab} = g_{ab} \splus h_{ab}$),
with $h_{ab}$ sourced by the stress-energy tensor $T_{ab}$ of a small
``particle'' of mass $\mu M$, with $0 < \mu \ll 1$.
I assume that both the background and physical spacetimes
(exterior to any event horizon) are globally hyperbolic,
foliated by the 3-dimensional spacelike hypersurfaces $\Sigma_t$,
and that (again exterior to any event horizon)
any $t \seq \constant$ hypersurface is a Cauchy surface.

Linearizing the physical-metric Einstein equations about the background
metric gives~(\textcite[section~7.5]{Wald-1984}, BL05)
\begin{align}
G_{ab} := \,\,\,
	& \boxop h_{ab} - g_{ab} \boxop h
								\nonumber\\
	& + \del_a \del_b h
	  + g_{ab} \del^c \del^d h_{cd}
								\nonumber\\
	& - \del_b \del^c h_{ac} - \del_a \del^c h_{bc}
								\nonumber\\
	& + 2R^c{}_a{}^d{}_b h_{cd}
								\nonumber\\
	& \qquad\qquad
		= -16\pi T_{ab}
									~ ,
				   \label{eqn-linearized-Einstein-generic-gauge}
\end{align}
where $\del$, $\boxop$, the Riemann tensor, and index raising/lowering
are all defined with respect to the background metric~$g_{ab}$.

Defining the trace-reversed metric perturbation
$\hbar_{ab} := h_{ab} - \half h g_{ab}$
(where $h := h_a{}^a$)
and adopting the Lorenz gauge condition
\begin{equation}
\del^b \hbar_{ab} = 0
									~ ,
					 \label{eqn-Lorenz-gauge-generic-coords}
\end{equation}
the linearized Einstein equations~\eqref{eqn-linearized-Einstein-generic-gauge}
simplify to
\begin{equation}
\boxop \hbar_{ab} + 2R^c{}_a{}^d{}_b \hbar_{cd}
	= -16\pi T_{ab}
									~ .
				    \label{eqn-linearized-Einstein-Lorenz-gauge}
\end{equation}


\subsection{Effective-Source Scheme}
\label{sesct-theory:generic/esrc}

The Barack-Golbourn-Vega-Detweiler effective-source scheme allows
modelling a point particle without requiring a $\delta$-function
stress-energy tensor $T_{ab}$.
First introduced by~\textcite{Barack-Golbourn-2007}
and~\textcite{Vega-Detweiler-2008:self-force-regularization},
the effective source method is now widely used;
see~\textcite[section~3.3]{Wardell-2015:self-force-cmpt-strategies-review}
for a review.

The basic concept of the effective-source scheme is to first choose
(compute) a ``puncture'' metric perturbation $\hbar_{ab}^\puncture$ which
approximates (in a manner to be described below) the Detweiler-Whiting
singular metric perturbation $\hbar_{ab}^\singular$
(\textcite{Detweiler-Whiting-2003}) near the particle,
so that the ``residual'' metric perturbation
\begin{equation}
\hbar_{ab}^\residual := \hbar_{ab} - \hbar_{ab}^\puncture
						    \label{eqn-hbar-ab-residual}
\end{equation}
is finite in a neighborhood of the particle
and ``somewhat differentiable'' (here, $C^2$) at the particle.
Substituting this definition into the $\O(\mu)$ Lorenz-gauge perturbed
Einstein equations~\eqref{eqn-linearized-Einstein-Lorenz-gauge}
then gives
\begin{widetext}
\begin{subequations}
						   \label{eqn-Einstein-puncture}
\begin{align}
\boxop \hbar_{ab}^\residual + 2R^c{}_a{}^d{}_b \hbar_{cd}^\residual
	& = -16\pi T_{ab}
	    - \boxop \hbar_{ab}^\puncture
	    - 2R^c{}_a{}^d{}_b \hbar_{cd}^\puncture
								\\
	& = \begin{cases}
	    0		& \text{at the particle}	\\
	    - \boxop \hbar_{ab}^\puncture
	    - 2R^c{}_a{}^d{}_b \hbar_{cd}^\puncture
			& \text{elsewhere}
	    \end{cases}
							   \label{eqn-Seff-defn}
								\\
	& =: S_{ab}^\eff
									~ ,
							\nonumber
\end{align}
\end{subequations}
\end{widetext}
where the effective source $S_{ab}^\eff$
is defined to be the RHS of~\eqref{eqn-Seff-defn}.

There is considerable freedom in the precise choice of the puncture.
My choice here is that embodied in Wardell's
\LorenzGaugeSeff{} open-source effective source code
(\textcite{Wardell-LorenzGauge1DEffectiveSource}).
This chooses $\hbar_{ab}^\puncture$ to match the first $4$~terms
of the Laurent series expansion of $\hbar_{ab}^\singular$
near the particle in powers of the distance from the particle.
This puncture is only defined in a convex normal neighborhood of
the particle, but that's not a problem for present purposes.

The simplest form of the effective-source scheme
-- which is the one I use here --
then chooses a finite worldtube $W$ such that the particle is contained
in the worldtube's interior and the puncture and effective source are
both defined in a neighborhood of the worldtube,
and defines the ``numerical'' metric perturbation
\begin{subequations}
\begin{align}
\hbar_{ab}^\numerical
	& = \begin{cases}
	    \hbar_{ab}^\residual
				& \text{inside the worldtube}	\\
	    \hbar_{ab}		& \text{outside the worldtube}	
	    \end{cases}
									\\
	& = \begin{cases}
	    \hbar_{ab} - \hbar_{ab}^\puncture
				& \text{inside the worldtube}	\\
	    \hbar_{ab}		& \text{outside the worldtube}	
	    \end{cases}
									~ .
					      \label{eqn-hbar-ab-numerical-defn}
\end{align}
\end{subequations}
By construction,
$\hbar_{ab}^\numerical$ is finite in a neighborhood of the particle,
$C^2$ at the particle and as smooth as $\hbar_{ab}$ elsewhere,
obeys the same far-field boundary conditions as $\hbar_{ab}$, 
and satisfies the evolution equations
\begin{equation}
\boxop \hbar_{ab}^\numerical + 2R^c{}_a{}^d{}_b \hbar_{cd}^\numerical
	= \begin{cases}
	  S_{ab}^\eff		& \text{inside the worldtube}	\\
	  0			& \text{outside the worldtube}	
	  \end{cases}
									~ .
					 \label{eqn-hbar-ab-numerical-evolution}
\end{equation}
$\hbar_{ab}^\numerical$ has a jump of $\hbar_{ab}^\puncture$
across the worldtube boundary $\partial W$,
i.e., for any worldtube-boundary point $x^i_\bndry \in \partial W$, 
\begin{equation}
\lim_\ScriptsizeAtop{x^i \to x^i_\bndry}
     {x^i \in W}                         \!\! (\hbar_{ab}^\numerical)
	= \lim_\ScriptsizeAtop{x^i \to x^i_\bndry}
			      {x^i \not\in W}      \!\! (\hbar_{ab}^\numerical)
	  - \hbar_{ab}^\puncture(x^i_\bndry)
									~ .
					      \label{eqn-hbar-ab-numerical-jump}
\end{equation}
After numerically solving
the evolution equation~\eqref{eqn-hbar-ab-numerical-evolution} for
$\hbar_{ab}^\numerical$, the physical metric perturbation is given by
\begin{equation}
\hbar_{ab}
	= \begin{cases}
	  \hbar_{ab}^\numerical + \hbar_{ab}^\puncture
				& \text{inside the worldtube}	\\
	  \hbar_{ab}^\numerical
				& \text{outside the worldtube}	
	  \end{cases}
									~ .
					  \label{eqn-hbar-ab(hbar-ab-numerical)}
\end{equation}

Notice that despite the fact that the puncture is only an approximation
to the true Detweiler-Whiting singular field, the $\hbar^\numerical$
evolution equations~\eqref{eqn-hbar-ab-numerical-evolution} (if solved exactly)
and the physical-metric-perturbation
reconstruction~\eqref{eqn-hbar-ab(hbar-ab-numerical)}
together yield an \emph{exact} solution of the $\O(\mu)$ perturbed
Einstein equations~\eqref{eqn-linearized-Einstein-Lorenz-gauge}.
In a practical computational scheme, the dominant errors
come from the numerical solution of the $\hbar^\numerical$
evolution equations~\eqref{eqn-hbar-ab-numerical-evolution}


\subsection{The Unstable Gauge Mode}
\label{sect-theory-generic/unstable-gauge-mode}

DB13's analytical solution and numerical experiments show that $m \seq 1$
time-domain Lorenz-gauge evolutions generically include an unstable gauge mode.
This gauge mode induces a metric perturbation~$h_{ab}^\unstable$ that
grows linearly with time while (for a suitable numerical evolution scheme)
still satisfying
the Lorenz gauge conditions~\eqref{eqn-Lorenz-gauge-generic-coords}
up to numerical accuracy everywhere away from the particle.
DB13's numerical experiments only included sourced evolutions,
but it's clear from their analytical solution (their equation~(131))
that nontrivial homogeneous (source-free) evolutions should exhibit
the same instability.

I will use $\ell \seq m \seq 1$ metric perturbation of Schwarzschild
spacetime as numerical examples of this behavior.
As discussed in section~\ref{sect-Schw}, I parameterize
such perturbations by the Barack-Lousto-Sago modes~$\hbarI$.
I define a (complex) inner product
$\InnerProduct{\hbarI_\A}{\hbarI_\B}$~($\nosum^\I$)
and associated norm $\Norm{\hbarI}$,
and a pointwise norm $\PointwiseNorm{\hbarI}$,
on Barack-Lousto-Sago modes.

\begin{figure*}
\begin{center}
\includegraphics[scale=1.00]{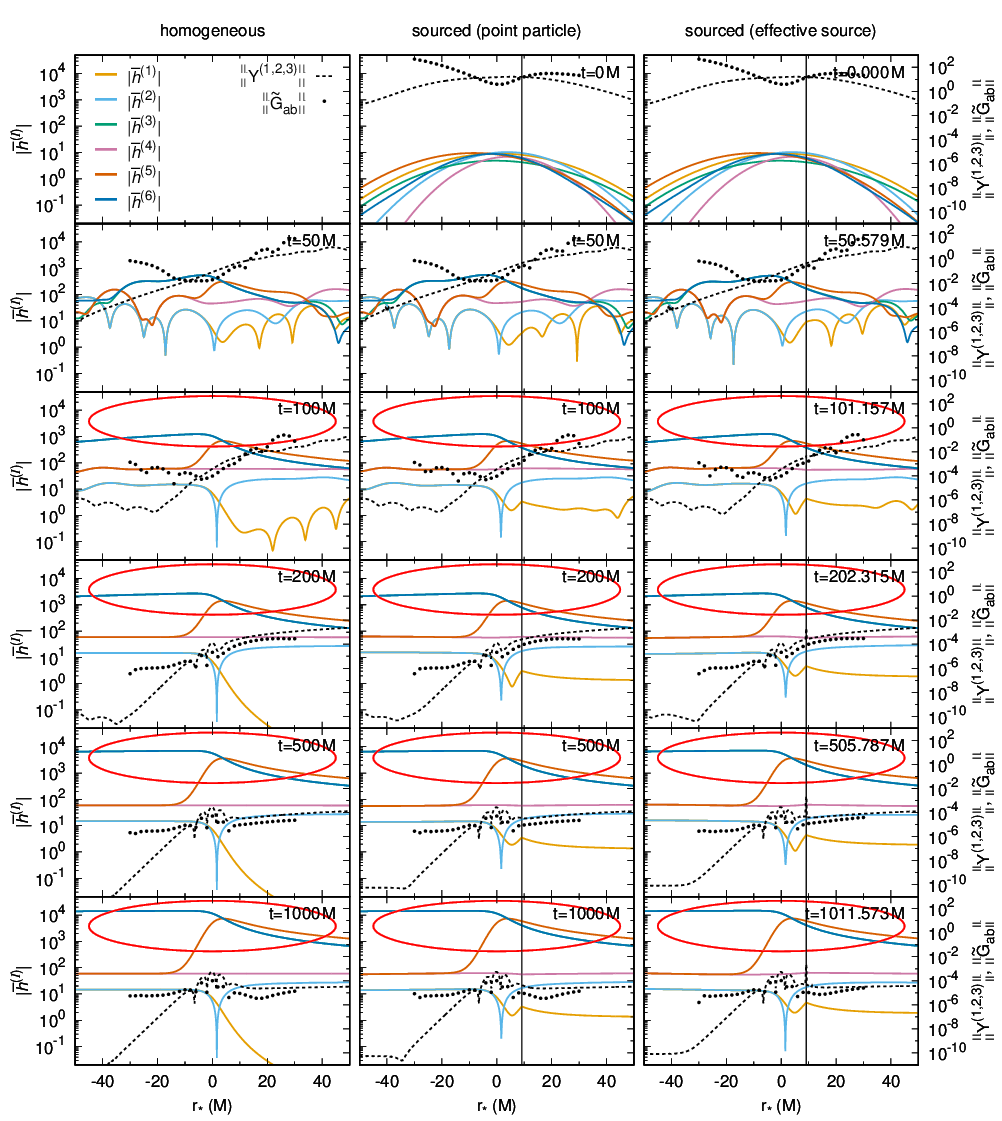}
\end{center}
\caption[Snapshots and Movie of the unstable gauge mode]
	{
	This figure shows snapshots and a movie of the growth of the
	unstable gauge mode in three sample $\ell \seq m \seq 1$ evolutions
	of Schwarzschild metric perturbations.
	(The evolutions are described in detail in
	section~\protect\ref{sect-numerical-tests}.)
	The left column shows a homogeneous evolution,
	the center column shows a sourced evolution
	using a point-particle scheme,
	and the right column shows a sourced evolution
	using an effective-source scheme;
	the movie shows the sourced point-particle evolution.
	The same legend (shown in the top left subplot) applies
	to each subplot.
	In each subplot and movie frame, the absolute values of the nontrivial
	Barack-Lousto-Sago metric-perturbation fields
	$\hbar^\nti{1}$, $\hbar^\nti{2}$, $\hbar^\nti{3}$,
	$\hbar^\nti{4}$, $\hbar^\nti{5}$, and $\hbar^\nti{6}$
	are plotted in color on the left scale,
	and pointwise norms of the nontrivial
	Barack-Lousto-Sago Lorenz gauge constraints,
	$\PointwiseNorm{Y^\nti{1}, Y^\nti{2}, Y^\nti{3}}$
	and of the independently-computed rescaled Einstein tensor,
	$\PointwiseNorm{\rescaledG_{ab}}$,
	are plotted in black on the right scale
	($\PointwiseNorm{\rescaledG_{ab}}$ is only plotted every $2M$ in $r_*$).
	For the sourced evolutions, the particle position is marked
	by a vertical line.
	Notice that for $t \gtsim 100M$
	all three evolutions are dominated by the growing unstable mode
	(shown circled in red),
	while the Lorenz gauge constraints norm
	$\PointwiseNorm{Y^\nti{1}, Y^\nti{2}, Y^\nti{3}}$
	and the rescaled Einstein tensor norm $\PointwiseNorm{\rescaledG_{ab}}$
	are small everywhere except within a few grid points of the particle
	in the effective-source evolution (where the constraints and the
	rescaled Einstein tensor are not computed accurately, for reasons
	described in section~\protect\ref{sect-Schw/diagnostics/nops}).
	}
\label{fig-unstable-mode/snapshots-and-movie}
\end{figure*}

\begin{figure}
\begin{center}
\begin{picture}(80,103)
\put(0,103.0){\textbf{(a)}}
\put(0,70.0){\textbf{(b)}}
\put(0,37.0){\textbf{(c)}}
\put(0,0){\includegraphics[scale=1.00]{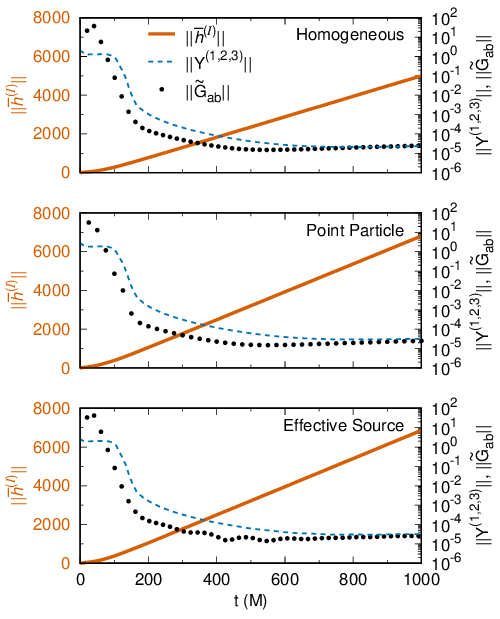}}
\end{picture}
\vspace*{-5mm}
\end{center}
\caption[Time evolution of norms of the unstable gauge mode]
	{
	This figure shows the time evolution of
	norms over grid points and components in the same three
	sample evolutions shown in
	figure~\protect\ref{fig-unstable-mode/snapshots-and-movie}.
	Part~(a) shows the homogeneous evolution,
	part~(b) shows shows the sourced evolution
	using a point-particle scheme,
	and part~(c) shows the sourced evolution
	using an effective-source scheme.
	The same legend (shown in part~(a)) applies to each subplot.
	In each subplot, a norm of the Barack-Lousto-Sago
	metric-perturbation fields, $\Norm{\hbarI}$,
	is plotted on the left scale,
	and norms of the Barack-Lousto-Sago Lorenz gauge constraints,
	$\Norm{Y^\nti{1}, Y^\nti{2}, Y^\nti{3}}$,
	and the rescaled Einstein tensor, $\Norm{\rescaledG_{ab}}$,
	are plotted on the right (logarithmic) scale.
	Notice that at late times all three evolutions are dominated
	by the unstable mode ($\Norm{\hbarI}$ is large
	and growing roughly linearly with time),
	while the Lorenz gauge constraint norm
	and the rescaled Einstein tensor norm are both small.
	}
\label{fig-unstable-mode/norms-of-t}
\end{figure}
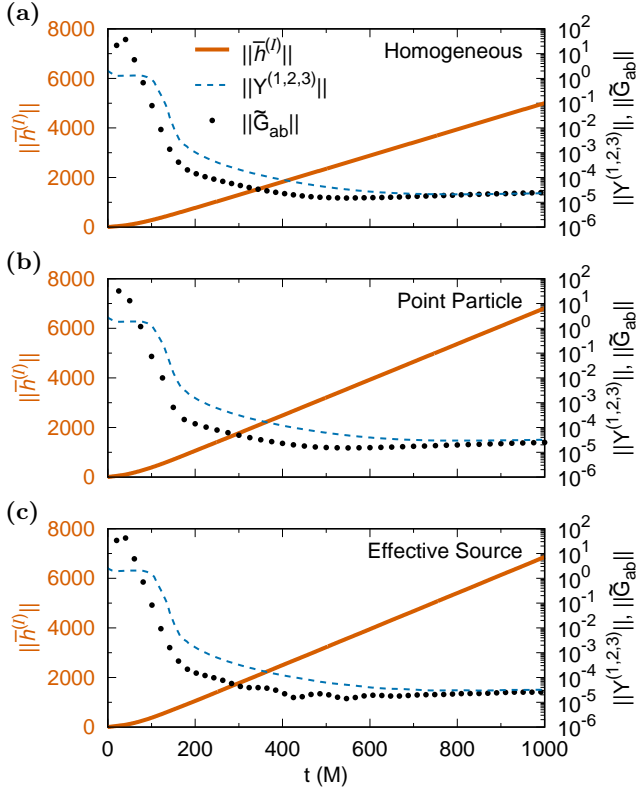

Figures~\ref{fig-unstable-mode/snapshots-and-movie}
and~\ref{fig-unstable-mode/norms-of-t} show the growth
of the unstable gauge mode in three sample $\ell \seq m \seq 1$
time-domain Lorenz-gauge evolutions of metric perturbations
of Schwarzschild spacetime.
All three sample evolutions use the same (arbitrary) initial data,
but differ in their source terms: they are, respectively,
a homogeneous evolution,
a sourced evolution using a point-particle scheme,
and a sourced evolution using an effective-source scheme.
\footnote{
	 These evolutions are described in detail
	 in section~\ref{sect-numerical-tests};
	 they are, respectively, the
	 \EvolutionName{homogeneous},
	 \EvolutionName{sourced-esrc}, and
	 \EvolutionName{sourced-ppart} evolutions
	 in table~\ref{tab-test-evolutions}.
	 }

For $t \gtsim 100M$, $\hbarI$ in each of these sample evolutions
is dominated by the (growing) unstable mode,
with the particle-orbit-period oscillations hardly visible.
Despite $\hbarI$ being large, at late times
pointwise norms of the Lorenz gauge constraints
($\PointwiseNorm{Y^\nti{1}, Y^\nti{2}, Y^\nti{3}}$)
and the independently-computed rescaled Einstein tensor
($\PointwiseNorm{\rescaledG_{ab}}$)
are both small at all spatial positions
except within a few grid points of the particle
(where neither $Y^\nti{I}$ nor $\rescaledG_{ab}$ are computed accurately),
demonstrating that the unstable mode
(i) satisfies the Lorenz gauge constraints and
(ii) is in fact a solution of the $\O(\mu)$~perturbed Einstein equations.
\footnote{
	 The Lorenz-gauge-constraints norm is large early in
	 each evolution because the (arbitrary) initial data
	 doesn't satisfy the Lorenz gauge constraints.
	 The BL05 evolution system's gauge constraint damping effectively
	 ``radiates away'' these early-time constraint violations.
	 }
${}^,$
\footnote{
	 $\PointwiseNorm{\rescaledG_{ab}}$
	 is large at early times, but this isn't surprising:
	 the BL05 evolution system was derived assuming the Lorenz gauge,
	 so there's no particular reason for it to be correct
	 ($\PointwiseNorm{\rescaledG_{ab}}$ small) in the presence
	 of large Lorenz gauge constraint violations (as are present
	 early in each evolution).
	 }
{}

\subsection{Orthogonalization}
\label{sect-theory-generic/ortho}


\subsubsection{Basic concept}
\label{sect-theory-generic/ortho/basic-concept}

Figure~\ref{fig-unstable-mode/snapshots-and-movie} shows that the
homogeneous and sourced evolutions have very similar-looking unstable
modes at late times.  This similarity can be quantified via the
inner product and norm of ``unit-vector'' Barack-Lousto-Sago modes
$\UnitVector{\hbarI} := \hbarI \big/ \Norm{\hbarI}$~($\nosum^\I$).
By construction, 
$\InnerProduct{\UnitVector{\hbarI_\A}}{\UnitVector{\hbarI_\B}}$
lies between $0$ and~$1$, reaching the latter only when
$\hbarI_\A$ and $\hbarI_\B$ are scalar multiples of each other.

\begin{figure}
\begin{center}
\includegraphics[scale=0.60]{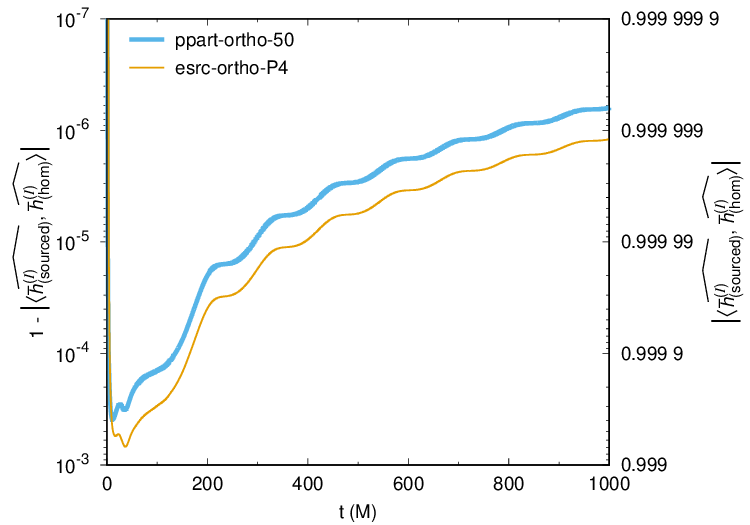}
\end{center}
\vspace*{-4mm}
\caption[Time evolution of unit-vector inner product
	 of $\hbarI_\sourced$ and $\hbarI_\hom$]
	{
	This figure shows the time evolution of the
	unit-vector inner products
	$\InnerProduct{\UnitVector{\hbarI_\EvolutionName{sourced-ppart}}}
		      {\UnitVector{\hbarI_\hom}}$
	(thick light-blue curve)
	and
	$\InnerProduct{\UnitVector{\hbarI_\EvolutionName{sourced-esrc}}}
		      {\UnitVector{\hbarI_\hom}}$
	(thin orange curve)
	in the same three sample evolutions shown in
	figure~\protect\ref{fig-unstable-mode/snapshots-and-movie}.
	Notice that at late times both unit-vector inner products
	are very close to~$1$.
	}
\label{fig-unstable-mode/uvip-sourced-hom}
\end{figure}

Figure~\ref{fig-unstable-mode/uvip-sourced-hom} shows that
$\InnerProduct{\UnitVector{\hbarI_\EvolutionName{sourced-ppart}}}
	      {\UnitVector{\hbarI_\hom}}$
and
$\InnerProduct{\UnitVector{\hbarI_\EvolutionName{sourced-esrc}}}
	      {\UnitVector{\hbarI_\hom}}$
are both very close to~$1$ at late times,
\footnote{
	 Although it's difficult to see on the scale of
	 figure~\protect\ref{fig-unstable-mode/uvip-sourced-hom},
	 both unit-vector inner products are also
	 very close to~$1$ at early times, but this is
	 ``just'' a consequence of using the same initial data
	 for the homogeneous and source evolutions,
	 and is not important for my arguments here.
	 }
{} i.e., at late times each sourced evolution is very close
to a scalar multiple of the homogeneous evolution.
In other words, it appears that at late times each of these evolutions
is dominated by (a scalar multiple of) the \emph{same} unstable mode,
i.e., there appears to be only a \emph{single} unstable mode.

This suggests the following ``orthogonalization'' scheme for constructing
a sourced Lorenz-gauge evolution which is mostly free of the unstable mode.
(Here I describe this scheme for a generic background spacetime;
in section~\ref{sect-Schw/ortho} I specialize to a Schwarzschild background.)

First, choose (define) an inner product
$\InnerProduct{h_{ab}^\A(t)}{h_{ab}^\B(t)}$
on metric perturbations $h_{ab}(t)$ in a constant-time slice;
this implicitly also defines a norm on metric perturbations
in a slice,
$\Norm{h_{ab}(t)}^2 := \InnerProduct{h_{ab}(t)}{h_{ab}(t)}$.
\footnote{
	 Recall that indices repeated between the two arguments
	 of the inner product are not subject to the Einstein
	 summation convention, so there is no summation over
	 $ab$ implied in this definition, nor in the definition
	 of $\Norm{h_{ab}(t)}^2$.
	 }
{}

Now consider the sourced Lorenz-gauge evolution $h_{ab}^\sourced$
of some nonzero initial data,
together with the homogeneous Lorenz-gauge evolution $h_{ab}^\hom$
of some (possibly-different) nonzero initial data.
\footnote{
	 In my numerical tests of the orthogonalization
	 scheme I have always used the same (nonzero) initial data
	 for the homogeneous and sourced evolutions, but the
	 orthogonalization scheme doesn't actually require this.
	 }
{}  At late times, $h_{ab}^\hom$ is dominated by the unstable gauge mode,
i.e., $h_{ab}^\hom$ is approximately some scalar multiple of
$h_{ab}^\unstable$.  This suggests defining
\begin{equation}
h_{ab}^\ortho := h_{ab}^\sourced + \lambda h_{ab}^\hom
									~ ,
							   \label{eqn-hab-ortho}
\end{equation}
where the (real) scalar~$\lambda$ is chosen so that
\begin{subequations}
							    \label{eqn-hab-perp}
\begin{equation}
h_{ab}^\ortho \perp h_{ab}^\hom
									~ ,
\end{equation}
or, equivalently,
\begin{equation}
\Norm{h_{ab}^\ortho}~\text{is minimized}
									~ .
\end{equation}
\end{subequations}
(These are preliminary definitions; I will redefine $h_{ab}^\ortho$
in~\eqref{eqn-hab-ortho(lambda-occasionally-updated)} below.)

That is, $h_{ab}^\ortho$ is obtained by Gram-Schmidt orthogonalizing
$h_{ab}^\sourced$ with respect to $h_{ab}^\hom$.
It's easy to show that
the value of $\lambda$ satisfying~\eqref{eqn-hab-perp} is
\begin{equation}
\lambda^\instantaneous
	 = - \, \frac{\InnerProduct{h_{ab}^\hom}{h_{ab}^\sourced}}
		     {\Norm{h_{ab}^\hom}^2}
					\quad
					\left(\nosum_{ab}\right)
									~ .
						\label{eqn-lambda-instantaneous}
\end{equation}

At late times,
$h_{ab}^\hom$ is approximately a scalar multiple of $h_{ab}^\unstable$,
so orthogonalizing $h_{ab}^\sourced$ with respect to $h_{ab}^\hom$
is approximately the same as
orthogonalizing $h_{ab}^\sourced$ with respect to $h_{ab}^\unstable$.
Therefore, at late times,
the orthogonalization~\eqref{eqn-hab-perp}
(which by definition makes $\InnerProduct{h_{ab}^\ortho}{h_{ab}^\hom} = 0$
and hence
$\InnerProduct{\UnitVector{h_{ab}^\ortho}}{\UnitVector{h_{ab}^\hom}} = 0$)
should make
$\InnerProduct{\UnitVector{h_{ab}^\ortho}}{\UnitVector{h_{ab}^\unstable}}$
``small'' at late times.
In other words, $h_{ab}^\ortho$ should be approximately
free of the unstable mode at late times.

If $\lambda$ were \emph{not} defined by~\eqref{eqn-hab-perp}
and~\eqref{eqn-lambda-instantaneous}, but were instead a fixed constant
(independent of time),
then because
\begin{enumerate}
\item[(i)]
	the Lorenz gauge
	condition~\eqref{eqn-Lorenz-gauge-generic-coords}
	and the $O(\mu)$~Lorenz-gauge perturbed
	Einstein equations~\eqref{eqn-linearized-Einstein-Lorenz-gauge}
	are both \emph{linear} in the metric perturbation,
\item[(ii)]
	$h_{ab}^\sourced$ is an inhomogeneous solution
	of the $O(\mu)$~Lorenz-gauge perturbed Einstein
	equations~\eqref{eqn-linearized-Einstein-Lorenz-gauge},
	and
\item[(iii)]
	$h_{ab}^\hom$ is a \emph{homogeneous} solution of the
	$O(\mu)$~Lorenz-gauge perturbed Einstein
	equations~\eqref{eqn-linearized-Einstein-Lorenz-gauge},
\end{enumerate}
it would necessarily follow that for any fixed $\lambda$,
$h_{ab}^\ortho$ as defined by the linear combination~\eqref{eqn-hab-ortho}
\begin{enumerate}
\item[(1)]	
	would be an inhomogeneous solution
	of the Lorenz-gauge $O(\mu)$~perturbed
	Einstein equations~\eqref{eqn-linearized-Einstein-Lorenz-gauge}
	with the same source terms as $h_{ab}^\sourced$,
and
\item[(2)]
	would satisfy the Lorenz gauge
	conditions~\eqref{eqn-Lorenz-gauge-generic-coords}.
\end{enumerate}

However, in general, $\lambda$ as defined by~\eqref{eqn-hab-perp}
and~\eqref{eqn-lambda-instantaneous} is time-dependent,
so substituting $h_{ab}^\ortho$ into the
Lorenz gauge condition~\eqref{eqn-Lorenz-gauge-generic-coords}
and the $O(\mu)$~perturbed
Einstein equations~\eqref{eqn-linearized-Einstein-Lorenz-gauge}
gives extra $\partial_t \lambda$ and $\partial_{tt} \lambda$ terms,
causing $h_{ab}^\ortho$ to not satisfy either of the conditions~(1) or~(2)
above.
\footnote{
\label{footnote-update-lambda-continuously}
	 I have confirmed numerically that if $\lambda$
	 is updated ``continuously'' (i.e., at each time step)
	 then $h_{ab}^\ortho$ doesn't satisfy the
	 $O(\mu)$~Lorenz-gauge perturbed Einstein
	 equations~\protect\eqref{eqn-linearized-Einstein-Lorenz-gauge}.
	 In particular, for the \EvolutionName{ppart-ortho-cont}
	 and \EvolutionName{esrc-ortho-cont} evolutions
	 in table~\protect\ref{tab-test-evolutions}),
	 $\PointwiseNorm{\rescaledG_{ab}}$ is typically $4$--$6$~orders
	 of magnitude larger than in the other orthogonalized evolutions
	 in table~\protect\ref{tab-test-evolutions},
	 and these large values do \emph{not} decrease at
	 higher numerical resolutions.
	 }

Therefore, my orthogonalization scheme actually uses a
\emph{piecewise-constant-in-time} $\lambda$ which is ``occasionally''
updated from $\lambda^\instantaneous$ at some discrete times
$0 < t_1 < t_2 < t_3 < \dots$ during the evolution.
That is, I define
\begin{equation}
\lambda^\OccasionallyUpdated\!(t)
	= \lambda^\instantaneous \bigl( \FloorWRTSet{t}{\{t_k\}} \bigr)
					 \label{eqn-lambda-occasionally-updated}
\end{equation}
where $\FloorWRTSet{t}{\{t_k\}}$ is the largest $t_k \sle t$.
I then redefine $h_{ab}^\ortho$ as
\begin{equation}
h_{ab}^\ortho(t)
	= h_{ab}^\sourced(t)
	  + \lambda^\OccasionallyUpdated\!(t) \, h_{ab}^\hom(t)
			      \label{eqn-hab-ortho(lambda-occasionally-updated)}
									~ .
\end{equation}

With this scheme, in general, $\lambda^\OccasionallyUpdated$ and
$h_{ab}^\ortho$ have jump discontinuities at each time $t \seq t_k$,
but the conditions~(1) and~(2) above are satisfied
within each time interval~$t_k \slt t \slt t_{k{+}1}$.

The orthogonality conditions~\eqref{eqn-hab-perp}
are satisfied exactly at each time $t \seq t_k$,
so the arguments given above that
$\InnerProduct{\UnitVector{h_{ab}^\ortho}}{\UnitVector{h_{ab}^\unstable}}$
should be ``small'' still apply at late times in the sequence $t \in \{t_k\}$.
Hence,
if $t_{k{+}1} \sminus t_k$ (the time between orthogonalizations)
is bounded as $t \to \infty$, and
$\InnerProduct{\UnitVector{h_{ab}^\ortho}}{\UnitVector{h_{ab}^\unstable}}$
doesn't grow too rapidly within each time interval $t_k \slt t \slt t_{k{+}1}$,
then
$\InnerProduct{\UnitVector{h_{ab}^\ortho}}{\UnitVector{h_{ab}^\unstable}}$
should be ``small''
(i.e., $h_{ab}^\ortho$ should be ``mostly'' free of the unstable mode)
at \emph{all} late times.

One concern with the orthogonalization scheme is that it requires a
numerical cancellation of the unstable mode between the (independent)
sourced and homogeneous evolutions, and this cancellation becomes
more difficult to achieve numerically -- i.e., the cancellation becomes
more sensitive to small (numerical) relative errors in the evolutions --
as the unstable mode grows larger at late times.
Fortunately, the unstable mode grows only linearly with time;
if it grew exponentially with time the required numerical cancellation
would soon become impractical.  In practice, I find that the unstable
mode grows sufficiently slowly that the numerical cancellation is
achievable to a reasonable accuracy.
\footnote{
	 In numerical analysis, the Gram-Schmidt process is
	 considered to be numerically somewhat unstable
	 (see, e.g.,
\protect\textcite[section~5.2.7--5.2.9]{Golub-Van-Loan:Matrix-Computations-4th-Ed}),
	 but this instability doesn't seem to significantly affect
	 my application.  That is, in practice, I find that
	 immediately after updating $\lambda^\OccasionallyUpdated$,
	 the numerically-computed unit-vector inner product magnitude
	 $
	 \Bigl|
	 \InnerProduct{\UnitVector{\hbarI_\ortho}}{\UnitVector{\hbarI_\hom}}
	 \Bigr|
	 $
	 (which would be $0$ in an exact computation
	 and is nonzero only due to floating-point roundoff
	 in the Gram-Schmidt process) is as small as could be
	 expected, i.e. $\O\left(\varepsilon \Norm{\hbarI_\ortho}\right)$.
	 This means that the numerical inaccuracy of Gram-Schmidt
	 orthogonalization makes only a minute contribution to the
	 value of
	 $
	 \Bigl|
	 \InnerProduct{\UnitVector{\hbarI_\ortho}}{\UnitVector{\hbarI_\hom}}
	 \Bigr|
	 $
	 at typical times between $\lambda^\OccasionallyUpdated$
	 updates (shown, e.g.,
	 in figures~\protect\ref{fig-ppart-ortho-50/uvip-ortho-hom},
	 \protect\ref{fig-esrc-ortho-P4/uvip-ortho-hom},
	 and~\protect\ref{fig-ppart-ortho-oa-50/uvip-ortho-hom}).
	 }


\subsubsection{Design choices}
\label{sect-theory-generic/ortho/design-choices}

There are two major design choices in the orthogonalization scheme:
\begin{itemize}
\item	The choice of the inner product, which in turn defines
	the meaning of ``$\perp$'' and ``$\Norm{\cdot}$''
	in~\eqref{eqn-hab-perp}.
	I discuss the choice of inner product
	in section~\ref{sect-Schw/ortho/choice-of-inner-product}.
\item	The choice of the orthogonalization times $\{t_k\}$.
	The simplest choice, and the one I have used for all
	the numerical tests reported here, is to make the
	orthogonalization times uniformly spaced in time,
	$t_k = k \, \Delta t^\ortho$.
\footnote{
	 Another possibility would be to choose the $\{t_k\}$
	 adaptively, monitoring
	 $\InnerProduct{\UnitVector{\hbarI_\ortho}}{\UnitVector{\hbarI_\hom}}$
	 during the evolution
	 and re-orthogonalizing each time this inner product
	 exceeds some chosen threshold.  I haven't tried this.
	 }

	If the particle orbit is periodic
	(e.g., any geodesic on a Schwarzschild background,
	or any equatorial geodesic on a Kerr background),
	it may be useful to choose the orthogonalization
	time spacing $\Delta t^\ortho$ to integrally divide
	the orbital period~$P$, so that $\lambda$ is updated
	at the same set of orbital phases in each orbit.
	This makes orbit-to-orbit comparisons of
	$\lambda^\OccasionallyUpdated$ and $h_{ab}^\ortho$ easier,
	and is particularly useful if used in combination with
	the orbit-averaging scheme discussed in
	appendix~\ref{app-ortho-variants/orbit-averaging}).
\end{itemize}


\subsubsection{Physical Meaning of Re-Orthogonalization}
\label{sect-theory-generic/ortho/physical-meaning}

What is the physical meaning of re-orthogonalization, i.e.,
of the jump discontinuity in $h_{ab}^\ortho$ at each time $t \seq t_k$?
In particular, is re-orthogonalization purely a change of gauge
(allowing $h_{ab}^\ortho$ to be interpreted as the continuous
time-evolution of the slices $\Sigma_t$ of a single physical spacetime),
or
does $h_{ab}^\ortho$ immediately (infinitesimally) after $t \seq t_k$
represent a physically distinct $t \seq \constant$ hypersurface
from $h_{ab}^\ortho$ immediately (infinitesimally) before $t \seq t_k$?

To answer this question, observe first that the DB13 unstable mode
is a pure Lorenz gauge mode.
By virtue of the definition~\eqref{eqn-hab-ortho(lambda-occasionally-updated)},
the jump in $h_{ab}^\ortho$ induced by re-orthogonalizing
(i.e., by updating $\lambda^\OccasionallyUpdated$)
is a scalar multiple of $h_{ab}^\hom$.
Thus, to the extent that the late-time $h_{ab}^\hom$
is dominated by the unstable mode,
the jump in $h_{ab}^\ortho$ induced by re-orthogonalization
is approximately a scalar multiple of the unstable mode,
and thus this jump corresponds approximately to a change of (Lorenz) gauge.

It would be interesting to try to explicitly verify this numerically,
either by comparing curvature invariants across re-orthogonalization,
and/or by attempting to explicitly construct
the re-orthogonalization gauge transformation.
I have not done either of these.

However, the numerical results presented here
\footnote{
	 See, for example, the left column of
	 figures~\protect\ref{fig-ppart-ortho-50/ortho-snapshots-and-movie}
	 and~\protect\ref{fig-esrc-ortho-P4/ortho-snapshots-and-movie}.
	 }
{} do establish (for the tests cases considered here) that at late times,
re-orthogonalization preserves the Lorenz gauge condition,
i.e., re-orthogonalization does \emph{not} significantly increase
the numerical violations of the $h_{ab}^\ortho$ Lorenz gauge constraints.
This is consistent with the above argument that at late times,
re-orthogonalization is approximately a change of gauge within
the Lorenz gauge family.


\section{Schwarzschild spacetime}
\label{sect-Schw}

For an initial test of the orthogonalization scheme,
I take the background spacetime to be Schwarzschild spacetime
of mass $M$ with the usual Schwarzschild coordinates $(t,r,\theta,\phi)$,
so that the line element is
\begin{equation}
ds^2 = -f\,dt^2 + f^{-1}\,dr^2 + r^2(d\theta^2 + \sin^2\theta\,d\phi^2)
									~ ,
\end{equation}
where $f = 1 - 2M/r$.
I introduce the usual Schwarzschild tortise coordinate
\begin{equation}
r_* = r + 2M \ln \left| \frac{r}{2M} - 1 \right|
									~ ,
							      \label{eqn-r_*(r)}
\end{equation}
as well as the null coordinates
\begin{subequations}
\begin{align}
u	& =	t - r_*							\\
v	& =	t + r_*							~ .
\end{align}
\end{subequations}

My numerical scheme requires inverting~\eqref{eqn-r_*(r)} to find
$r$ as a function of~$r_*$; I describe this computation in
appendix~\ref{app-numerical/r(rstar)}.


\subsection{\allbf{$\ell m$~decomposition}}
\label{sect-Schw/ell-emm-decomposition}

Following BL05 and BS07, I decompose
the trace-reversed metric perturbation $\hbar_{ab}$
using the Barack-Lousto-Sago basis of
tensor spherical-harmonic modes~$\YabIellm$,
\begin{equation}
\hbar_{ab}(t,r,\theta,\phi)
	= \frac{\mu}{r}
	  \sum_{\ell,m} \sum_I
	  a^\nti{I}
	  \hbarIellm(t,r)
	  \YabIellm(r,\theta,\phi)
									~ ,
				      \label{eqn-hbar-ab-expansion-in-BLS-modes}
\end{equation}
where
the fixed coefficients~$a^\nti{I}$
(given explicitly by BL05's equation~(9)) are chosen for convenience,
and the Barack-Lousto-Sago tensor-spherical-harmonic
basis functions~$\YabIellm$ are given explicitly
by BL05's equations~(A1), (A2), and~(A3), with the definitions of
$\hbar^\nti{3\ell m}$ and $Y_{ab}^\nti{3\ell m}$ modified as per
BS07's note~37,
\begin{widetext}
\begin{subequations}
					  \label{eqn-Barack-Sago-Yab3-and-hbar3}
\begin{align}
\left(
\text{$\displaystyle \hbar^\nti{3 \ell m}$ in BS07 and this work}
\right)
	& = f^{-1} \left(
		   \text{$\displaystyle \hbar^\nti{3 \ell m}$ in BL05}
		   \right)
									\\
\left(
\text{$\displaystyle Y_{ab}^\nti{3 \ell m}$ in BS07 and this work}
\right)
	& = f \left(
	      \text{$\displaystyle Y_{ab}^\nti{3 \ell m}$ in BL05}
	      \right)
									~ .
\end{align}
\end{subequations}
\end{widetext}
The metric perturbation~$h_{ab}$ and the trace-reversed
metric perturbation~$\hbar_{ab}$ are (by definition) real,
but the fields $\hbarIellm$ are complex fields.

I numerically evolve the fields $\hbarIellm$ using a Cauchy evolution
scheme in $(t,r_*)$ coordinates.  BL05 write the evolution and Lorenz
gauge constraint equations in terms of the partial derivatives
$\partial_r \hbarIellm$ (at constant~$t$)
and $\partial_v \hbarIellm$ (at constant~$u$);
for my numerical scheme I transform these into
$\partial_{r_*} \hbarIellm$ and $\partial_{r_*r_*} \hbarIellm$
(both at constant~$t$)
and $\partial_t \hbarIellm$ (at constant~$\{r,r_*\}$)
via
\begin{subequations}
					   \label{eqn-rv-derivs-to-t-r_*-derivs}
\begin{align}
\text{$\partial_r \hbar$ at constant~$t$}
	& =	f^{-1} \partial_{r_*} \hbar				\\
\text{$\partial_v \hbar$ at constant~$u$}
	& =	\thalf \partial_t \hbar + \thalf \partial_{r_*} \hbar
									~ .
\end{align}
\end{subequations}

For each $(\ell,m)$, the $(t,r_*)$-coordinate evolution equations
for the fields $\hbarIellm$ are
\begin{subequations}
						     \label{eqn-hbarI-evolution}
\begin{align}
\partial_{tt} \hbar^\I
	& = \RHS_\vacuum^\I(\hbar^\J) + 4 \Sterm^\I
					 \label{eqn-hbarI-evolution-with-source}
									\\
	& =: \RHS^\total{}^\I(\hbar^\J)
									~ ,
\end{align}
where both $\RHS_\vacuum^\I$  and $\RHS^\total{}^\I$
are spatial differential operators,
and where the factor of~$4$ multiplying the source term
in~\eqref{eqn-hbarI-evolution-with-source}
accounts for my use of a different normalization convention
for the $\boxop$ operator than BL05.

In more detail, the vacuum right-hand-side operator is given by
\begin{align}
\RHS_\vacuum^\I(\hbar^\J)
	= & \,\, \partial_{r_*r_*} \hbar^I - V \hbar^\I
								\nonumber\\
	  & - 4 \M^I(\hbar^J, \partial_{r_*} \hbar^\J, \partial_t \hbar^\J)
									~ ,
							  \label{eqn-RHS-vacuum}
\end{align}
where the potential $V = V(r)$ is given by
\begin{equation}
V(r) = f \left[ \frac{\partial_r f}{r} + \frac{\ell(\ell{+}1)}{r^2} \right]
									~ ,
\end{equation}
\end{subequations}
and where $\M$ is a linear coupling operator which mixes the different
Barack-Lousto-Sago modes indexed by $\I$ but doesn't mix the
tensor-spherical-harmonic indices~$\nti{\ell m}$.
$\M$ is given explicitly by BL05's equation~(18)
if BL05's main gauge constraint damping terms are included,
\footnote{
	 BL05's equation~(18) incorporates a slightly different
	 gauge-constraint damping scheme than that described by BL05
	 in the two paragraphs immediately prior to their equation~(17).
	 All the results presented here use BL05's equation~(18).
	 }
{} or by BL05's equation~(C1) if gauge constraint damping is omitted.
The source term $\Sterm$ models the particle and is
described in section~\ref{sect-Schw/modelling-particle}.

Since the evolution equations~\eqref{eqn-hbarI-evolution}
don't mix the $(\ell,m)$, each $(\ell,m)$ may be evolved independently.
For $\ell \seq m \seq 1$,
$\YabIellm$ is nontrivial only for $I \in \{1,2,3,4,5,6\}$.

In terms of the fields $\hbarIellm(t,r)$, the Lorenz gauge
conditions (constraints)~\eqref{eqn-Lorenz-gauge-generic-coords}
are given by BL05's equation~(16), again modified to use
the BS07 definition~\eqref{eqn-Barack-Sago-Yab3-and-hbar3}
of $\hbar^\nti{3 \ell m}$.
I refer to the mode-decomposed Lorenz gauge conditions
(BL05's $H_a^\nti{\ell m}$, defined as the left-hand-side fields
in BL05's equation~(16)) as $Y^\nti{a} = Y^\nti{a}(t,r_*)$,
with the dependence on $\ell m$ being implicit.
For $\ell \seq m \seq 1$, $Y^\nti{a}$ is nontrivial only for
$a \in \{1,2,3\}$.

In the usual terminology of numerical relativity
(see, e.g., \textcite[chapter~3]{Bona-Masso-book-2005}
or \textcite[section~2.6]{Alcubierre-book-2008}),
the BL05/BS07 evolution system is a free evolution with (Lorenz-gauge)
constraint damping.

Appendix~\ref{app-numerical} describes my numerical scheme for
(approximately) time-integrating the evolution
equations~\eqref{eqn-hbarI-evolution}.


\subsection{Modelling the Particle}
\label{sect-Schw/modelling-particle}

I take the particle to be in a circular geodesic orbit
at coordinate radius $r = r_p$,
with angular velocity $\omega = \sqrt{M/r_p^3}$
and (coordinate) orbital period $P = 2\pi/\omega$.

All the numerical results presented here use $r_p = 7.2M$,
so that $P \approx 121.389M$.


\subsubsection{Point-Particle Scheme}
\label{sect-Schw/modelling-particle/ppart}

The simplest way to model the particle is as a point particle with
the $\delta$-function stress-energy tensor given by BL05's equation~(5).
The corresponding source terms in the BL05 evolution equation~(17)
are given by BL05's equations~(29) and~(30).

Because of the $\delta$-function source term,
the fields $\hbarIellm$ and their time derivatives
are only $C^0$ across the particle,
\footnote{
	 This uses the fact that the particle is in a
	 \emph{circular} orbit; for a noncircular orbit
	 the time derivatives would, in general, also have
	 jumps across the particle.
	 }
{}
while spatial derivatives of $\hbarIellm$
have (time-dependent) jumps across the particle.
I calculate these jumps using the scheme described in
appendix~\ref{app-point-particle-jump-conditions}.
Once these jumps are known, I use them in the finite differencing
scheme as described in appendix~\ref{app-numerical/source/adjusted-FD-ppart}.


\subsubsection{Effective Source}
\label{sect-Schw/modelling-particle/esrc}

Alternatively, the particle may be modelled using an effective-source
scheme.  Compared to a point-particle scheme, an effective-source scheme
is more complicated, but generalizes readily to $2{+}1$- or $3{+}1$-dimensional
numerical evolutions.

Using an effective-source scheme, for each $(\ell,m)$
the $\hbarIellm$~evolution equations~\eqref{eqn-hbarI-evolution} become
\begin{align}
\partial_{tt} \hbarI_\numerical
	= {} & \RHS_\vacuum^\I
								\nonumber\\
	     & + \begin{cases}
		 4 S_\eff^\I	& \text{inside the worldtube}	\\
		 0		& \text{outside the worldtube}	
		 \end{cases}
									~ ,
\end{align}
with the corresponding jump conditions across the worldtube boundary
\begin{subequations}
					      \label{eqn-hbarI-numerical(hbarI)}
\begin{align}
\hbarI_\numerical
	& = \begin{cases}
	    \hbarI_\residual
				& \text{inside the worldtube}	\\
	    \hbarI{}		& \text{outside the worldtube}	
	    \end{cases}
									\\
	& = \begin{cases}
	    \hbarI{} - \hbarI_\puncture
				& \text{inside the worldtube}	\\
	    \hbarI{}		& \text{outside the worldtube}	
	    \end{cases}
									~ .
				     \label{eqn-hbarI-numerical(hbarI,puncture)}
\end{align}
\end{subequations}
I describe the implementation of these jump conditions
at the finite-differencing level
in appendix~\ref{app-numerical/source/adjusted-FD-esrc}.

If the initial data doesn't already satisfy the
jump conditions~\eqref{eqn-hbarI-numerical(hbarI)}
across the worldtube boundary,
the dynamical evolution quickly drives the $\hbarI$
(and $\partial_t \hbarI$) into a configuration satisfying the jump conditions.
However, this process tends to generate high-spatial-frequency noise
in the $\hbarI$, which can reduce the accuracy of the numerical computation.
In appendix~\ref{app-ortho-variants/gradual-turnon}
I discuss a possible scheme (which turns out not to work well)
for reducing the amount of this high-spatial-frequency noise by
gradually turning on the puncture at the start of the evolution.


\subsection{Orthogonalization}
\label{sect-Schw/ortho}

DB13 found that only the $\ell \seq m \seq 1$ evolutions suffer
from the unstable gauge mode, so hereinafter I restrict my attention
to the $\ell \seq m \seq 1$ case, and I drop the indices $\ell m$.

Rather than orthogonalizing the metric perturbation $h_{ab}$ itself,
my code instead actually orthogonalizes the Barack-Lousto-Sago $\hbarI$
fields.  Since the $\hbarI$ are complex fields, I choose (define)
a complex inner product $\InnerProduct{\hbarI_\A}{\hbarI_\B}$
on $\hbarI$ fields.
\footnote{
	 Recall that indices repeated between the two
	 arguments of the inner product are \emph{not} subject to the
	 Einstein summation convention, e.g., there is no summation
	 over $I$ implied in this definition, or in the
	 definition of $\Norm{\hbarI}^2$.
	 }
${}^,$
\footnote{
	 To be clear, this is the inner product between
	 one set of the 6~nontrivial $\hbarI$ fields,
	 and another set of the 6~nontrivial $\hbarI$ fields,
	 $\InnerProduct{\hbar^\nti{1,2,3,4,5,6}_\A}
		       {\hbar^\nti{1,2,3,4,5,6}_\B}$.
	 }
{}
This implicitly also defines a norm on $\hbarI$ fields,
$\Norm{\hbarI}^2 := \InnerProduct{\hbarI}{\hbarI}$.

The remaining description of section~\ref{sect-theory-generic/ortho}
still applies, i.e., I initially define
\begin{equation}
\hbarI_\ortho
	:= \hbarI_\sourced + \lambda \hbarI_\hom
							 \label{eqn-hbarI-ortho}
									~ ,
\end{equation}
where $\lambda$ is now a \emph{complex} scalar, chosen so that
\begin{subequations}
							  \label{eqn-hbarI-perp}
\begin{equation}
\hbarI_\ortho \perp \hbarI_\hom
									~ ,
\end{equation}
or, equivalently,
\begin{equation}
\Norm{\hbarI_\ortho}~\text{is minimized}
									~ .
\end{equation}
\end{subequations}
The value of $\lambda $ satisfying~\eqref{eqn-hbarI-perp} is
\begin{equation}
\lambda^\instantaneous
	= - \, \frac{\InnerProduct{\hbarI_\hom}{\hbarI_\sourced}}
		    {\Norm{\hbarI_\hom}^2}
					\quad
					\left(\nosum^\I\right)
									~ .
					  \label{eqn-hbarI-lambda-instantaneous}
\end{equation}
$\lambda^\OccasionallyUpdated$ is still defined
by~\eqref{eqn-lambda-occasionally-updated}, and
my final redefinition of $\hbarI_\ortho$ is
\begin{equation}
\hbarI_\ortho(t)
	:= \hbarI_\sourced(t)
	   + \lambda^\OccasionallyUpdated(t) \, \hbarI_\hom(t)
			    \label{eqn-hbarI-ortho(lambda-occasionally-updated)}
									~ .
\end{equation}


\subsubsection{Choice of Inner Product}
\label{sect-Schw/ortho/choice-of-inner-product}

I have experimented with several different choices of inner product.

The fundamental requirement for the inner product is that it should
``see'' the unstable mode.
Figure~\ref{fig-unstable-mode/snapshots-and-movie} shows that the
unstable mode has a large amplitude
throughout the plotted range $r_* \in [-50M, +50M]$,
so the inner product should include (at least most of) this region.
The $\hbarI$ near the inner/outer grid boundaries are strongly
influenced by my non-physical inner/outer boundary conditions
(described in appendix~\ref{app-numerical/BCs}), so I exclude
these regions from the inner produt.

An important consideration in choosing the inner product is how the
region near the particle should be handled.  The physical metric perturbation
may be singular at the particle, which both makes it difficult to define
a sensible inner product there and runs the risk of the near-particle
contributions dominating the remainder of the computational domain.
Therefore, I always exclude the near-particle region from the inner
product.

In practice, I choose the inner product to depend on the $\hbarI$
(only) within the region $r_* \in X-Y$, where $X = [-100M, 100M]$
and $Y$ is a region near the particle, as described below.
\footnote{
\label{footnote-other-inner-product-interval}
	 It would be interesting to try values other
	 than $r_* = \pm 100M$ for the endpoints of the $X$ interval,
	 and/or to try making $X$ non-symmetric about $r_* = 0$.
	 I haven't tried either of these.
	 }
$^,$
\footnote{
\label{footnote-inner-product-gradual-cutoff}
	 It might also be interesting to try making the
	 cutoff of the inner product outside $X$ gradual
	 rather than abrupt, i.e., weighting the inner product
	 with a weight function which smoothly decreases from
	 $1$ for $r_* \in X$ to very small values at large $|r_*|$.
	 Such an inner product should respond more smoothly
	 than my present inner product when outgoing radiation
	 crosses the boundary of the $X$ region.
	 I haven't tried this.
	 }

To this end, a simple definition for the inner product is the (complex)
Euclidean inner product on Barack-Lousto-Sago $\hbarI$ fields,
\begin{equation}
\BigInnerProduct{\hbarI_\A}{\hbarI_\B}
	:= \frac{1}{N} \sum_{\ii \in X{-}Y}
		       \sum_I
		       \conjugate{\left(\hbarI_\A\right)_\ii}
				  \left(\hbarI_\B\right)_\ii
									~ ,
					     \label{eqn-Euclidean-inner-product}
\end{equation}
where $\ii$ indexes grid points,
$N$ is the actual number of grid points in the set $X{-}Y$,
and
$Y$ (the near-particle region excluded from the inner product)
is the set of all grid points which either
(i) are within the worldtube (if an effective source is being used),
or
(ii) are within a finite-difference molecule radius of the particle
     (if a point-particle model is being used).

Unfortunately, when doing convergence tests which compare orthogonalized
evolutions at different finite-differencing resolutions $\Delta r_*$
(discussed in detail in appendix~\ref{app-convergence-tests}),
using the inner product~\eqref{eqn-Euclidean-inner-product}
results in $\lambda$ and $\hbarI_\ortho$ showing only slow
\big($\O(\Delta r_*)$\big) convergence to a continuum limit.
There are two reasons for this slow convergence:
\begin{itemize}
\item	The set of near-particle grid points~$Y$ varies with $\Delta r_*$.
\item	Viewing the sum over grid points in~\eqref{eqn-Euclidean-inner-product}
	as an approximation to the integral~\ref{eqn-integral-inner-product}
	below, \eqref{eqn-Euclidean-inner-product} is the rectangle-rule
	approximation (the Riemann sum), which generically has an
	$\O(\Delta r_*)$ error in approximating the integral.
\end{itemize}

To allow faster (higher order) convergence, I now instead define
the inner product as the integral
\begin{equation}
\BigInnerProduct{\hbarI_\A}{\hbarI_\B}
	:= \int_{X{-}Y} \left(
			\sum_I \conjugate{\left(\hbarI_\A\right)}
					  \left(\hbarI_\B\right)
			\right) dr_*
									~ ,
					      \label{eqn-integral-inner-product}
\end{equation}
where the set $Y$ is now chosen to be independent of $\Delta r_*$.
In particular, I choose $Y$ as follows:
\begin{itemize}
\item	Start with an $r_*$ interval which is the worldtube
	(if an effective-source scheme is being used),
	or which is a single point at the particle
	(if a point-particle scheme is being used).
\item	Then, widen this interval by moving each of its endpoints
	outwards by a macroscopic distance $(\delta r_*)_\widen$
	(chosen to be larger than any finite-difference molecule radius).
\item	Finally, quantize the widened interval by rounding
	each of its endpoints outwards to an integer multiple
	of some macroscopic distance $(\delta r_*)_\quantize$,
	chosen to be an integer multiple of the least common multiple
	of all the grid spacings $\Delta r_*$ used in convergence tests
	(cf.~appendix~\ref{app-convergence-tests}).
\end{itemize}

With this definition, the inner product only ``sees'' the
physical metric perturbation outside a neighborhood of the particle
(and outside a neighborhood of the worldtube if an effective-source
scheme is being used),
$X$ and $Y$'s endpoints are (i.e., coincide with) grid points,
and $X$ and $Y$'s endpoints are both independent of the grid spacing
$\Delta r_*$.
Table~\ref{tab-common-pars} gives the values of $(\delta r_*)_\widen$
and $(\delta r_*)_\quantize$ used in my numerical tests of the
orthogonalization scheme.

For numerical calculation, I approximate
the integral~\eqref{eqn-integral-inner-product} using Simpson's rule,
which generically has an $\O\bigl((\Delta r_*)^4\bigr)$ error term,
allowing for up to 4th~order convergence of $\lambda$ and $\hbarI_\ortho$.
I discuss convergence tests in detail
in appendix~\ref{app-convergence-tests}.


\subsubsection{Choice of Orthogonalization times}
\label{sect-Schw/ortho/choice-of-ortho-times}

I have experimented with a number of choices for the orthogonalization
time spacing $\Delta t^\ortho$.  Here I present results for
$\Delta t^\ortho = 50M$, $\Delta t^\ortho = P/4 \approx 30.347M$,
and $\Delta t^\ortho = P/12 \approx 10.116M$.


\subsection{Diagnostics}

Because the Lorenz gauge condition~\eqref{eqn-Lorenz-gauge-generic-coords}
doesn't fully specify the gauge, $\hbarI_\ortho$ is generally not in
the same gauge as an $\hbarI$ computed by some other method
(e.g., a frequency-domain method).  This makes a direct cross-method
comparison of $\hbarI$ difficult.

Instead, I assess the orthogonalization scheme by numerically testing
that at late times (when $\hbarI_\hom$ is dominated by the unstable mode),
the following criteria are satisfied:
\begin{itemize}
\item	In between orthogonalization times,
	$\hbarI_\ortho$ should satisfy the
	$O(\mu)$~Lorenz-gauge perturbed Einstein
	equations~\protect\eqref{eqn-linearized-Einstein-Lorenz-gauge}.
	I assess this by testing that the independently-computed
	$\O(\mu)$ perturbed Einstein tensor~$\rescaledG_{ab}$
	(described in detail in
	section~\ref{sect-Schw/diagnostics/Einstein-tensor})
	is small
	and converges towards zero as the numerical resolution is increased.
	(I discuss convergence tests in appendix~\ref{app-convergence-tests}.)
\item	In between orthogonalization times,
	$\hbarI_\ortho$ should satisfy the $O(\mu)$~Lorenz gauge
	condition~\eqref{eqn-Lorenz-gauge-generic-coords},
	i.e., the Lorenz gauge constraints $Y^\nti{1}$, $Y^\nti{2}$,
	and $Y^\nti{3}$ should all be small,
	and should converge towards zero
	as the numerical resolution is increased.
\item	$\hbarI_\ortho$ should contain only a relatively small
	component of the unstable mode.  Since $\hbarI_\hom$
	is dominated by the unstable mode at late times,
	I operationalize this criterion as the requirement that
	$\InnerProduct{\UnitVector{\hbarI_\ortho}}{\UnitVector{\hbarI_\hom}}$
	should be ``small'', say $\ltsim 0.4$.
\end{itemize}


\subsubsection{Independent computation of the Einstein tensor}
\label{sect-Schw/diagnostics/Einstein-tensor}

To assist in verifying the correctness of the evolution equations,
and to verify that the orthogonalization scheme yields correct evolutions,
I find it very useful to have an independently-programmed computation
of the $\O(\mu)$ perturbed Einstein tensor~$G_{ab}$
defined (everywhere away from the particle)
by~\eqref{eqn-linearized-Einstein-generic-gauge}.

$G_{ab}$ depends on all 4~spatial coordinates $(t,r,\theta,\phi)$,
but for $\ell \seq m \seq 1$ the angular dependence of $G_{ab}$ can be
separated by multiplying each $(t,r,\theta,\phi)$-coordinate tensor
component of $G_{ab}$ by the corresponding component of
\begin{equation}
S^{(\theta\phi)}_{ab}
	= \left[
	  \begin{array}{cccc}
	  S^{-1}	& S^{-1}	& C^{-1}	& S^{-1}	\\
	  *		& S^{-1}	& C^{-1}	& S^{-1}	\\
	  *		& *		& S^{-1}	& 1		\\
	  *		& *		& *		& S^{-3}	
	  \end{array}
	  \right]
									~ ,
\end{equation}
where
\begin{subequations}
\begin{align}
S	& = e^{i\phi} \sin\theta
									\\
C	& = e^{i\phi} \cos\theta
									~ ,
\end{align}
\end{subequations}
and where $*$ denotes components determined by symmetry.
That is, the component-by-component product
$S_{ab}^{(\theta\phi)} \times G_{ab}$ has no angular dependence.

To obtain components which neither blow up nor vanish at the horizon,
I then further rescale $S_{ab}^{(\theta\phi)} \times G_{ab}$ by
multiplying each of its components by the corresponding component
of
\begin{equation}
S_{ab}^{(r)}
	= \left[
	  \begin{array}{cccc}
	  R		& R^2		& R		& R		\\
	  *		& R^3		& R^2		& R^2		\\
	  *		& *		& R^2		& 1		\\
	  *		& *		& *		& R^2		
	  \end{array}
	  \right]
									~ ,
\end{equation}
where $R = r - 2M$.
I define the resulting ``rescaled'' Einstein tensor as
\begin{equation}
\rescaledG_{ab}
	= S_{ab}^{(r)} \times S_{ab}^{(\theta\phi)} \times G_{ab}
							\label{eqn-rescaled-Gab}
									~ ,
\end{equation}
and use this, and norms over coordinate components of $\rescaledG_{ab}$
at an event ($\PointwiseNorm{\rescaledG_{ab}}$)
and over components and grid points ($\Norm{\rescaledG_{ab}}$),
as diagnostics.

For the work described here, I generated formulas for each
$\rescaledG_{ab}$ component as a function of the $\hbarI$
and their 1st and 2nd~spacetime partial derivatives,
directly from~\eqref{eqn-linearized-Einstein-generic-gauge}
by using the \Software{SageManifolds} package
(\textcite{Gourgoulhon-Bejger-Mancini-2015,Gourgoulhon-Mancini-2018})
in the open-source \Software{Sage} symbolic computation system
(\textcite{Stein-Joyner-2005,Eroecal-Stein-2010,SageMath}).
In my implementation, this computation suffers from numerical
cancellations for $r_* \ll 0$ or $r_* \gg 0$,
so I limit the Einstein-tensor computation
to the range $r_* \in [-30M, +30M]$.


\subsubsection{Diagnostic Norms}
\label{sect-Schw/diagnostics/norms}

I use pointwise norms $\PointwiseNorm{\,\cdot\,}$
(i.e., norms over coordinate components,
computed independently at each grid point)
to assess the spatial variation of diagnostics
(e.g., in figures~\ref{fig-unstable-mode/snapshots-and-movie},
\ref{fig-ppart-ortho-50/ortho-snapshots-and-movie},
and~\ref{fig-esrc-ortho-P4/ortho-snapshots-and-movie}).
In particular,
\begin{itemize}
\item	I define $\PointwiseNorm{Y^\nti{1,2,3}}$
	to be the RMS of $Y^\nti{1}$, $Y^\nti{2}$, and~$Y^\nti{3}$.
\item	I define $\PointwiseNorm{\rescaledG_{ab}}$
	to be $\sqrt{|\rescaledG_{ab}| \, |\rescaledG^{ab}|}$.
\end{itemize}

I also use gridwise norms $\Norm{\,\cdot\,}$
(i.e., norms across an entire set of grid points and coordinate components),
particularly when assessing the temporal variation of diagnostics
and for convergence tests (appendix~\ref{app-convergence-tests}).
In particular,
\begin{itemize}
\item	I define $\Norm{Y^\nti{1,2,3}}$ and $\Norm{\hbarI_\ortho}$
	as the inner-product norm (described in detail in
	section~\ref{sect-Schw/ortho/choice-of-inner-product}).
\item	For historical reasons, my code only computes
	$\rescaledG_{ab}$ on a fixed and relatively coarsely-spaced
	set of grid points
	(spacing $1M$ in $r_*$ for all the results reported here),
	independent of the numerical resolution.
	This set doesn't provide enough information
	to accurately compute the inner-product
	integral~\ref{eqn-integral-inner-product}.
	Instead, I define $\Norm{\rescaledG_{ab}}$ to be
	the RMS over grid points of $\PointwiseNorm{\rescaledG_{ab}}$.
\end{itemize}


\subsubsection{Computation of Time Derivatives for Diagnostics}
\label{sect-Schw/diagnostics/cmpt-time-derivs}

In order for the calculation of diagnostics
(in particular, the Lorenz gauge constraints $Y^a$
and the rescaled Einstein tensor $\rescaledG_{ab}$)
to be fully independent of my implementation of the Barack-Lousto-Sago
evolution system for the $\hbarI$,
my code calculates $Y^a$ and $\rescaledG_{ab}$ using
$\partial_t \hbarI$ and $\partial_{tt} \hbarI$ obtained
by explicitly time--finite-differencing
the sequence of $\hbarI$ geneated by the numerical evolution,
rather than using the time derivatives computed by the
Barack-Lousto-Sago evolution equations~\eqref{eqn-hbarI-evolution}.

For computing diagnostics, the time finite-differencing uses
off-centered (1-sided) 4th~order molecules, using the current time slice
and the 4~(5)~immediately preceding time slices
for $\partial_t \hbarI$ ($\partial_{tt} \hbarI$) respectively.
My code suppresses computation of these time derivatives
and the diagnostics $Y^a$ and $\rescaledG_{ab}$ at any time
where the time--finite-differencing would not have sufficient
time levels of continuously-changing data.
In practice, this means that diagnostics aren't computed
for the initial data or the first 4 or 5~time steps of an evolution,
nor are they computed for 4 or 5 time steps
after $\lambda^\OccasionallyUpdated$ is updated
(since this causes $\hbarI$ to change discontinuously).
\footnote{
	 It would be possible to avoid this gap in the
	 diagnostics after each $\lambda^\OccasionallyUpdated$
	 update, by keeping a duplicate copy of the $\hbarI_\ortho$
	 computed using the pre-update $\lambda^\OccasionallyUpdated$.
	 This would somewhat complicate the software, and
	 possibly significantly increase memory usage, but
	 would likely have only a minor CPU-time cost since
	 only a small fraction of time steps would require
	 the extra computation.
	 I haven't implemented this technique.
	 }


\subsubsection{No-Puncture-Subtraction Diagnostics}
\label{sect-Schw/diagnostics/nops}

My formulas for $\rescaledG_{ab}$ don't take into account the presence
of any source terms in the Einstein equations, and hence give wrong
results within the worldtube if an effective-source scheme is used.
To avoid this problem, my numerical code offers the option of computing
``no-puncture-subtracted'' (``nops'') diagnostics, where the code effectively
undoes the puncture subtraction of~\eqref{eqn-hbarI-numerical(hbarI)}
inside the worldtube by defining
\begin{equation}
\hbarI_\nops
	= \begin{cases}
	  \hbarI_\numerical + \hbarI_\puncture
				& \text{inside the worldtube}	\\
	  \hbarI_\numerical	& \text{outside the worldtube}	
	  \end{cases}
									~ ,
\end{equation}
and computing all diagnostics using the $\hbarI_\nops$.
(Since the spatial finite differencing still doesn't ``know'' about
the particle, these diagnostics are still inaccurate within a finite
difference molecule radius of the particle.  These diagnostics also
have some numerical cancellations near the particle due to the
particle's Coulomb-type $1/(r-r_p)$ singularity.)


\section{Numerical Tests of the Orthogonalization Scheme}
\label{sect-numerical-tests}

In this section I present numerical tests of the basic orthogonalization
scheme for point-particle and effective-source particle models.
(I present numerical tests of some variants of the orthogonalization
scheme in appendix~\ref{app-ortho-variants}.)
Tables~\ref{tab-test-evolutions} and~\ref{tab-common-pars}
give parameters for my test evolutions.

To study the convergence of the finite-difference numerical results
to a continuum limit
(discussed in detail in appendix~\ref{app-convergence-tests}),
I have run many evolutions at a number of different
finite-difference resolutions
between $\Delta r_* = M/2$ and $\Delta r_* = M/32$.
Unless noted otherwise,
all results presented here are for $\Delta r_* = M/8$.


\subsection{Initial Data}
\label{sect-numerical-tests/initial-data}

All the evolutions presented here use the same (arbitrary) initial data,
optionally modified as described in the next paragraph.
For each $I$, this initial data sets each of
$\RealPart{\hbarI}$, $\ImagPart{\hbarI}$,
$\RealPart{\partial_t \hbarI}$, and $\ImagPart{\partial_t \hbarI}$
to an independently-chosen Gaussian in $r_*$.
Each Gaussian amplitude is randomly chosen
from a uniform distribution on $[-10,+10]$,
each Gaussian mean $r_*$ is randomly chosen
from a uniform distribution on $[-10M,+10M]$,
and each Gaussian standard deviation in $r_*$ is randomly chosen
from a uniform distribution on $[10M,20M]$.

For evolutions using an effective-source scheme
(and not using the gradual turnon scheme
discussed in appendix~\ref{app-ortho-variants/gradual-turnon}),
it may be useful to explicitly subtract the puncture from the initial data
within the worldtube, so that the (updated) initial data then satisfies
the jump conditions~\eqref{eqn-hbar-ab-numerical-jump}.
Table~\ref{tab-test-evolutions} explicitly states whether or not
this subtraction is done for each test evolution.

\begin{turnpage}
\begin{table*}
\begin{center}
\begin{tabular}{llclcccl@{\hspace*{1em}}l}
			&
			& puncture
			& evolution
			& Courant
			&
			&
			&
			&
									\\
			&
			& subtracted from
			& duration
			& factor
			&
			&
			&
			&
									\\
Evolution name		& particle?
			& iniital data?
			& \cwso{$t_{\max}$}{duration}
			& $\Delta t \big/ \Delta r_*$
			& nops?
			& orthogonalized?
			& $\smash{\lambda^\OccasionallyUpdated}$ 
			& notes
									\\
\hline 
\EvolutionName{homogeneous}
			& no
			& no
			& $1000M$
			& 1.0
			& no
			& ---
			& ---
			& shown in figures~\protect\ref{fig-unstable-mode/snapshots-and-movie}--\protect\ref{fig-unstable-mode/uvip-sourced-hom}
									\\[1ex]
\EvolutionName{sourced-ppart}
			& point particle
			& no
			& $1000M$
			& 1.0
			& no
			& no
			& ---
			& shown in figures~\protect\ref{fig-unstable-mode/snapshots-and-movie}--\protect\ref{fig-unstable-mode/uvip-sourced-hom}
									\\[1ex]
\EvolutionName{sourced-esrc}
			& effective source
			& no
			& $1000M$
			& 1.0
			& yes
			& no
			& ---
			& shown in figures~\protect\ref{fig-unstable-mode/snapshots-and-movie}--\protect\ref{fig-unstable-mode/uvip-sourced-hom}
									\\[1ex]
\hline 
\EvolutionName{ppart-ortho-50}
			& point particle
			& no
			& $2000M$
			& 1.0
			& no
			& yes
			& updated every $50M$
			& shown in figures~\protect\ref{fig-ppart-ortho-50/lambda}--\protect\ref{fig-ppart-ortho-50/uvip-ortho-hom} and~\protect\ref{fig-convergence/hbar}--\protect\ref{fig-convergence/Einstein}
									\\
			&
			&
			&
			&
			&
			&
			& from $\lambda^\instantaneous$
			&
									\\[1ex]
\EvolutionName{esrc-ortho-P4}
			& effective source
			& yes
			& $1011.573M$
			& $P/120 \approx 1.012$
			& yes
			& yes
			& updated every
			& shown in figures~\protect\ref{fig-esrc-ortho-P4/lambda}--\protect\ref{fig-esrc-ortho-P4/uvip-ortho-hom} and~\protect\ref{fig-convergence/hbar}--\protect\ref{fig-convergence/Einstein}
									\\
			&
			&
			&
			&
			&
			&
			& $P/4 \approx 30.347M$
			&
									\\
			&
			&
			&
			&
			&
			&
			& from $\lambda^\instantaneous$
			&
									\\[1ex]
\EvolutionName{esrc-ortho-P12}
			& effective source
			& yes
			& $1011.573M$
			& $P/120 \approx 1.012$
			& yes
			& yes
			& updated every
			& shown in figures~\protect\ref{fig-esrc-ortho-P12/uvip-ortho-hom}
			  and~\protect\ref{fig-esrc-ortho-P12/norms-of-t}
									\\
			&
			&
			&
			&
			&
			&
			& $P/12 \approx 10.116M$
			&
									\\
			&
			&
			&
			&
			&
			&
			& from $\lambda^\instantaneous$
			&
									\\[1ex]
\hline 
\EvolutionName{ppart-ortho-oa-50}
			& point particle
			& no
			& $2000M$
			& 1.0
			& no
			& yes
			& updated every $50M$
			& shown in figures~\protect\ref{fig-ppart-ortho-oa-50/lambda}--\protect\ref{fig-ppart-ortho-oa-50/uvip-ortho-hom}
									\\
			&
			&
			&
			&
			&
			&
			& from $\lambda^\OrbitAveraged$
			&
									\\[1ex]
\EvolutionName{esrc-gto-ortho-P4}
			& effective source
			& no
			& $1011.573M$
			& $P/120 \approx 1.012$
			& yes
			& yes
			& updated every
			& shown in figures~\protect\ref{fig-esrc-gto-ortho-P4/lambda},
			  and~\protect\ref{fig-esrc-gto-ortho-P4/norms-of-t}
									\\
			& (gradual turnon)
			&
			&
			&
			&
			&
			& $P/4 \approx 30.347M$
			&
									\\
			&
			&
			&
			&
			&
			&
			& from $\lambda^\instantaneous$
			&
									\\[1ex]
\hline 
\rlap{\EvolutionName{ppart-ortho-fixed-from-oa-50-t=500}}
			&
			&
			&
			&
			&
			&
			&
			&
									\\
			& point particle
			& no
			& $2000M$
			& 1.0
			& no
			& yes
			& fixed (never updated)
			& shown in figure~\protect\ref{fig-ppart-ortho-fixed/cmp-uvip-ortho-hom-from-t=various}
									\\
			&
			&
			&
			&
			&
			&
			& $= \lambda^\OrbitAveraged$
			&
									\\
			&
			&
			&
			&
			&
			&
			& at $t_{\max}^\base=500M$ in
			&
									\\
			&
			&
			&
			&
			&
			&
			& \rlap{\EvolutionName{ppart-ortho-oa-50} evolution}
			&
									\\
\rlap{\EvolutionName{ppart-ortho-fixed-from-oa-50-t=1000}}
			&
			&
			&
			&
			&
			&
			&
			&
									\\
			& point particle
			& no
			& $2000M$
			& 1.0
			& no
			& yes
			& fixed (never updated)
			& shown in figure~\protect\ref{fig-ppart-ortho-fixed/cmp-uvip-ortho-hom-from-t=various}
									\\
			&
			&
			&
			&
			&
			&
			& $= \lambda^\OrbitAveraged$
			&
									\\
			&
			&
			&
			&
			&
			&
			& at $t_{\max}^\base=1000M$ in
			&
									\\
			&
			&
			&
			&
			&
			&
			& \rlap{\EvolutionName{ppart-ortho-oa-50} evolution}
			&
									\\
\rlap{\EvolutionName{ppart-ortho-fixed-from-oa-50-t=2000}}
			&
			&
			&
			&
			&
			&
			&
			&
									\\
			& point particle
			& no
			& $5000M$
			& 1.0
			& no
			& yes
			& fixed (never updated)
			& shown in figures~\protect\ref{fig-ppart-ortho-fixed/cmp-uvip-ortho-hom-from-t=various} and~\protect\ref{fig-ppart-ortho-fixed/norms-of-t}
									\\
			&
			&
			&
			&
			&
			&
			& $= \lambda^\OrbitAveraged$
			&
									\\
			&
			&
			&
			&
			&
			&
			& at $t_{\max}^\base=2000M$ in
			&
									\\
			&
			&
			&
			&
			&
			&
			& \rlap{\EvolutionName{ppart-ortho-oa-50} evolution}
			&
									\\
\hline 
\EvolutionName{ppart-ortho-cont}
			& point particle
			& no
			& $1000M$
			& 1.0
			& no
			& yes
			& updated every
			& discussed in footnote~\protect\ref{footnote-update-lambda-continuously}
									\\
			&
			&
			&
			&
			&
			&
			& time step
			&
									\\
			&
			&
			&
			&
			&
			&
			& from $\lambda^\instantaneous$
			&
									\\
\EvolutionName{esrc-ortho-cont}
			& effective source
			& yes
			& $1011.573M$
			& $P/120 \approx 1.012$
			& yes
			& yes
			& updated every
			& discussed in footnote~\protect\ref{footnote-update-lambda-continuously}
									\\
			&
			&
			&
			&
			&
			&
			& time step
			&
									\\
			&
			&
			&
			&
			&
			&
			& from $\lambda^\instantaneous$
			&
									\\[1ex]
\hline 
\end{tabular}
\end{center}
\caption[Test evolutions]
	{
	This table describes parameters specific to each test evolution.
	The column headed "nops?" specifies whether or not
	the diagnostics are computed using a no-puncture-subtracted
	copy of the $\hbarI$, as described in
	section~\protect\ref{sect-Schw/diagnostics/nops}.
	See table~\protect\ref{tab-common-pars} for
	parameters common to all the test evolutions.
	}
\label{tab-test-evolutions}
\end{table*}
\end{turnpage}

\begin{table*}
\begin{threeparttable}
\begin{tabular}{ll}
Parameter name		& value
									\\
\hline 
spatial domain		& $r_* \in [-100M \sminus 1.5t_{\max},
				    +100M \splus  1.5t_{\max}]$
\tnote{$\dagger$}							\\
$(\delta r_*)_\widen$	& $1M$						\\
$(\delta r_*)_\quantize$& $1M$						\\
particle (if present)	& located at $r \seq 7.2M$ ($r_* \sapprox 9.111M$);
			  orbital period $P \sapprox 121.389M$		\\
worldtube (if present)	& radius $\pm 2M$ in $r_*$,
			  centered on grid point closest to particle
\end{tabular}
\caption[Common parameters for test evolutions]
	{
	This table describes parameters common to all the test evolutions.
	See table~\protect\ref{tab-test-evolutions} for
	parameters specific to each test evolution.
	}
\label{tab-common-pars}
\begin{tablenotes}
\item[$\dagger$]
	The factor of $1.5$ is chosen to give a
	safety margin against possible numerical modes
	which may travel inwards from the spatial boundaries
	at superluminal speeds.
\end{tablenotes}
\end{threeparttable}
\end{table*}


\subsection{Orthogonalized Evolution (Point Particle)}
\label{sect-numerical-tests/ppart-ortho-50}

I begin my presentation of numerical tests of the orthogonalization scheme
with the simplest case: a point-particle model.

\begin{figure}
\begin{center}
\begin{picture}(80,157)
%
%
\put(0,120)
  {
  \begin{picture}(0,0)
  \put(0,34){\textbf{(a)}}
  \put(5,0){\includegraphics[scale=0.90]{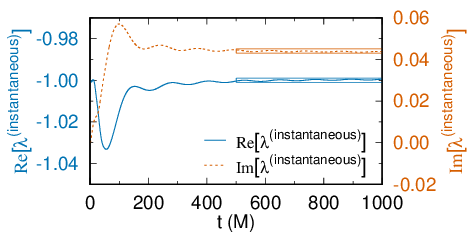}}
  \end{picture}
  }
\put(0,70)
  {
  \begin{picture}(0,0)
  \put(0,45.0){\textbf{(b)}}
  \put(5,0){\includegraphics[scale=0.90]{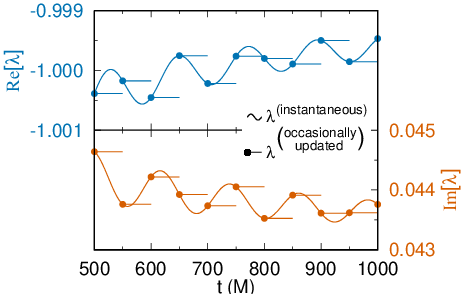}}
  \end{picture}
  }
\put(0,0)
  {
  \begin{picture}(0,0)
  \put(0,63.0){\textbf{(c)}}
  \put(5,0){\includegraphics[scale=0.90]{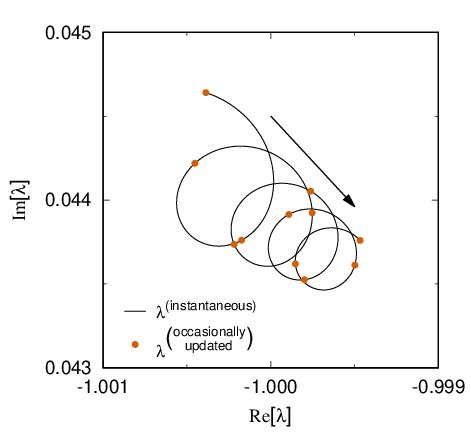}}
  \end{picture}
  }
\end{picture}
\vspace*{-5mm}
\end{center}
\caption[Time evolution of instantaneous and occasionally-updated $\lambda$
	 for the \EvolutionName{ppart-ortho-50} evolution]
	{
	This figure shows the time evolution of $\lambda^\instantaneous$
	and $\lambda^\OccasionallyUpdated$
	for the \EvolutionName{ppart-ortho-50} evolution.
	Part~(a) shows the real and imaginary parts of
	$\lambda^\instantaneous$ as functions of time.
	The rectangular regions are shown at an enlarged scale
	in parts~(b) and~(c).
	Part~(b) shows, at an enlarged scale, the real and imaginary parts
	of both $\lambda^\instantaneous$ and $\lambda^\OccasionallyUpdated$
	as functions of time for late times ($t \ge 500M$).
	The solid dots and horizontal lines show the sample-and-hold
	behavior of $\lambda^\OccasionallyUpdated$.
	The legend applies to both real and imaginary parts.
	Part~(c) shows, at an enlarged scale,
	the trajectories in the complex plane
	of both $\lambda^\instantaneous$ (the spiral curve)
	and $\lambda^\OccasionallyUpdated$ (the solid dots)
	for late times (again $t \ge 500M$).
	The arrow shows the direction of the time evolution.
	}
\label{fig-ppart-ortho-50/lambda}
\end{figure}
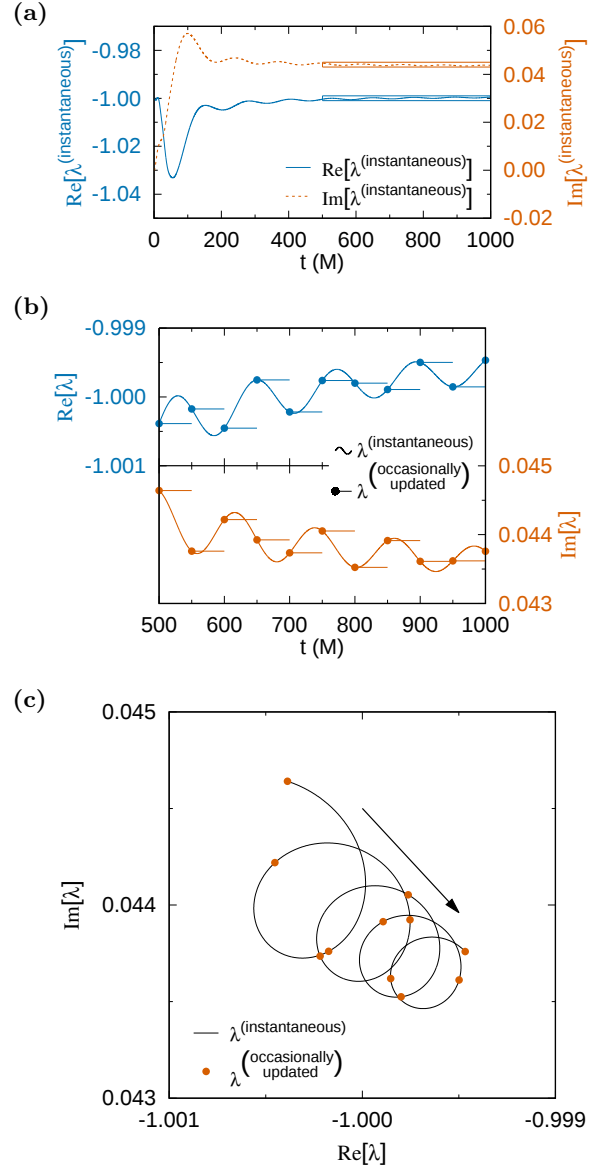

Figure~\ref{fig-ppart-ortho-50/lambda} shows the time evolution of
$\lambda^\instantaneous$ and $\lambda^\OccasionallyUpdated$
for the \EvolutionName{ppart-ortho-50} evolution.
After an initial transient, $\lambda^\instantaneous$ shows
gradually-decaying spiral oscillations in the complex plane,
superimposed on a gradually-slowing drift.
The oscillations have the same period as the particle orbit,
$P \approx 121.389M$.
$\lambda^\OccasionallyUpdated$ samples-and-holds
$\lambda^\instantaneous$ every $\Delta t^\ortho = 50M$
(figure~\ref{fig-ppart-ortho-50/lambda}(b) and~(c)).
Since this sampling period doesn't integrally divide the orbital period~$P$,
the $\lambda^\OccasionallyUpdated$ updates sample the oscillations
at a different phase at each oscillation, so $\lambda^\OccasionallyUpdated$
moves somewhat irregularly in the complex plane.

\begin{figure}
\begin{center}
\includegraphics[scale=1.00]{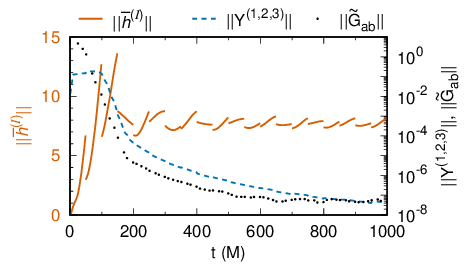}
\end{center}
\caption[Time evolution of norms of the orthogonalized
	 \EvolutionName{ppart-ortho-50} evolution]
	{
	This figure shows the time evolution of
	norms over grid points and components
	in the \EvolutionName{ppart-ortho-50} evolution.
	The inner-product norm of $\hbarI_\ortho$
	is plotted on the left scale,
	and the inner-product norm of
	the Lorenz gauge constraints $Y^\nti{1,2,3}$
	and the RMS-norm of the rescaled Einstein tensor $\rescaledG_{ab}$
	are plotted on the right (logarithmic) scale.
	Notice that unlike the non-orthogonalized evolutions
	shown in figure~\protect\ref{fig-unstable-mode/norms-of-t},
	here $\Norm{\hbarI_\ortho}$ remains bounded at late times,
	showing no secular growth with time.
	}
\label{fig-ppart-ortho-50/norms-of-t}
\end{figure}

Figure~\ref{fig-ppart-ortho-50/norms-of-t} shows the time evolution
of norms over grid points and tensor components of the resulting
$\hbarI_\ortho$, $Y^\nti{1,2,3}$, and $\rescaledG_{ab}$.
Notice that unlike the non-orthogonalized evolutions
shown in figures~\ref{fig-unstable-mode/snapshots-and-movie}
and~\ref{fig-unstable-mode/norms-of-t},
here $\hbarI$ (more precisely, $\hbarI_\ortho$)
remains bounded at late times, showing no secular growth with time,
while $Y^\nti{1,2,3}$ and $\rescaledG_{ab}$ still remain small.

\begin{figure*}
\begin{center}
\includegraphics[scale=1.00]{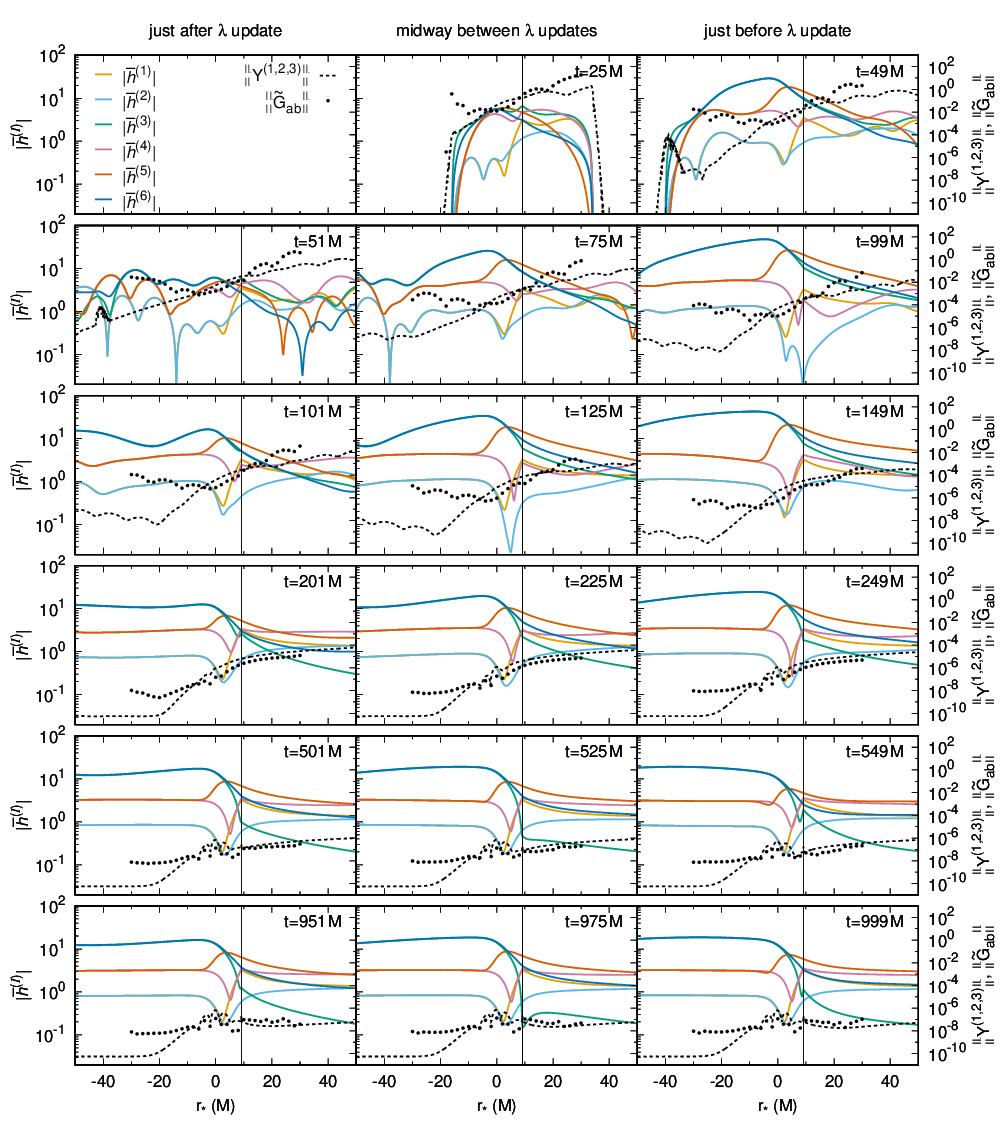}
\end{center}
\caption[Snapshots and movie of the \EvolutionName{ppart-ortho-50}
	 orthogonalized evolution]
	{
	This figure shows snapshots and a movie
	of the \EvolutionName{ppart-ortho-50} orthogonalized evolution.
	The same legend (shown in the top left subplot) applies
	to all the subplots.
	Time runs across each row, then down.
	Within each row,
	the left subplot is at a time just after
	a $\lambda^\OccasionallyUpdated$ update;
	the center subplot is midway between
	$\lambda^\OccasionallyUpdated$ updates (which happen every $50M$),
	and the right subplot is at a time just before
	the next $\lambda^\OccasionallyUpdated$ update.
	In each subplot, the absolute values of the nontrivial
	Barack-Lousto-Sago metric-perturbation fields
	$\hbar^\nti{1,2,3,4,5,6}$ are plotted in color on the left scale
	(which is expanded relative to
	figure~\protect\ref{fig-unstable-mode/snapshots-and-movie}),
	and pointwise norms of the nontrivial Barack-Lousto-Sago
	Lorenz gauge constraints, $\PointwiseNorm{Y^\nti{1,2,3}}$,
	and of the independently-computed rescaled Einstein tensor,
	$\PointwiseNorm{\rescaledG_{ab}}$ are plotted in black
	on the right scale
	(which is the same as
	figure~\protect\ref{fig-unstable-mode/snapshots-and-movie});
	$\PointwiseNorm{\rescaledG_{ab}}$ is only plotted
	at $r_*$ intervals of~$2M$.
	In the movie, notice that unlike the movie of
	figure~\protect\ref{fig-unstable-mode/snapshots-and-movie},
	here the particle-orbit-period oscillations are clearly visible;
	the evolution is \emph{not} dominated by the unstable mode.
	}
\label{fig-ppart-ortho-50/ortho-snapshots-and-movie}
\end{figure*}

Figure~\ref{fig-ppart-ortho-50/ortho-snapshots-and-movie}
shows snapshots and a movie of the resulting orthogonalized evolution
at various times.  Unlike figure~\ref{fig-unstable-mode/snapshots-and-movie},
here the particle-orbit-period oscillations are clearly visible
in the movie; the evolution is \emph{not} dominated by the unstable mode.
There are some low-level oscillations in the
rescaled Einstein tensor $\PointwiseNorm{\rescaledG_{ab}}$
for $r_* \ltsim -25M$ and $r_* \gtsim 20M$;
these are likely due to numerical cancellations in my computation
of $\rescaledG_{ab}$ and don't grow with time.
At late times $\PointwiseNorm{Y^\nti{1,2,3}}$
and $\PointwiseNorm{\rescaledG_{ab}}$ are small everywhere.

\begin{figure}
\begin{center}
\includegraphics[scale=0.60]{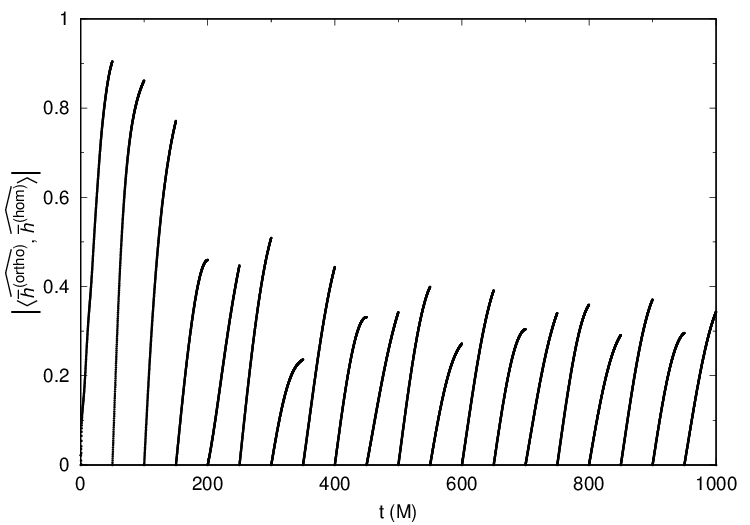}
\end{center}
\caption[Time evolution of unit-vector inner product
	 of $\hbarI_\ortho$ and $\hbarI_\hom$
	 for the \EvolutionName{ppart-ortho-50} evolution.]
	{
	This figure shows the time evolution of the
	unit-vector inner product
	$\InnerProduct{\UnitVector{\hbarI_\ortho}}{\UnitVector{\hbarI_\hom}}$
	for the \EvolutionName{ppart-ortho-50} evolution.
	This inner product grows due to the unstable mode,
	but resets to zero each time $\lambda^\OccasionallyUpdated$
	is updated.  As conjectured in
	section~\protect\ref{sect-theory-generic/ortho/basic-concept},
	this inner product is relatively small at \emph{all} late times.
	}
\label{fig-ppart-ortho-50/uvip-ortho-hom}
\end{figure}

In section~\ref{sect-theory-generic/ortho/basic-concept},
I conjectured that at late times the unit-vector inner product
$\InnerProduct{\UnitVector{h_{ab}^\ortho}}{\UnitVector{h_{ab}^\unstable}}$
should remain small.  While I do not have direct access to
$h_{ab}^\unstable$, I can use $\hbarI_\hom$ as a proxy for
$h_{ab}^\unstable$ at late times.
Figure~\ref{fig-ppart-ortho-50/uvip-ortho-hom} shows the
time evolution of the unit-vector inner product
$\InnerProduct{\UnitVector{\hbarI_\ortho}}{\UnitVector{\hbarI_\hom}}$
for the \EvolutionName{ppart-ortho-50} evolution.  Notice that,
as conjectured
in section~\ref{sect-theory-generic/ortho/basic-concept}
(and in contrast to the non-orthogonalized evolutions shown
in figures~\ref{fig-unstable-mode/snapshots-and-movie}
and~\ref{fig-unstable-mode/norms-of-t}, where the inner product
is ${>}\,0.999$ at late times), at late times this inner product
stays relatively small ($\le 0.4$) at late times, with no secular
growth with time.

The success of the orthogonalization scheme in this evolution
despite sampling the $\lambda^\instantaneous$ oscillations
at irregular phases suggests that the orthogonalization scheme
should also work well for non-periodic particle orbits.


\subsection{Orthogonalized Evolution (Effective Source)}
\label{sect-numerical-tests/esrc-ortho-P4}

I now present a numerical test of the orthogonalization scheme
using an effective-source model for the particle.  Here I also
make use of the particle orbit being periodic, and choose the
orthogonalization time spacing $\Delta t^\ortho$ to integrally divide
the particle orbit period ($\Delta_t^\ortho = P/4$), so that
$\lambda^\OccasionallyUpdated$ is updated at the same set
of orbital phases in each orbit.

\begin{figure}
\begin{center}
\begin{picture}(80,157)
%
%
\put(0,120)
  {
  \begin{picture}(0,0)
  \put(0,34){\textbf{(a)}}
  \put(5,0){\includegraphics[scale=0.90]{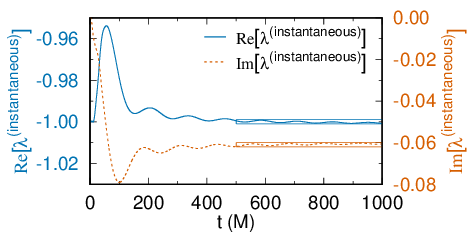}}
  \end{picture}
  }
\put(0,70)
  {
  \begin{picture}(0,0)
  \put(0,45.0){\textbf{(b)}}
  \put(5,0){\includegraphics[scale=0.90]{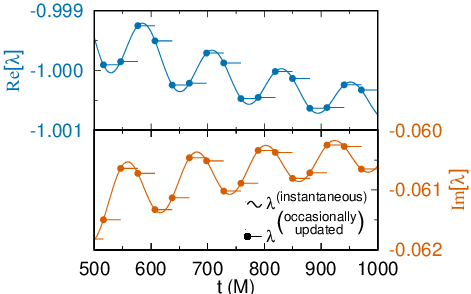}}
  \end{picture}
  }
\put(0,0)
  {
  \begin{picture}(0,0)
  \put(0,64.0){\textbf{(c)}}
  \put(5,0){\includegraphics[scale=0.90]{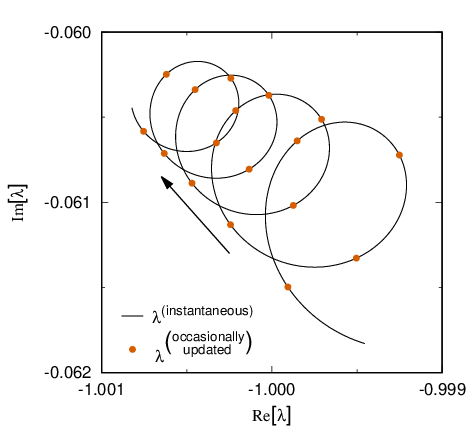}}
  \end{picture}
  }
\end{picture}
\vspace*{-5mm}
\end{center}
\caption[Time evolution of instantaneous and occasionally-updated $\lambda$
	 for the \EvolutionName{esrc-ortho-P4} evolution]
	{
	This figure shows the time evolution of $\lambda^\instantaneous$
	and $\lambda^\OccasionallyUpdated$
	for the \EvolutionName{esrc-ortho-P4} evolution.
	Part~(a) shows the real and imaginary parts of
	$\lambda^\instantaneous$ as functions of time.
	The rectangular regions are shown at an enlarged scale
	in parts~(b) and~(c).
	Part~(b) shows, at an enlarged scale, the real and imaginary parts
	of both $\lambda^\instantaneous$ and $\lambda^\OccasionallyUpdated$
	as functions of time for late times ($t \ge 500M$).
	The solid dots and horizontal lines show the sample-and-hold
	behavior of $\lambda^\OccasionallyUpdated$.
	The legend applies to both real and imaginary parts.
	Part~(c) shows, at an enlarged scale,
	the trajectories in the complex plane
	of both $\lambda^\instantaneous$ (the spiral curve)
	and $\lambda^\OccasionallyUpdated$ (the solid dots)
	for late times ($t \ge 500M$).
	The arrow shows the direction of the time evolution.
	}
\label{fig-esrc-ortho-P4/lambda}
\end{figure}

Figure~\ref{fig-esrc-ortho-P4/lambda} shows the time evolution
of $\lambda^\instantaneous$ and $\lambda^\OccasionallyUpdated$
for the \EvolutionName{esrc-ortho-P4} evolution.
$\lambda^\instantaneous$ shows similar oscillations to those
observed in the point-particle case (figure~\ref{fig-ppart-ortho-50/lambda}).
Because of the orbit-congruent sampling,
$\lambda^\OccasionallyUpdated$ displays more regular ($\approx$~periodic)
behavior than in the \EvolutionName{ppart-ortho-50} evolution
(figure~\ref{fig-ppart-ortho-50/lambda}).

\begin{figure}
\begin{center}
\includegraphics[scale=1.00]{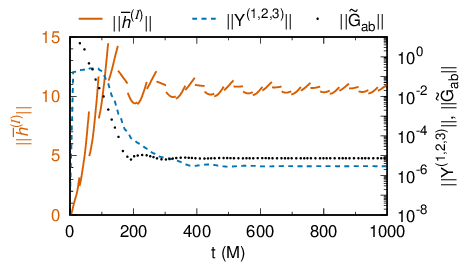}
\end{center}
\caption[Time evolution of norms of the orthogonalized \EvolutionName{esrc-ortho-P4} evolution]
	{
	This figure shows the time evolution of
	norms over grid points and components
	in the \EvolutionName{esrc-ortho-P4} evolution.
	The inner-product norm of $\hbarI_\ortho$
	is plotted on the left scale,
	and the inner-product norm of
	the Lorenz gauge constraints $Y^\nti{1,2,3}$
	and the RMS-norm of the rescaled Einstein tensor $\rescaledG_{ab}$
	are plotted on the right (logarithmic) scale.
	Notice that unlike the non-orthogonalized evolutions
	shown in figure~\protect\ref{fig-unstable-mode/norms-of-t},
	here all the norms remain bounded at late times,
	showing no secular growth with time.
	}
\label{fig-esrc-ortho-P4/norms-of-t}
\end{figure}

Figure~\ref{fig-esrc-ortho-P4/norms-of-t} shows the time evolution
of the norms 
$\Norm{\hbarI_\ortho}$, $\Norm{Y^\nti{1,2,3}}$, and $\Norm{\rescaledG_{ab}}$.  
Notice that at late times, all the norms remain bounded,
showing no secular growth with time.

\begin{figure*}
\begin{center}
\includegraphics[scale=1.00]{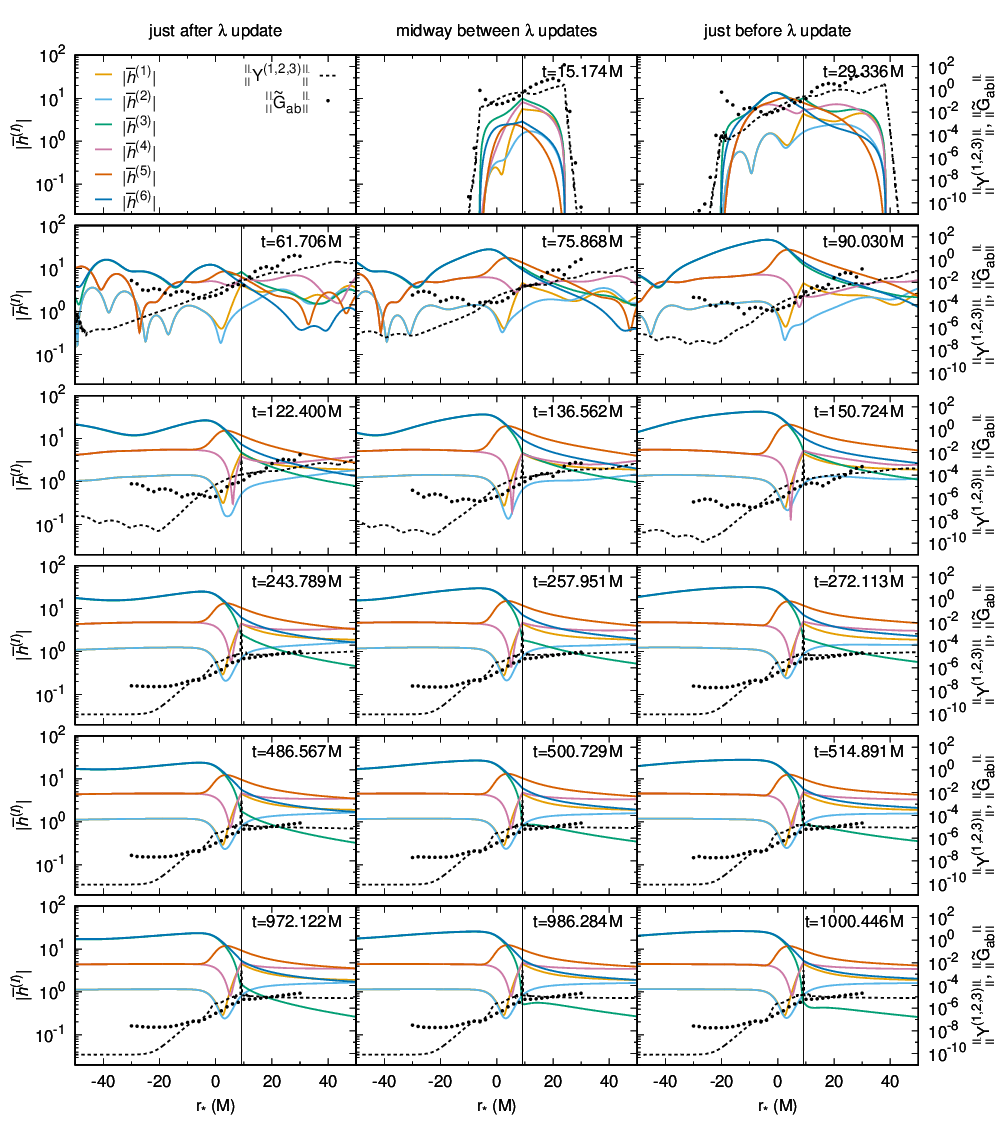}
\end{center}
\caption[Snapshots and movie of the
	 \EvolutionName{esrc-ortho-P4} orthogonalized evolution]
	{
	This figure shows snapshots and a movie
	of the \EvolutionName{esrc-ortho-P4} orthogonalized evolution.
	The same legend applies to all the subplots;
	note that the vertical scale is expanded relative
	to figure~\protect\ref{fig-unstable-mode/snapshots-and-movie}.
	Time runs across each row, then down.
	Within each row, the left subplot is at a time
	just after a $\lambda^\OccasionallyUpdated$ update;
	the center subplot is midway between
	$\lambda^\OccasionallyUpdated$ updates
	(which happen every $\Delta t^\ortho \approx 30.347M$),
	and the right subplot is at a time
	just before the next $\lambda^\OccasionallyUpdated$ update.
	Notice that,
	like
	figure~\protect\ref{fig-ppart-ortho-50/ortho-snapshots-and-movie}
	and unlike
	figure~\protect\ref{fig-unstable-mode/snapshots-and-movie},
	here the $\hbarI$ all remain bounded throughout the evolution.
	In the movie, notice that like the movie of
	figure~\protect\ref{fig-ppart-ortho-50/ortho-snapshots-and-movie}
	and unlike the movie of
	figure~\protect\ref{fig-unstable-mode/snapshots-and-movie},
	here the evolution is \emph{not} dominated by the unstable mode.
	}
\label{fig-esrc-ortho-P4/ortho-snapshots-and-movie}
\end{figure*}

Figure~\ref{fig-esrc-ortho-P4/ortho-snapshots-and-movie} shows snapshots
and a movie of the resulting \EvolutionName{esrc-ortho-P4} orthogonalized
evolution at various times.
These are qualitatively very similar to the snapshots and movie
of the point-particle \EvolutionName{ppart-ortho-50}
orthogonalized snapshots and movie shown in
figure~\ref{fig-ppart-ortho-50/ortho-snapshots-and-movie}.
In particular,
the orthogonalized $\hbarI$ remain bounded throughout the evolution,
and their Lorenz gauge constraints $Y^\nti{1,2,3}$
and rescaled Einstein tensor $\rescaledG_{ab}$
are smsll at late times.
The particle-orbit-period oscillations are clearly visible in the movie;
the evolution is \emph{not} dominated by the unstable mode.

\begin{figure}
\begin{center}
\includegraphics[scale=0.60]{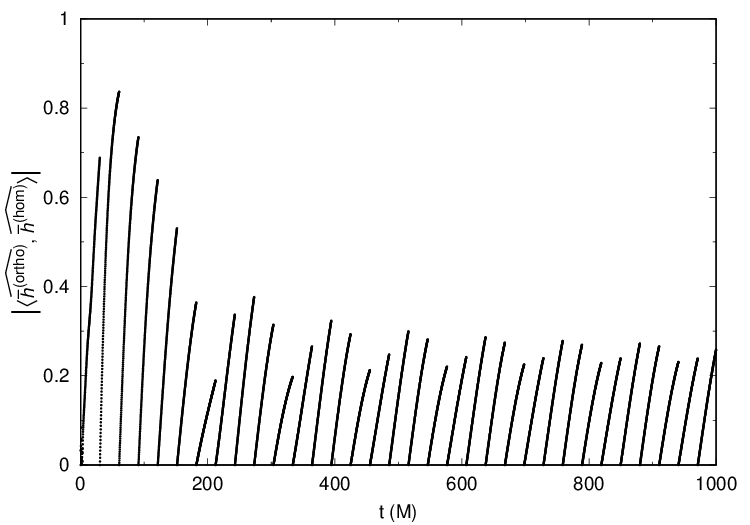}
\end{center}
\caption[Time evolution of unit-vector inner product
	 of $\hbarI_\ortho$ and $\hbarI_\hom$
	 for the \EvolutionName{esrc-ortho-P4} evolution.]
	{
	This figure shows the time evolution of the
	unit-vector inner product
	$\InnerProduct{\UnitVector{\hbarI_\ortho}}{\UnitVector{\hbarI_\hom}}$
	for the \EvolutionName{esrc-ortho-P4} evolution.
	This inner product grows due to the unstable mode,
	but resets to zero each time $\lambda^\OccasionallyUpdated$
	is updated.  As conjectured in
	section~\protect\ref{sect-theory-generic/ortho/basic-concept},
	this inner product is relatively small at \emph{all} late times.
	}
\label{fig-esrc-ortho-P4/uvip-ortho-hom}
\end{figure}

Figure~\ref{fig-esrc-ortho-P4/uvip-ortho-hom} shows the
time evolution of the unit-vector inner product
$\InnerProduct{\UnitVector{\hbarI_\ortho}}{\UnitVector{\hbarI_\hom}}$
for the \EvolutionName{esrc-ortho-P4} evolution.  This shows very
similar behavior to that shown in
figure~\ref{fig-ppart-ortho-50/uvip-ortho-hom}
for the \EvolutionName{ppart-ortho-50} evolution,
with the inner product being relatively small at \emph{all} late times.

\section{Discussion and Conclusions}
\label{sect-discussion+conclusions}

In this paper I present and test a scheme for mostly eliminating
the $\ell \seq m \seq 1$ gauge instability in time-domain Lorenz-gauge
evolutions of small metric perturbations of a background
(typically Schwarzschild or Kerr) spacetime.
\footnote{
	 Here I consider only 1st-order metric perturbations,
	 i.e., I am considering (only) evolution and gauge
	 effects which are linear in the metric perturbations.
	 }

My scheme is based on \emph{orthogonalizing} the metric perturbation
with respect to an auxiliary homogeneous metric-perturbation
evolved in parallel with the main (sourced) evolution.
That is, given an (unstable) sourced metric perturbation $h_{ab}$,
I define the orthogonalized metric perturbation
$h_{ab}^\ortho := h_{ab} + \lambda h_{ab}^\hom$,
where $h_{ab}^\hom$ is the (unstable) auxiliary homogeneous metric perturbation
and the scalar $\lambda$ is chosen so that
$h_{ab}^\ortho \perp h_{ab}^\hom$
with respect to a specified inner product on metric perturbations.

If $\lambda$ were updated continuously
(i.e., at each time step of a numerical evolution),
then $\partial_t h_{ab}^\ortho$ and $\partial_{tt} h_{ab}^\ortho$
would contain $\partial_t \lambda$ and $\partial_{tt} \lambda$ terms,
causing $h_{ab}^\ortho$ to violate the
perturbed Einstein equations and the Lorenz gauge conditions.
Therefore, my scheme instead uses a \emph{piecewise-constant-in-time} $\lambda$
which is only updated ``occasionally'' (typically every $10M$ to $50M$
\footnote{
	 $M$ is the mass of the background Schwarzshild
	 or Kerr spacetime.
	 }
) during the evolution.
At late times in the evolution
(i.e., at times when $h_{ab}^\hom$ is dominated by the unstable mode),
the resulting $h_{ab}^\ortho$ satisfies
the perturbed Einstein equations and the Lorenz gauge conditions
during each time interval where $\lambda$ is constant.
$h_{ab}^\ortho$ has a jump discontinuity each time $\lambda$ is
updated; at late times in the evolution this jump should be purely
a gauge transformation (within the family of Lorenz gauges).

I have tested the orthogonalization scheme for a Schwarzschild
background spacetime, using both point-particle and
``puncture'' (effective-source) sources for the metric perturbation.
My choice for the orthogonalization inner product is a spatial
integral of the usual Euclidean inner product of the Barak-Lousto-Sago
metric-perturbation variables.

I find that the orthogonalization scheme works well, in that the
orthogonalized metric perturbation satisfies the perturbed Einstein equations
and the Lorenz gauge conditions (up to finite-differencing accuracy),
and contains only a small component of the unstable gauge mode;
this ``small component'' can be made smaller by updating the
orthogonalization more frequently.

To make this research more accessible,
the full source code (about 60K~lines of \Cplusplus{} and 2K~lines of Sage)
and parameter files
used to generate all the data in this manuscript
are included as online supplemental material.
I am also submitting the code and parameter files
to the Black Hole Perturbation Toolkit~(\textcite{BHPToolkit}).


\subsection{Possible Extensions to this Work}
\label{sect-discussion+conclusions/possible-extensions}

There are a number of directions in which this work could be extended.

As discussed in appendix~\ref{app-convergence-tests}, my present
numerical scheme shows fairly good convergence to a continuum limit
for the orthogonalized metric perturbation (more precisely the
orthogonalized Barack-Lousto-Sago metric-perturbation fields $\hbarI_\ortho$),
but the convergence is poorer for the
independently-computed Lorenz gauge constraints and rescaled
Einstein tensor.  Changes to the numerical scheme to improve
this convergence would be very valuable.

My choices for inner product and orthogonalization update times
seem reasonable, but many other choices are possible.  For example,
footnotes~\ref{footnote-other-inner-product-interval}
and~\ref{footnote-inner-product-gradual-cutoff}
suggest possible variants of the inner product.
Alternatively, the ideas of
\textcite{Green-etal-Kerr-conserved-currents-2023}
might offer a route to constructing a geometrically-based
inner product for metric perturbations of Kerr spacetime.

It would be useful to test the orthogonalization scheme with more
general metric perturbations, notably those sourced by a small body
moving on an eccentric orbit in Schwarzschild spacetime (my numerical tests
here consider only circular orbits), and/or source-free evolutions.

If the extra $\partial_t \lambda$ and $\partial_{tt} \lambda$ terms
in $\partial_t h_{ab}^\ortho$ and $\partial_{tt} h_{ab}^\ortho$
could somehow be cancelled or subtracted out, then the orthogonalization
could be updated continuously, allowing an orthogonalized evolution
completely free of the unstable mode (up to numerical accuracy).

My present numerical evolution scheme is relatively simple.
I expect that the orthogonalization scheme would work equally well
with more sophisticated numerical evolution schemes (e.g., adaptive
mesh refinement, spectral, and/or discontinuous Galerkin methods),
and/or using hyperboloidal compactification at the horizon and $\Scri^+$,
but it would be useful to explicitly verify this.

The numerical results presented here establish that $h_{ab}^\ortho$ is
in the Lorenz gauge at all late times, including both immediately before
and immediately after re-orthogonalizations.
In section~\ref{sect-theory-generic/ortho/physical-meaning}
I argue that at late times,
the change in $h_{ab}^\ortho$ when re-orthogonalizing
should be \emph{only} a (Lorenz) gauge transformation, i.e.,
that $h_{ab}^\ortho$ immediately after a late-time re-orthogonalization
represents the \emph{same} physical $t=\constant$ hypersurface
as $h_{ab}^\ortho$ immediately before (the same) re-orthogonalization.
However, I have not explicitly verified this numerically.
It would be valuable to do this, i.e.,
to explicitly construct the late-time re-orthogonalization gauge
transformation and demonstrate that the change in $h_{ab}^\ortho$
induced by a late-time re-orthogonalization is indeed purely a (Lorenz)
gauge transformation.

The present work only calculates metric perturbations (or more
precisely the Barack-Lousto-Sago variables $\hbarI$); it would be
interesting to extend this to calculating the emitted gravitational waves,
and then investigate whether or not late-time re-orthogonalizations
significantly perturbs the gravitational waves and/or other gauge invariants.

And, of course, it would be very interesting to apply the orthogonalization
scheme to metric perturbations on a Kerr background; this could provide
a route to calculating (1st-order) radiation-reaction effects for
highly-eccentric EMRIs.


\section{Acknowledgements}

It is a pleasure to thank some of the many people who have contributed
to this research project.

I thank S.~Dolan
for many informative discussions,
for hosting my research visits in 2014 and 2016,
for deriving the ``$\TT\XX\VV$'' equations described
in appendix~\ref{app-point-particle-jump-conditions/general-formalism}
for calculating jumps in the Barack-Lousto-Sago fields across a
point particle,
and for sharing unpublished research results on the Barack-Lousto-Sago
evolution system.

I thank L.~Barack
for many informative discussions,
and for his generosity in answering various questions
about details of the BL05 evolution scheme.
I thank
M.~Colleoni for valuable assistance with code comparisons
to validate my implementation of the BL05 evolution system.

I thank B.~Wardell for his
\LorenzGaugeSeff{} open-source effective source code
(\textcite{Wardell-LorenzGauge1DEffectiveSource}),
and for many informative discussions about effective-source schemes.

I thank W.~Stein for initiating and leading the development of
the open-source \Software{Sage} symbolic computation system,
and E.~Gourgoulhon for developing and making available the open-source
\Software{SageManifolds} software for tensor calculus and other computations.

I thank A.~Buonanno and the Max-Planck-Institut f\"{u}r Gravitationsphysik
for financial support and for hosting a research visit in 2018,
and
the Alexander von Humboldt Foundation
and the Indiana University
Center for Spacetime Symmetries,
Department of Astronomy, and
Office of the Vice-Provost for Research
for financial support.

\appendix

\section{Point-Particle Jump Conditions}
\label{app-point-particle-jump-conditions}

In this appendix I describe my scheme for calculating the jumps
in $\hbarI$ across the particle.


\subsection{General Formalism}
\label{app-point-particle-jump-conditions/general-formalism}

Slightly generalizing the BL05 evolution equations,
consider a $1{+}1$-dimensional linear wave equation
with a $\delta$-function source,
\begin{equation}
-\partial_{tt} \uu + \partial_{xx} \uu
+ \TT(x) \partial_t \uu + \XX(x) \partial_x \uu + \VV(x) \uu
	= \ss(t)
									~ ,
						       \label{eqn-TXV-evolution}
\end{equation}
where for the remainder of this appendix only,
$x := r_*$,
$N > 0$ is an integer,
$\ss(t)$ is an $N$-element column vector,
$\TT(x)$, $\XX(x)$, and $\VV(x)$ are
$N{\times}N$~coefficient matrices,
and for any variable or expression $q$,
$q' := \partial_x q$
and $[q]_p$ is the jump in $q$ across the particle.

Integrating~\eqref{eqn-TXV-evolution} from $x_p - \epsilon$ to $x_p + \epsilon$
and taking the limit $\epsilon \rightarrow 0$ gives the jump in
$\partial_x \uu$ as
\begin{equation}
\left[ \uu' \right]_p
	= \lim_{\epsilon \rightarrow 0}
	  \bigl\{ \uu'(x_p+\epsilon) - \uu'(x_p-\epsilon) \bigr\}
	= \ss(t)
									~ .
							     \label{eqn-jump-u'}
\end{equation}

To find the jump in $\partial_{xx} \uu$,
take the difference of~\eqref{eqn-TXV-evolution} evaluated at
$x_p + \epsilon$ and $x_p - \epsilon$,
then consider the limit of this difference as $\epsilon \to 0$.
This gives
\begin{equation}
[ \uu'' ]_p = - \XX(x_p) [\uu']_p = - \XX_p \ss(t)
									~ .
							    \label{eqn-jump-u''}
\end{equation}

To find the jump in $\partial_{xxx} \uu$,
take an $x$~derivative of~\eqref{eqn-TXV-evolution}
and then repeat the above procedure.
This yields
\begin{align}
-\partial_{tt} [\uu']_p	& + [\uu''']_p
								\nonumber\\
			& + \XX_p' [\uu']_p + \XX_p [\uu'']_p
			  + \TT_p \partial_t [\uu']_p
			  + \VV_p [\uu']_p
								\nonumber\\
			& \qquad\qquad
				= 0
									~ ,
\end{align}
so that
\begin{equation}
[\uu''']_p = \partial_{tt} \ss
	     - \TT_p \partial_t \ss
	     - (\XX_p' - \XX_p^2 + \VV_p) \ss
									~ .
							   \label{eqn-jump-u'''}
\end{equation}
(If desired, jumps in higher-order derivatives can be found by
iterating this procedure.)


\subsection{Application to the $\hbarIellm$ evolution equations}
\label{app-point-particle-jump-conditions/application-to-hbarIellm}

To apply the above formalism to
the BL05 evolution equations~\eqref{eqn-hbarI-evolution},
take $N$ to be the number of nontrivial~$\hbarIellm$
($N \seq 6$ for the $\ell \seq m \seq 1$ case which is my focus here),
$\uu$ to be an $N$-element column vector of those nontrivial~$\hbarIellm$,
and take $\ss(t)$ to be an $N$-element column vector
of the nontrivial components of
the coefficient of $\delta(r - r_0)$ in BL05's equation~(29),
multiplied by~$4$
(because the source term in~\eqref{eqn-hbarI-evolution} is $4\Sterm^\I$).

The $\TT$, $\XX$, and $\VV$ matrices together encode
the right hand side of
the $\hbarIellm$ evolution equation~\eqref{eqn-hbarI-evolution},
\begin{subequations}
\begin{align}
\TT_\nti{IJ}
	& = \frac{\partial \, \RHS^\vacuum_\nti{I}(\uu,
						   \partial_{r_*} \uu,
						   \partial_{r_*r_*} \uu,
						   \partial_t \uu)}
		 {\partial \bigl( (\partial_t \uu)_\nti{J} \bigr)}
									\\
\XX_\nti{IJ}
	& = \frac{\partial \, \RHS^\vacuum_\nti{I}(\uu,
						   \partial_{r_*} \uu,
						   \partial_{r_*r_*} \uu,
						   \partial_t \uu)}
		 {\partial \bigl( (\partial_{r_*} \uu)_\nti{J} \bigr)}
									\\
\VV_\nti{IJ}
	& = \frac{\partial \, \RHS^\vacuum_\nti{I}(\uu,
						   \partial_{r_*} \uu,
						   \partial_{r_*r_*} \uu,
						   \partial_t \uu)}
		 {\partial \uu_\nti{J}}
									~ ,
\end{align}
\end{subequations}
where $\RHS^\vacuum$ is regarded as a function of the independent variables
$\uu$, $\partial_{r_*} \uu$, $\partial_{r_*r_*} \uu$, and $\partial_t \uu$.

From a software-engineering perspective, it's desirable to have
$\RHS^\vacuum$, and in particular the somewhat-complicated $\M$~operator,
defined in only a single place in the software
(this is often called the ``don't repeat yourself'' rule,
widely popularized by \textcite{Pragmatic-Programmer-book}).
To this end, my numerical code computes the $\TT$, $\XX$, and $\VV$ matrices
by directly finite differencing the $\RHS^\vacuum$ function.
Because $\RHS^\vacuum$ is \emph{linear} in $\hbarIellm$,
$\partial_{r_*} \hbarIellm$, $\partial_{r_*r_*} \hbarIellm$,
and $\partial_t \hbarIellm$,
this finite differencing does \emph{not} need to use a small step size
for accuracy.
Instead, the $\TT$, $\XX$, and $\VV$ matrices can be computed exactly
\footnote{
	 ``Exactly'' here means modulo floating-point rounding
	 errors.
	 }
{}
(albeit slightly inefficiently) via
\begin{widetext}
\begin{subequations}
							  \label{eqn-TXV-via-FD}
\begin{align}
\TT_\nti{IJ}
	& =   \RHS^\vacuum_\nti{I}(\ZZ,\ZZ,\ZZ,\dd_\nti{J})
	    - \RHS^\vacuum_\nti{I}(\ZZ,\ZZ,\ZZ,\ZZ  )
									\\
\XX_\nti{IJ}
	& =   \RHS^\vacuum_\nti{I}(\ZZ,\dd_\nti{J},\ZZ,\ZZ)
	    - \RHS^\vacuum_\nti{I}(\ZZ,\ZZ  ,\ZZ,\ZZ)
									\\
\VV_\nti{IJ}
	& =   \RHS^\vacuum_\nti{I}(\dd_\nti{J},\ZZ,\ZZ,\ZZ)
	    - \RHS^\vacuum_\nti{I}(\ZZ  ,\ZZ,\ZZ,\ZZ)
									~ ,
\end{align}
\end{subequations}
\end{widetext}
where $\ZZ$ is the $N$-element zero vector,
and $\dd_\nti{J}$ is the $N$-element Kronecker-delta vector
\begin{equation}
\bigl.(\dd_\nti{J})\bigr._\nti{K}
	 = \begin{cases}
	   1	& \text{if $J = K$}	\\
	   0	& \text{if $J \ne K$}	
	   \end{cases}
									~ .
\end{equation}
Since $\TT$, $\XX$, and $\VV$ are only needed at one spatial position
(the particle position),
the actual computational cost of the $3N{+}1$ $\RHS^\vacuum$ evaluations
needed to compute all components of $\TT_p$, $\XX_p$, and $\VV_p$
via~\eqref{eqn-TXV-via-FD} is minor compared to the overall cost
of the evolution.

In my implementation the numerical grid is chosen such that the
particle position does \emph{not} coincide with any grid point.
This implies that the source term vanishes at every grid point;
the source appears only through the jumps across the particle.

This scheme does have one other slightly awkward requirement,
namely that $\RHS^\vacuum$ must be evaluated at a non-grid point
(the particle position).
This means that $\RHS^\vacuum$ must be evaluated \emph{without}
using any coefficients precomputed or cached at the grid points.
I found that this requires a certain amount of special programming,
but isn't too difficult in practice.

Given $\TT_p$, $\XX_p$, $\VV_p$, and $\ss(t)$,
the jumps in $\partial_{r_*} \hbarIellm$, $\partial_{r_*r_*} \hbarIellm$,
and $\partial_{r_*r_*r_*} \hbarIellm$ across the particle position
are then given by~\eqref{eqn-jump-u'}, \eqref{eqn-jump-u''},
and~\eqref{eqn-jump-u'''} respectively.


\section{Numerical Methods}
\label{app-numerical}

The focus of this work is on testing and exploring the behavior of the
orthogonalization scheme, not on developing a highly efficient code.
Therefore, I use a relatively simple finite-differencing numerical scheme.


\subsection{\allbf{Computing $r(r_*)$}}
\label{app-numerical/r(rstar)}

Since I use a numerical grid uniform in~$r_*$,
computing various coefficients at the grid points requires knowing
the $r$ coordinates of the grid points.
I compute these by numerically inverting
the tortise-coordinate definition~\eqref{eqn-r_*(r)}.
To do this, I define
\begin{align}
y	& = \ln \left( \frac{r}{2M} - 1 \right)				\\
x_*	& = \frac{r_*}{2M}
									~ ,
\end{align}
so that $r_* = r + 2M y$ and $r = 2M \left( 1 + e^y \right)$.
\eqref{eqn-r_*(r)} then becomes
\begin{equation}
x_* = 1 + y + e^y
									~ .
							      \label{eqn-x_*(y)}
\end{equation}
I numerically solve~\eqref{eqn-x_*(y)} by Newton's method,
finding a zero of the function
\begin{equation}
q(y) = 1 + y + e^y - x_*
									~ .
								\label{eqn-q(y)}
\end{equation}
The initial guess is based on neglecting either $y$ or $e^y$
in~\eqref{eqn-q(y)}, giving
\begin{equation}
y_\initial
	 = \begin{cases}
	   \log(x_* - 1)& \text{if $x_* >   1$ ($y \gtsim -0.577$)}	\\
	   x_* - 1	& \text{if $x_* \le 1$ ($y \ltsim -0.577$)}	
    \end{cases}
\end{equation}

The Newton's-method solution is moderately expensive (it typically
requires $3$--$10$ iterations at each grid point, with each iteration
needing a couple of transcendental functions), so my code keeps a
cache of the radia of all the grid points.


\subsection{Notation for the remainder of this Appendix}
\label{app-numerical/notation}

Throughout the remainder of this appendix,
I suppress the indices ${}_{ab}^\Iellm$,
and I use a lower-case Latin typewriter-font~$\ii$ to index grid functions.

When describing finite-difference molecules (stencils), I use
a lower-case Latin typewriter-font~$\mm$ to index the molecule coefficients,
so that a generic finite-difference operator~$\op$ is defined by
\begin{equation}
\bigl(\op(q)\bigr)_\ii
	= \sum_{\mm=\mm_{\min}}^{\mm_{\max}} \op_\mm \, q_{\ii{+}\mm}
									~ ,
							      \label{eqn-FD-sum}
\end{equation}
where $\Delta r_*$ is the grid spacing,
$q$ is the grid function being finite-differenced,
and the molecule has nonzero coefficients (only)
in the interval $\mm \in [\mm_{\min}, \mm_{\max}]$.
I refer to the molecule as having
the evaluation point~$\ii$
and (the set of) input points
$\bigl\{\ii{+}\mm \,\big|\, \mm \in [\mm_{\min}, \mm_{\max}]\bigr\}$.

I write molecule coefficients as a small matrix,
with the coefficient at the evaluation point $\mm = 0$ underlined.
For example, I write the usual centered 4th-order $\partial_{r_*}$
finite difference molecule as
\begin{equation}
\ddd =	\frac{1}{12 \, \Delta r_*}
	\Bigl[
	\begin{array}{ccccc}
	+1	& -8 	& \U{\,0\,}	& +8	& -1	\\
	\end{array}
	\Bigr]
									~ ,
\end{equation}
where here $m_{\min} \seq -2$ and $m_{\max} \seq +2$.

I refer to a finite-difference molecule as
occupying the $r_*$~interval defined by its input points, i.e.,
\begin{equation}
\bigl[ (r_*)_\ii + m_{\min}\,\Delta r_*, (r_*)_\ii + m_{\max}\,\Delta r_* \bigr]
									~ .
\end{equation}
I say that a finite-difference molecule (operator)
``crosses the particle''
if and only if the particle is contained in that interval,
i.e., if and only if the sum~\eqref{eqn-FD-sum} includes both
terms with $(r_*)_{\ii{+}\mm} < (r_*)_p$
and terms with $(r_*)_{\ii{+}\mm} > (r_*)_p$.
When using an effective-source scheme with a worldtube,
I say that a finite-difference molecule (operator)
``crosses the worldtube boundary''
if and only if the sum~\eqref{eqn-FD-sum} includes both
terms with $\ii{+}\mm$ inside the worldtube
and terms with $\ii{+}\mm$ outside the worldtube.


\subsection{Main finite differencing}

For each $(\ell,m)$, I use a 4th-order Runge-Kutta method of lines scheme to
time-integrate of the $\hbarI$ evolution equations~\eqref{eqn-hbarI-evolution},
written in the 1st-order-in-time form
\begin{equation}
\partial_t \bigl( \hbarI, \partial_t \hbarI \bigr)
	= \bigl(
	  \partial_t \hbarI, \RHS^\total(\hbarI, \partial_t \hbarI)
	  \bigr)
									~ .
\end{equation}

I use 4th~order centered finite differencing on a uniform-in-$r_*$ grid
to approximate the spatial derivatives in the $\RHS^\vacuum$
definition~\eqref{eqn-RHS-vacuum} and the $\M$~operator.


\subsection{Dissipation}
\label{app-numerical/dissipation}

To reduce numerical noise, I add an additional dissipation term
$\epsilon \Diss(\hbar)$ to the right hand side of~\eqref{eqn-RHS-vacuum}
at selected grid points, where
the dissipation operator $\Diss$ is the 6th-order Kreiss-Oliger
dissipation operator (\textcite[appendix~C]{Rinne-2010})
with finite-difference molecule
\begin{equation}
\Diss = \frac{1}{64 \, \Delta r_*}
	\Bigl[
	\begin{array}{ccccccc}
	+1	& -6 	& +15 	& \U{-20}& +15 	& -6 	& +1	\\
	\end{array}
	\Bigr]
									~ .
							 \label{eqn-dissipation}
\end{equation}
All the results presented here use the dissipation coefficient
$\epsilon = 0.1$.

I add the dissipation term only at those grid points with
$r_* \in [0, 15M]$ which also satisfy the following condition:
\begin{itemize}
\item	For a point-particle scheme, the dissipation molecule
	must not cross the particle.
\item	For an effective-source scheme, the dissipation molecule
	must not intersect the worldtube.
\end{itemize}


\subsection{Source Terms}
\label{app-numerical/source}

Despite the major conceptual differences between the point-particle
and effective-source particle models, their implementations at the level
of finite differencing are actually mostly similar.

In both cases, for each finite-differencing operation at a grid point~$\ii$,
the code first tests
\begin{itemize}
\item	for a point-particle model, whether or not
	the finite difference molecule would cross the particle; or,
\item	for an effective-source model, whether or not
	the finite difference molecule would cross the worldtube boundary.
\end{itemize}
In each case, this is (or can be) a very quick test, requiring only
testing whether or not the grid-point index~$\ii$ is contained in a
previously-computed interval or pair of intervals of grid-point indices.

If this test finds that the molecule would \emph{not} cross the particle
or worldtube boundary, then the code does the usual finite differencing
operation on $\hbar$ with no further attention needed to the
source terms.
\footnote{
	 For a point particle, I choose the grid so that
	 the particle is not at a grid point.
	 This implies that at every grid point, the $\delta$-function
	 source term in BL05's equation~(29) vanishes.
	 }
$^,$
\footnote{
	 For the effective source used here, $\hbar^\numerical$
	 is only $C^2$ at the particle position, so it would be
	 more accurate -- and would quite likely improve
	 convergence (appendix~\protect\ref{app-convergence-tests})
	 -- to special-case finite differencing operations
	 which cross the particle, perhaps in the manner described
 in~\textcite[appendix B.10]{Thornburg-Wardell-2017:Kerr-scalar-self-force}.
	 I haven't (yet) made this improvement.
	 }

If this test finds that the molecule \emph{would} cross the particle
or worldtube boundary, then the code instead finite-differences an
``adjusted'' temporary copy of a molecule-sized region of the
$\hbar$ grid function, where the ``adjustment'' corrects for
the source model:
\begin{itemize}
\item	for a point-particle model, the adjustment (described in
	detail in appendix~\ref{app-numerical/source/adjusted-FD-ppart})
	corrects for the jumps in spatial derivatives of $\hbar$
	across the particle;
\item	for an effective-source model, the adjustment (described in
	detail in appendix~\ref{app-numerical/source/adjusted-FD-esrc})
	corrects for the jump~\eqref{eqn-hbar-ab-numerical-jump}
	in~$\hbar_{ab}^\numerical$ across the
	worldtube boundary.
\end{itemize}
Because the ``adjustment'' is only done at a few grid points in each
slice, the overall computational cost of this scheme is quite low.


\subsubsection{Adjusted Finite Differencing: Point Particle}
\label{app-numerical/source/adjusted-FD-ppart}

For a finite difference molecule which would cross the particle,
the ``adjustment'' corrects for the jumps in spatial derivatives of
$\hbar$ across the particle by making each point of the temporary copy
match the inside/outside-the-particle semantics of the evaluation point,
\begin{subequations}
						   \label{eqn-adjusted-FD/ppart}
\begin{widetext}
\begin{equation}
\hbar^\adjusted_{\ii{+}\mm}
	= \begin{cases}
	  \hbar_{\ii{+}\mm}
		& \text{\begin{tabular}[t]{@{}l@{}}
			if $(r_*)_\ii < (r_*)_p$
			and $(r_*)_{\ii{+}\mm} < (r_*)_p$	\\
			or if $(r_*)_\ii > (r_*)_p$
			and $(r_*)_{\ii{+}\mm} > (r_*)_p$	
			\end{tabular}}
									\\[4ex]
	  \hbar_{\ii{+}\mm} - J
		& \text{if $(r_*)_\ii < (r_*)_p$
			and $(r_*)_{\ii{+}\mm} > (r_*)_p$}
									\\[1ex]
	  \hbar_{\ii{+}\mm} + J
		& \text{if $(r_*)_\ii > (r_*)_p$
			and $(r_*)_{\ii{+}\mm} < (r_*)_p$}
	  \end{cases}
									~ ,
\end{equation}
\end{widetext}
where $J$ is obtained by Taylor-expanding the jump in $\hbar$
in the distance~$d$ from the particle,
\begin{equation}
J = J_{r_*} d + \frac{1}{2!} J_{r_*r_*} d^2 + \frac{1}{3!} J_{r_*r_*r_*} d^3
    + \O(d^4)
									~ ,
						   \label{eqn-ppart-jump-series}
\end{equation}
\end{subequations}
where $d = (r_*)_{\ii+\mm} - (r_*)_p$
and where $J_{r_*}$, $J_{r_*r_*}$, and $J_{r_*r_*r_*}$ are the jumps
in $\partial_{r_*} \hbar$, $\partial_{r_*r_*} \hbar$,
and $\partial_{r_*r_*r_*} \hbar$ respectively.

\begin{figure}
\begin{center}
\includegraphics[scale=0.85]{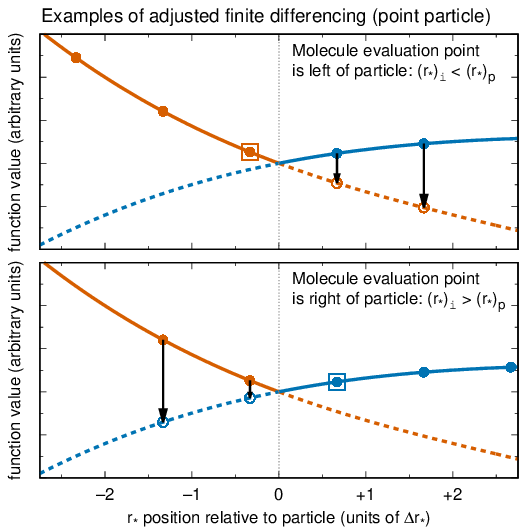}
\end{center}
\caption[Examples of adjusted finite differencing (point particle)]
	{
	This figure shows two examples of adjusted finite differencing
	using the scheme~\protect\eqref{eqn-adjusted-FD/ppart}
	for the case of a point particle.
	In each panel, a sample $\hbarIellm$ is plotted as a solid line
	(red for $r_* \slt (r_*)_p$ and blue for $r_* \sgt (r_*)_p$),
	with the grid-point values shown as filled circles.
	This sample $\hbarIellm$ has jump discontinuities in each
	of its spatial derivatives across the particle position
	(marked by a dashed vertical line).
	In each panel, the molecule-evaluation point is marked by
	a box around the filled circle.
\pseudopar
	The upper panel shows an example where the molecule evaluation point
	is to the left of the particle ($(r_*)_\ii \slt (r_*)_p$).
	In this case, the adjustment (vertical arrows)
	is applied to the molecule input points
	to the right of the particle ($(r_*)_{\ii{+}\mm} \sgt (r_*)_p$),
	so that when combined with the molecule input points
	to the left of the particle (solid red circles)
	the adjusted $\hbarIellm$ values (open red circles)
	fall on a smooth curve (red curve, both solid and dashed)
	for finite differencing.
\pseudopar
	The lower panel shows an example where the molecule evaluation point
	is to the right of the particle ($(r_*)_\ii \sgt (r_*)_p$).
	In this case, the adjustment (vertical arrows)
	is applied to the molecule input points
	to the left of the particle ($(r_*)_{\ii{+}\mm} \slt (r_*)_p$),
	so that when combined with the molecule input points
	to the right of the particle (solid blue circles),
	the adjusted $\hbarIellm$ values (open blue circles)
	again fall on a smooth curve (blue curve, both solid and dashed)
	for finite differencing.
	}
\label{fig-adjusted-FD/point-particle}
\end{figure}

Because the evolution equation~\eqref{eqn-hbarI-evolution}
contains $\partial_{r_*r_*}$, this scheme is only
$\O\bigl(\Delta r_*\bigr)^2$~accurate,
but that's sufficient for present purposes.
If a higher order of accuracy were desired,
I describe in
section~\ref{sect-Schw/modelling-particle/ppart}
how jumps in higher derivatives could be calculated,
allowing the jump series~\eqref{eqn-ppart-jump-series} to be extended
to higher orders of the distance~$d$.


\subsubsection{Adjusted Finite Differencing: Effective Source}
\label{app-numerical/source/adjusted-FD-esrc}

For a finite difference molecule which would cross the particle,
the ``adjustment'' corrects for the jump~\eqref{eqn-hbar-ab-numerical-jump}
in $\hbar^\numerical$ across the worldtube boundary,
\begin{widetext}
\begin{equation}
\hbar^\adjusted_{\ii{+}\mm}
	= \begin{cases}
	  \hbar_{\ii{+}\mm}^\numerical
		& \text{\begin{tabular}[t]{@{}l@{}}
			if $\ii \in \WW$ and $\ii{+}\mm \in \WW$
								\\
			or if $\ii \not\in \WW$ and $\ii{+}\mm \not\in \WW$
			\end{tabular}}
									\\[4ex]
	  \hbar_{\ii{+}\mm}^\numerical + \hbar^\puncture_{\ii{+}\mm}
		& \text{if $\ii \in \WW$ and $\ii{+}\mm \not\in \WW$}
									\\[1ex]
	  \hbar_{\ii{+}\mm}^\numerical - \hbar^\puncture_{\ii{+}\mm}
		& \text{if $\ii \not\in \WW$ and $\ii{+}\mm \in \WW$}
	  \end{cases}
									~ ,
						    \label{eqn-adjusted-FD-esrc}
\end{equation}
\end{widetext}
where $\WW$ is the set of grid-point indices in the worldtube.


\subsection{Boundary Conditions}
\label{app-numerical/BCs}

Since the goal of the present work is solely to test the
orthogonalization scheme, rather than to construct an efficient
production code, I choose the spatial domain to be sufficiently
large that the spatial grid boundaries are causally disconnected
with the inner-product region.
(Table~\ref{tab-common-pars} gives the spatial grid boundary positions
for each evolution.)
This scheme is very inefficient
(it evolves many grid points -- in fact the vast majority of them --
which are outside the inner-product region),
but it avoids any concern about the boundary conditions
possibly affecting my tests of the orthogonalization scheme.

Since the actual physical boundary condition
doesn't affect the evolution within the inner-product region,
I use an arbitrary boundary condition at the actual physical boundaries:
before each evaluation of the right-hand-side function,
each state-vector spatial ghost zone is zeroed.


\section{Tests of Variants of the Orthogonalization Scheme}
\label{app-ortho-variants}

In this appendix I present numerical tests of four variants
of the basic orthogonalization scheme:
using a shorter orthogonalization time spacing,
orbit-averaging~$\lambda$,
gradual turnon of the puncture and effective source,
and using a fixed (time-independent)~$\lambda$.


\subsection{Shorter Orthogonalization Time Spacing}
\label{app-ortho-variants/short-Delta-t-ortho}

To show the effect of changing
the orthogonalization time spacing $\Delta t^\ortho$,
here I briefly present the \EvolutionName{esrc-ortho-P12} evolution,
which is identical to the \EvolutionName{esrc-ortho-P4} evolution
presented in section~\ref{sect-numerical-tests/esrc-ortho-P4}
except that the orthogonalization time spacing is a factor of~$3$
shorter: $P/12 \approx 10.116M$ instead of $P/4 \approx 30.347M$.

\begin{figure}
\begin{center}
\includegraphics[scale=0.60]{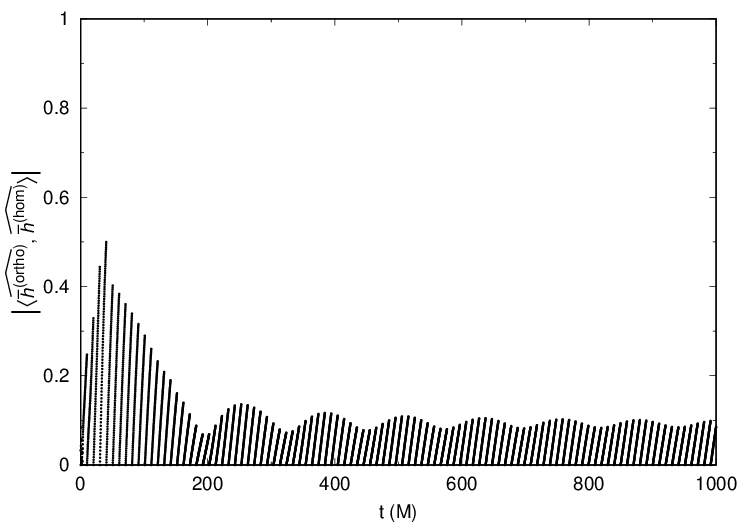}
\end{center}
\caption[Time evolution of unit-vector inner product
	 of $\hbarI_\ortho$ and $\hbarI_\hom$
	 for the \EvolutionName{esrc-ortho-P12} evolution.]
	{
	This figure shows the time evolution of the
	unit-vector inner product
	$\InnerProduct{\UnitVector{\hbarI_\ortho}}{\UnitVector{\hbarI_\hom}}$
	for the \EvolutionName{esrc-ortho-P12} evolution,
	where $\lambda^\OccasionallyUpdated$ is updated
	every $P/12 \approx 10.116M$.
	The scale is the same as that of
	figure~\ref{fig-esrc-ortho-P4/norms-of-t}
	(which shows the \EvolutionName{esrc-ortho-P4} evolution).
	In comparison to that evolution, here the unit-vector inner product
	is much smaller ($\ltsim 0.1$) at late times.
	}
\label{fig-esrc-ortho-P12/uvip-ortho-hom}
\end{figure}

Figure~\ref{fig-esrc-ortho-P12/uvip-ortho-hom} shows the
time evolution of the unit-vector inner product
$\InnerProduct{\UnitVector{\hbarI_\ortho}}{\UnitVector{\hbarI_\hom}}$
for the \EvolutionName{esrc-ortho-P12} evolution.
Because of the frequent orthogonalization updates,
at late times the unit-vector inner product is very small ($\ltsim 0.1$).
That is, at late times the \EvolutionName{esrc-ortho-P12} evolution
contains only a small component of the unstable mode.

\begin{figure}
\begin{center}
\includegraphics[scale=1.00]{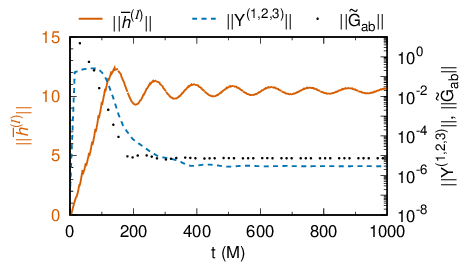}
\end{center}
\caption[Time evolution of norms of the orthogonalized \EvolutionName{esrc-ortho-P12} evolution]
	{
	This figure shows the time evolution of
	norms over grid points and components
	in the \EvolutionName{esrc-ortho-P12} evolution.
	The inner-product norm of $\hbarI_\ortho$
	is plotted on the left scale,
	and the inner-product norm of
	the Lorenz gauge constraints $Y^\nti{1,2,3}$
	and the RMS-norm of the rescaled Einstein tensor $\rescaledG_{ab}$
	are plotted on the right (logarithmic) scale.
	Both scales are the same as those of
	figure~\ref{fig-esrc-ortho-P4/norms-of-t}
	(which shows the \EvolutionName{esrc-ortho-P4} evolution).
	All the norms remain bounded at late times,
	showing no secular growth with time.
	}
\label{fig-esrc-ortho-P12/norms-of-t}
\end{figure}

Figure~\ref{fig-esrc-ortho-P12/norms-of-t} shows the time evolution
of the norms 
$\Norm{\hbarI_\ortho}$, $\Norm{Y^\nti{1,2,3}}$, and $\Norm{\rescaledG_{ab}}$.  
Notice that at late times, all the norms remain bounded,
showing no secular growth with time.

\subsection{Orbit Averaging}
\label{app-ortho-variants/orbit-averaging}

Figures~\ref{fig-ppart-ortho-50/lambda} and~\ref{fig-esrc-ortho-P4/lambda}
show that after an initial transient, $\lambda^\instantaneous$ --
and thus $\lambda^\OccasionallyUpdated$ -- tends to oscillate
in a slowly-decaying spiral pattern in the complex plane,
with oscillation period equal to the particle's orbital period~$P$.
Changing the definition of $\lambda^\OccasionallyUpdated$ to be
less oscillatory would be useful:
\begin{itemize}
\item	A less-oscillatory $\lambda^\OccasionallyUpdated$
	would reduce the magnitude of the jumps in $h_{ab}^\ortho$
	each time $\lambda^\OccasionallyUpdated$ is updated.
\item	A less-oscillatory $\lambda^\OccasionallyUpdated$
	would make a finite-time $\lambda^\OccasionallyUpdated$
	a better estimate of the $t \to \infty$ limit, which
	would improve the performance of the fixed-$\lambda$
	orthogonalization variant discussed in
	section~\ref{app-ortho-variants/fixed-lambda}.
\end{itemize}

This suggests averaging $\lambda$ over an orbital period,
i.e., defining \begin{subequations}
					       \label{eqn-lambda-orbit-averaged}
\begin{equation}
\lambda^\OrbitAveraged(t)
	= \bigAverage{\lambda^\instantaneous}{[t{-}P,t]}
									~ ,
\end{equation}
and then redefining~$\lambda^\OccasionallyUpdated$ as
\begin{equation}
\lambda^\OccasionallyUpdated\!(t)
	= \lambda^\OrbitAveraged\bigl( \FloorWRTSet{t}{\{t_k\}} \bigr)
									~ .
\end{equation}
\end{subequations}
(This scheme retains
the definitions~\eqref{eqn-hab-ortho(lambda-occasionally-updated)}
and~\eqref{eqn-hbarI-ortho(lambda-occasionally-updated)}
of $h_{ab}^\ortho$ and $\hbarI_\ortho$
in terms of $\lambda^\OccasionallyUpdated$.)

[In view of the orthogonality condition~\eqref{eqn-hab-perp},
an alternative possible definition for the orbit averaging would be to
define $\lambda^\OrbitAveraged(t)$ as that value of $\lambda$ for which
\begin{subequations}
				     \label{eqn-lambda-orbit-averaged-alternate}
\begin{equation}
\BigAverage{\InnerProduct{h_{ab}^\ortho}{h_{ab}^\hom}}{[t{-}P,t]} = 0
									~ ,
\end{equation}
or, if no such $\lambda$ exists, that value of $\lambda$ which
minimizes
\begin{equation}
\left|
\BigAverage{\InnerProduct{h_{ab}^\ortho}{h_{ab}^\hom}}{[t{-}P,t]}
\right|
									~ .
\end{equation}
\end{subequations}
I have not implemented this variant; indeed, it's not immediately
obvious how to efficiently compute the (a) $\lambda$
satisfying~\eqref{eqn-lambda-orbit-averaged-alternate}.]

To test the orbit-averaging scheme~\eqref{eqn-lambda-orbit-averaged},
I consider the \EvolutionName{ppart-ortho-oa-50} evolution,
which apart from the orbit averaging is identical
to the non--orbit-averaged \EvolutionName{ppart-ortho-50} evolution
described in section~\ref{sect-numerical-tests/ppart-ortho-50}.

\begin{figure}
\begin{center}
\begin{picture}(80,157)
%
%
\put(0,120)
  {
  \begin{picture}(0,0)
  \put(0,34){\textbf{(a)}}
  \put(5,0){\includegraphics[scale=0.90]{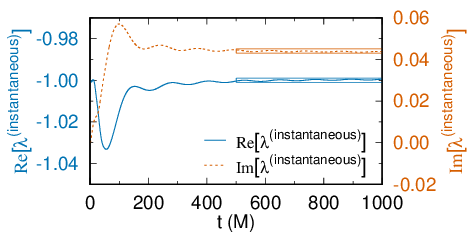}}
  \end{picture}
  }
\put(0,70)
  {
  \begin{picture}(0,0)
  \put(0,45.0){\textbf{(b)}}
  \put(5,0){\includegraphics[scale=0.90]{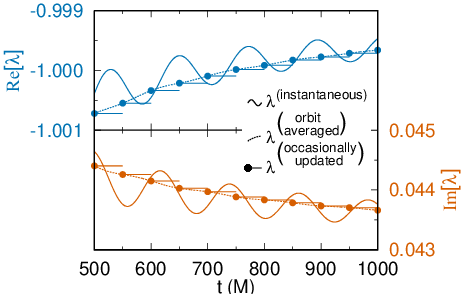}}
  \end{picture}
  }
\put(0,0)
  {
  \begin{picture}(0,0)
  \put(0,63.0){\textbf{(c)}}
  \put(5,0){\includegraphics[scale=0.90]{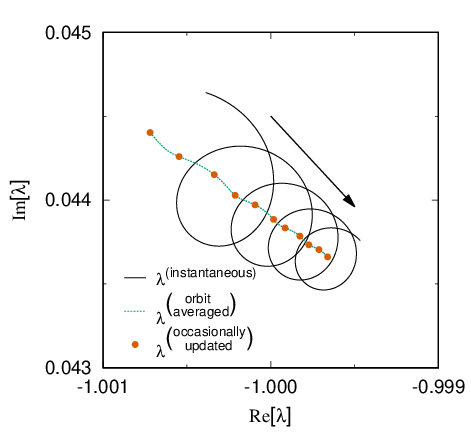}}
  \end{picture}
  }
\end{picture}
\vspace*{-5mm}
\end{center}
\caption[Time evolution of instantaneous and occasionally-updated $\lambda$
	 for the \EvolutionName{ppart-ortho-oa-50} evolution]
	{
	This figure shows the time evolution of $\lambda^\instantaneous$
	and $\lambda^\OccasionallyUpdated$
	for the \EvolutionName{ppart-ortho-oa-50} evolution.
	Part~(a)
	(which is identical to that of
	figure~\protect\ref{fig-ppart-ortho-50/lambda})
	shows the real and imaginary parts of
	$\lambda^\instantaneous$ as functions of time.
	The rectangular regions are shown at an enlarged scale
	in parts~(b) and~(c).
	Part~(b) shows, at an enlarged scale, the real and imaginary parts
	of $\lambda^\instantaneous$, $\lambda^\OrbitAveraged$,
	and $\lambda^\OccasionallyUpdated$
	as functions of time for late times ($t \ge 500M$).
	The solid dots and horizontal lines show the sample-and-hold
	behavior of $\lambda^\OccasionallyUpdated$.
	The legend applies to both real and imaginary parts.
	Part~(c) shows, at an enlarged scale,
	the trajectories in the complex plane
	of $\lambda^\instantaneous$ (the spiral curve),
	$\lambda^\OrbitAveraged$ (the wiggly dotted line),
	and $\lambda^\OccasionallyUpdated$ (the solid dots)
	for late times ($t \ge 500M$).
	The arrow shows the direction of the time evolution.
	}
\label{fig-ppart-ortho-oa-50/lambda}
\end{figure}

Figure~\ref{fig-ppart-ortho-oa-50/lambda} shows the time evolution
of $\lambda^\instantaneous$, $\lambda^\OrbitAveraged$,
and $\lambda^\OccasionallyUpdated$
for the \EvolutionName{ppart-ortho-oa-50} evolution.
Part~(a), showing the time evolution of $\lambda^\instantaneous$,
is identical to that of figure~\ref{fig-ppart-ortho-50/lambda}
(which shows the \EvolutionName{ppart-ortho-50} evolution).
Parts~(b) and~(c) show the effect of the orbit averaging.

Comparing figures~\ref{fig-ppart-ortho-50/lambda}
and~\ref{fig-ppart-ortho-oa-50/lambda},
the changes in $\lambda^\OccasionallyUpdated$
(and thus, the changes in the $\hbarI$) at each update are much smaller
in the orbit-averaged \EvolutionName{ppart-ortho-oa-50} evolution
(figure~\ref{fig-ppart-ortho-oa-50/lambda})
than in the non--orbit-averaged \EvolutionName{ppart-ortho-50} evolution
(figure~\ref{fig-ppart-ortho-50/lambda}),
showing the success of the orbit-averaging scheme.

\begin{figure}
\begin{center}
\includegraphics[scale=1.00]{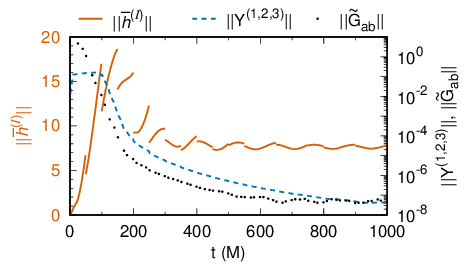}
\end{center}
\caption[Time evolution of norms of the orthogonalized \EvolutionName{esrc-ortho-oa-P4} evolution]
	{
	This figure shows the time evolution of
	norms over grid points and components
	in the \EvolutionName{ppart-ortho-oa-50} evolution.
	The inner-product norm of $\hbarI_\ortho$
	is plotted on the left scale,
	and the inner-product norm of
	the Lorenz gauge constraints $Y^\nti{1,2,3}$
	and RMS-norm of the rescaled Einstein tensor $\rescaledG_{ab}$
	are plotted on the right (logarithmic) scale.
	This figure should be compared with
	figure~\protect\ref{fig-ppart-ortho-50/norms-of-t}
	(which shows the same norms for the non--orbit-averaged
	\EvolutionName{ppart-ortho-50} evolution).
	}
\label{fig-ppart-ortho-oa-50/norms-of-t}
\end{figure}

Figure~\ref{fig-ppart-ortho-oa-50/norms-of-t}
shows the time evolution of the norms 
$\Norm{\hbarI_\ortho}$, $\Norm{Y^\nti{1,2,3}}$, and $\Norm{\rescaledG_{ab}}$.  
Notice that at late times
$\Norm{\hbarI_\ortho}$ remains bounded, showing no secular growth with time,
while $\Norm{Y^\nti{1,2,3}}$ and $\Norm{\rescaledG_{ab}}$ remain small.
Comparing with figure~\ref{fig-ppart-ortho-50/norms-of-t}
(which shows the non--orbit-averaged \EvolutionName{ppart-ortho-50}
evolution), here the changes in $\lambda^\OccasionallyUpdated$
with each update are much smaller.

The snapshots and movie of this evolution are visually quite similar
to those shown in figure~\ref{fig-ppart-ortho-50/ortho-snapshots-and-movie}
for the (non--orbit-averaged) \EvolutionName{ppart-ortho-50} evolution,
and are omitted here in the interests of brevity.

\begin{figure}
\begin{center}
\includegraphics[scale=0.60]{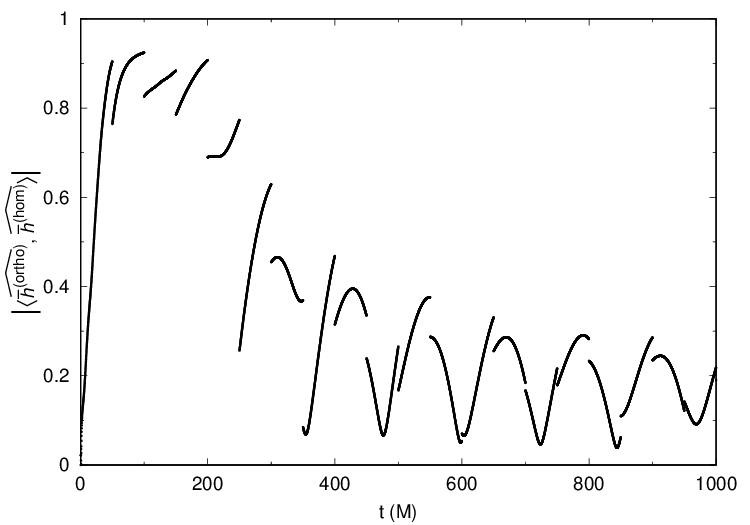}
\end{center}
\caption[Time evolution of unit-vector inner product
	 of $\hbarI_\ortho$ and $\hbarI_\hom$
	 for the \EvolutionName{ppart-ortho-oa-50} evolution.]
	{
	This figure shows the time evolution of the
	unit-vector inner product
	$\InnerProduct{\UnitVector{\hbarI_\ortho}}{\UnitVector{\hbarI_\hom}}$
	for the \EvolutionName{ppart-ortho-oa-50} evolution.
	Because of the orbit averaging, the unit-vector inner product
	doesn't reset to zero each time $\lambda^\OccasionallyUpdated$
	is updated.  However, at late times this inner product
	still stays relatively small.
	This figure should be compared with
	figure~\protect\ref{fig-ppart-ortho-50/uvip-ortho-hom}
	(which shows the same unit-vector inner product for the
	non--orbit-averaged \EvolutionName{ppart-ortho-50} evolution).
	}
\label{fig-ppart-ortho-oa-50/uvip-ortho-hom}
\end{figure}

Figure~\ref{fig-ppart-ortho-oa-50/uvip-ortho-hom} shows the
time evolution of the unit-vector inner product
$\InnerProduct{\UnitVector{\hbarI_\ortho}}{\UnitVector{\hbarI_\hom}}$
for the orbit-averaged \EvolutionName{ppart-ortho-oa-50} evolution.
Comparing with figure~\ref{fig-ppart-ortho-50/uvip-ortho-hom}
(which shows this same unit-vector inner product for the
non--orbit-averaged \EvolutionName{ppart-ortho-50} evolution),
the qualitative behavior is somewhat different.  Notably,
the unit-vector inner product does \emph{not} reset to zero each time
$\lambda^\OccasionallyUpdated$ is updated, and the jumps in the
unit-vector inner product at each update are quite small.
However, at late times the unit-vector inner product still stays
relatively small, which is a key diagnostic of the orthogonalization
scheme's success.

\subsection{Gradual Turnon of the Puncture and Effective Source}
\label{app-ortho-variants/gradual-turnon}

As noted in section~\ref{sect-Schw/modelling-particle/esrc},
if the initial data doesn't already satisfy the
jump conditions~\eqref{eqn-hbarI-numerical(hbarI)}
across the worldtube boundary, the dynamical evolution generates
high-spatial-frequency noise in the process of driving the fields
into a configuration satisfying the jump conditions.
This high-spatial-frequency noise tends to reduce the accuracy of the
numerical computation.

One way to reduce this noise is to use a
\emph{gradual turnon} of the puncture and effective source:
replace~$\hbar_{ab}^\puncture$ in~\eqref{eqn-Einstein-puncture}
and~$\hbarI_\puncture$ in~\eqref{eqn-hbarI-numerical(hbarI,puncture)}
by $\gto(t) \, \hbar_{ab}^\puncture$ and $\gto(t) \, \hbarI_\puncture$
respectively,
where the smooth ``gradual-turnon function'' $\gto$ is chosen to satisfy
$\gto(0) \approx 0$ at $t=0$ and $\gto(t) \approx 1$ at late times.

When using gradual turnon, I use the same gradual-turnon function $\gto$
as \textcite[appendix~B3]{Thornburg-Wardell-2017:Kerr-scalar-self-force},
namely
\begin{subequations}
\begin{align}
\gto(t)	= &	\dhalf \bigl( 1 + \erf(z) \bigr)	
						      \label{eqn-gradual-turnon}
									\\
	= &	\begin{cases}
		1 - \dhalf \erfc(z)	& \text{if $z \ge 0$}	\\
		    \dhalf \erfc(-z)	& \text{if $z < 0$}	
		\end{cases}
									~ .
					    \label{eqn-gradual-turnon-alternate}
\end{align}
\end{subequations}
where the scaled time $z := A + (t-t_\initial)/B$,
and the alternative definition~\eqref{eqn-gradual-turnon-alternate}
avoids the numerical cancellation in~\eqref{eqn-gradual-turnon} at
early times.

When using gradual turnon,
I use the parameters $A = -5$ and $B = 20M$,
so that $\gto(0) \approx 8 \times 10^{-13}$
(small enough that the initial high-frequency noise is negligible),
$\gto(100M) = \half$, and $|\gto(t) - 1| < 10^{-6}$ for $t \ge 168M$
(so that the effect of the gradual-turnon factor in the evolution equations
is less than a part per million for all later times).

To test the gradual-turnon scheme,
I consider the \EvolutionName{esrc-gto-ortho-P4} evolution,
which apart from the gradual turnon is identical
to the \EvolutionName{esrc-ortho-P4} evolution
described in section~\ref{sect-numerical-tests/esrc-ortho-P4}.

\begin{figure}
\begin{center}
\begin{picture}(80,157)
%
%
\put(0,120)
  {
  \begin{picture}(0,0)
  \put(0,34){\textbf{(a)}}
  \put(5,0){\includegraphics[scale=0.90]{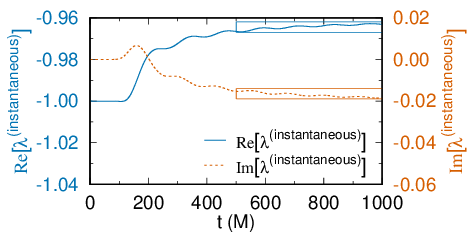}}
  \end{picture}
  }
\put(0,70)
  {
  \begin{picture}(0,0)
  \put(0,45.0){\textbf{(b)}}
  \put(5,0){\includegraphics[scale=0.90]{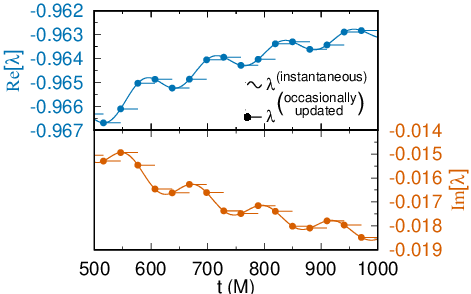}}
  \end{picture}
  }
\put(0,0)
  {
  \begin{picture}(0,0)
  \put(0,64.0){\textbf{(c)}}
  \put(5,0){\includegraphics[scale=0.90]{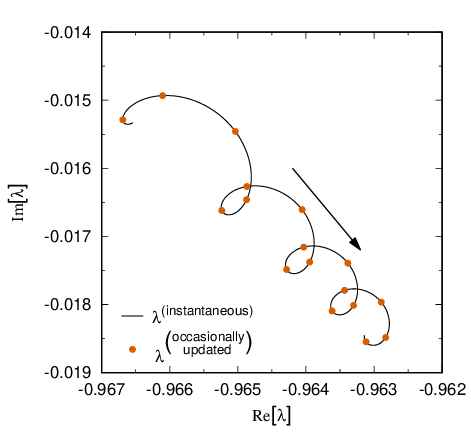}}
  \end{picture}
  }
\end{picture}
\vspace*{-5mm}
\end{center}
\caption[Time evolution of instantaneous and occasionally-updated $\lambda$
	 for the \EvolutionName{esrc-gto-ortho-P4} evolution]
	{
	This figure shows the time evolution of $\lambda^\instantaneous$
	and $\lambda^\OccasionallyUpdated$
	for the \EvolutionName{esrc-gto-ortho-P4} evolution.
	Part~(a) shows the real and imaginary parts of
	$\lambda^\instantaneous$ as functions of time.
	The rectangular regions are shown at an enlarged scale
	in parts~(b) and~(c).
	Part~(b) shows, at an enlarged scale, the real and imaginary parts
	of both $\lambda^\instantaneous$ and $\lambda^\OccasionallyUpdated$
	as functions of time for late times ($t \ge 500M$).
	The solid dots and dashed horizontal lines show the sample-and-hold
	behavior of $\lambda^\OccasionallyUpdated$.
	The legend applies to both real and imaginary parts.
	Part~(c) shows, at an enlarged scale,
	the trajectories in the complex plane
	of both $\lambda^\instantaneous$ (the spiral curve)
	and $\lambda^\OccasionallyUpdated$ (the solid dots)
	for late times ($t \ge 500M$).
	The arrow shows the direction of the time evolution.
	}
\label{fig-esrc-gto-ortho-P4/lambda}
\end{figure}
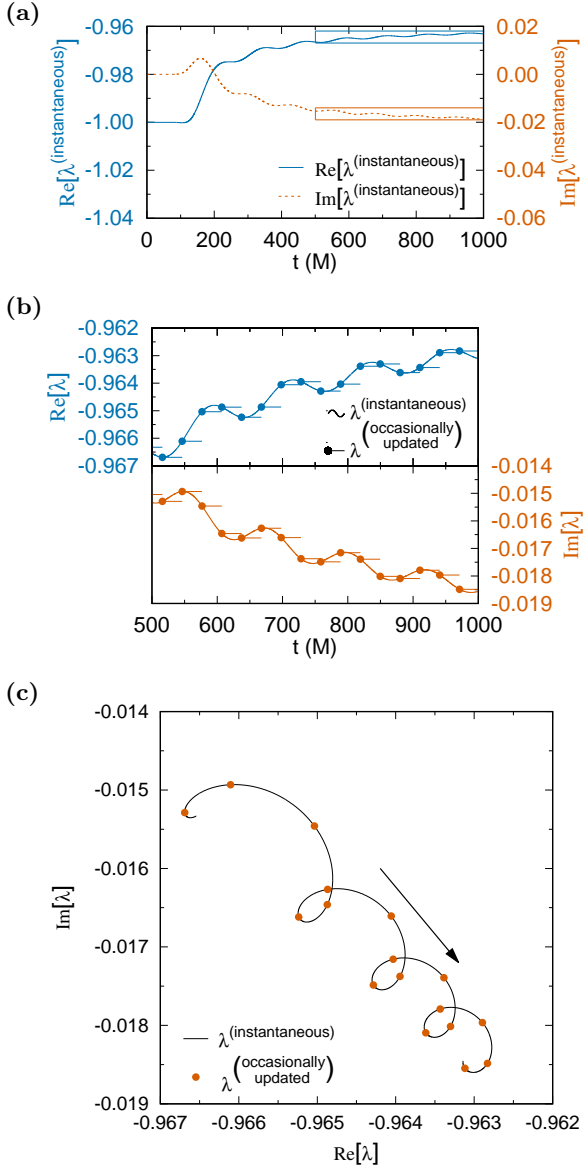

Figure~\ref{fig-esrc-gto-ortho-P4/lambda} shows
the time evolution of $\lambda^\instantaneous$
and $\lambda^\OccasionallyUpdated$ for the
\EvolutionName{esrc-gto-ortho-P4} evolution.
Compared to figure~\ref{fig-esrc-ortho-P4/lambda}
(which shows the non--gradual-turnon \EvolutionName{esrc-ortho-P4} evolution),
here $\lambda^\instantaneous$ (and hence $\lambda^\OccasionallyUpdated$)
varies much more rapidly.

\begin{figure}
\begin{center}
\includegraphics[scale=1.00]{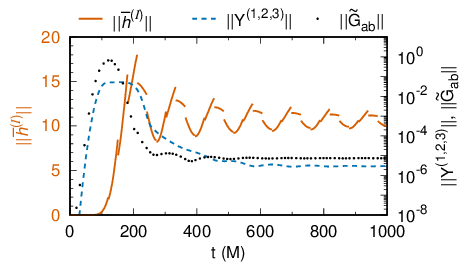}
\end{center}
\caption[Time evolution of norms of the orthogonalized \EvolutionName{esrc-ortho-oa-P4} evolution]
	{
	This figure shows the time evolution of
	norms over grid points and components
	in the \EvolutionName{esrc-gto-ortho-P4} evolution.
	The inner-product norm of $\hbarI_\ortho$
	is plotted on the left scale,
	and the inner-product norm of
	the Lorenz gauge constraints $Y^\nti{1,2,3}$
	and the RMS-norm of the rescaled Einstein tensor $\rescaledG_{ab}$
	are plotted on the right (logarithmic) scale.
	}
\label{fig-esrc-gto-ortho-P4/norms-of-t}
\end{figure}

This more-rapid variation doesn't seem to harm the overall
performance of the orthogonalization scheme.
Figure~\ref{fig-esrc-gto-ortho-P4/norms-of-t}
shows the time evolution of the norms 
$\Norm{\hbarI_\ortho}$, $\Norm{Y^\nti{1,2,3}}$, and $\Norm{\rescaledG_{ab}}$.  
Notice that at late times, all the norms remain bounded,
showing no secular growth with time.

\subsection{\allbf{Using a Fixed~$\lambda$}}
\label{app-ortho-variants/fixed-lambda}

Another way to reduce the jumps in $\lambda^\OccasionallyUpdated$
is to never update $\lambda^\OccasionallyUpdated$,
i.e., to choose $\lambda^\OccasionallyUpdated$
to actually be some fixed (time-independent) constant $\lambda^\fixed$.
In order for the unstable gauge mode to still be mostly cancelled,
$\lambda^\fixed$ should be fairly close to $\lambda^\instantaneous$.

This motivates the following scheme:
\begin{enumerate}
\item	Compute a ``base'' orthogonalized evolution
	(possibly using orbit-averaging)
	for some time period $t \in [0,t_{\max}^\base]$.
\item	Choose $\lambda^\fixed$ to be
	$\lambda^\instantaneous(t_{\max}^\base)$,
	$\lambda^\OrbitAveraged(t_{\max}^\base)$,
	or some series acceleration
	(e.g., the Aitken, Richardson, or Shanks transformation)
	of the time sequence of 
	$\lambda^\instantaneous$ or $\lambda^\OrbitAveraged$
	in the base orthogonalized evolution.
\item	Now compute a new ``fixed-$\lambda$''
	approximately-orthogonalized evolution
	(starting from $t=0$)
	using $\lambda^\OccasionallyUpdated(t) = \lambda^\fixed$.
	The goal is that the fixed-$\lambda$ metric perturbation
	should contain only a relatively small component of the unstable mode
	throughout some ``useful'' range of times about $t_{\max}^\base$.
	For a Schwarzschild background, we can assess this
	by determining the time interval over which
	the fixed-$\lambda$ unit-vector inner product
	$\InnerProduct{\UnitVector{\hbarI_\ortho}}{\UnitVector{\hbarI_\hom}}$
	is relatively small.
\end{enumerate}

Because
$\lambda^\instantaneous(t)$ and $\lambda^\OrbitAveraged(t)$
vary only relatively weakly with $t$ at late times
(see, e.g., figures~\ref{fig-ppart-ortho-50/lambda},
\ref{fig-esrc-ortho-P4/lambda},
and especially~\ref{fig-ppart-ortho-oa-50/lambda}),
$\lambda^\fixed$ will remain close to $\lambda^\instantaneous(t)$
-- and thus the fixed-$\lambda$ unit-vector inner product
$\InnerProduct{\UnitVector{\hbarI_\ortho}}{\UnitVector{\hbarI_\hom}}$
should remain relatively small --
throughout a significant range of times about $t \seq t_{\max}^\base$.

Moreover, since the variation of
$\lambda^\instantaneous(t)$ and $\lambda^\OrbitAveraged(t)$
with $t$ is slower at later times,
increasing $t_{\max}^\base$ should widen (increase the duration of)
the time interval about $t \seq t_{\max}^\base$ for which the fixed-$\lambda$
$\InnerProduct{\UnitVector{\hbarI_\ortho}}{\UnitVector{\hbarI_\hom}}$
is relatively small.

For an initial test of the $\lambda^\fixed$ scheme,
I use the \EvolutionName{ppart-ortho-oa-50} evolution as the base evolution,
and set $\lambda^\fixed := \lambda^\OrbitAveraged(t_{\max}^\base)$ with
$t_{\max}^\base \seq 500M$, $1000M$, or $2000M$
(the \EvolutionName{ppart-ortho-fixed-from-oa-50-t=500},
\EvolutionName{ppart-ortho-fixed-from-oa-50-t=1000},
and \EvolutionName{ppart-ortho-fixed-from-oa-50-t=2000}
evolutions, respectively).

\begin{figure}
\begin{center}
\includegraphics[scale=0.60]{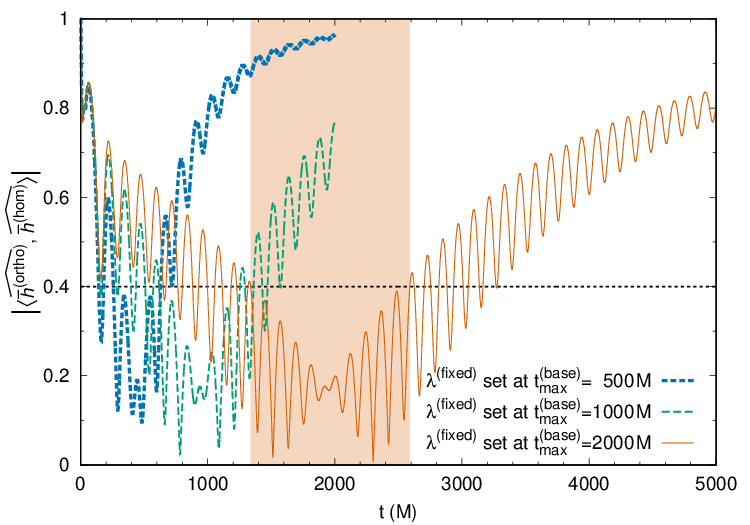}
\end{center}
\caption[Time evolution of unit-vector inner product
	 of $\hbarI_\ortho$ and $\hbarI_\hom$
	 for various \EvolutionName{ppart-ortho-fixed-*} evolutions.]
	{
	This figure shows the time evolution of the
	unit-vector inner product
	$\InnerProduct{\UnitVector{\hbarI_\ortho}}{\UnitVector{\hbarI_\hom}}$
	for fixed-$\lambda$ evolutions where
	$\lambda^\fixed$ is set from
	the \EvolutionName{ppart-ortho-oa-50} evolution's
	$\lambda^\OccasionallyUpdated$
	at $t_{\max}^\base \seq 500M$, $1000M$, and $2000M$.
	Notice that as the time ($t_{\max}^\base$)
	at which $\lambda^\fixed$ is set increases,
	the time intervals for which
	$\InnerProduct{\UnitVector{\hbarI_\ortho}}{\UnitVector{\hbarI_\hom}}$
	is ``relatively small''
	($\le 0.4$, shown as the dashed horizontal line)
	becomes longer.
	This time interval is shown as the shaded region
	for the $t_{\max}^\base \seq 2000M$ case.
	The precise time intervals for each case are given in
	table~\protect\ref{tab-ppart-ortho-fixed/time-intervals}.
	}
\label{fig-ppart-ortho-fixed/cmp-uvip-ortho-hom-from-t=various}
\end{figure}

Figure~\ref{fig-ppart-ortho-fixed/cmp-uvip-ortho-hom-from-t=various}
shows the time evolution of
$\InnerProduct{\UnitVector{\hbarI_\ortho}}{\UnitVector{\hbarI_\hom}}$
for these three $\lambda^\fixed$ evolutions.
As expected, in each case
$\InnerProduct{\UnitVector{\hbarI_\ortho}}{\UnitVector{\hbarI_\hom}}$
is relatively small (operationalized here as $\sle 0.4$) for a finite time
interval about the time $t_{\max}^\base$, and the duration of
this time interval increases as $t_{\max}^\base$ increases.
(The precise time intervals for each case are given in
table~\ref{tab-ppart-ortho-fixed/time-intervals}.)
That is, by increasing $t_{\max}^\base$, i.e., by setting
$\lambda^\fixed$ from the base-evolution $\lambda^\OccasionallyUpdated$
at a later time, the duration of the time interval for which the
fixed-$\lambda$ evolution is mostly free of the unstable mode
can be increased.

\begin{table*}
\begin{center}
\renewcommand{\arraystretch}{1.25}
\begin{tabular}{l@{\hspace*{1.0em}}c@{\hspace*{1.0em}}c@{\hspace*{1.0em}}c@{\hspace*{1.0em}}c}
Fixed-$\lambda$ evolution				&
$t_{\max}^\base$	& time interval
			& duration of time interval			\\
\hline 
\EvolutionName{ppart-ortho-fixed-from-oa-50-t=500}	&
$\Z{}500M$		& $t \in [\Z{}258,\Z{}630]M$
			& $\Z{}372M$					\\
\EvolutionName{ppart-ortho-fixed-from-oa-50-t=1000}	&
$1000M$			& $t \in [\Z{}617,1254]M$
			& $\Z{}637M$					\\
\EvolutionName{ppart-ortho-fixed-from-oa-50-t=2000}	&
$2000M$			& $t \in [1338,2591]M$
			& $1253M$					\\
\hline 
\end{tabular}
\end{center}
\caption[Time Intervals where
	$\InnerProduct{\UnitVector{\hbarI_\ortho}}
		      {\UnitVector{\hbarI_\hom}} \sle 0.4$]
	{
	For each fixed-$\lambda$ evolution,
	This table gives the $t_{\max}^\base$
	and the time intervals and durations for which
	$\InnerProduct{\UnitVector{\hbarI_\ortho}}{\UnitVector{\hbarI_\hom}}
		\sle 0.4$.
	All the fixed-$\lambda$ evolutions use the
	\EvolutionName{ppart-ortho-50} evolution
	as the base evolution.
	}
\label{tab-ppart-ortho-fixed/time-intervals}
\end{table*}

\begin{figure}
\begin{center}
\includegraphics[scale=1.00]{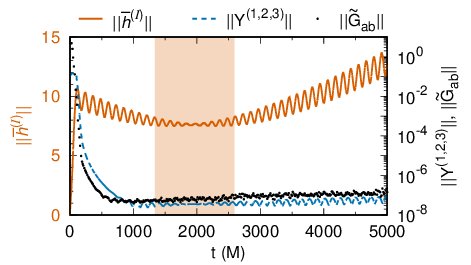}
\end{center}
\caption[Time evolution of norms of the orthogonalized \EvolutionName{esrc-ortho-oa-P4} evolution]
	{
	This figure shows the time evolution of
	norms over grid points and components
	in the \EvolutionName{ppart-ortho-fixed-from-oa-at-t=2000} evolution.
	The inner-product norm of $\hbarI_\ortho$
	is plotted on the left scale,
	and the inner-product norm of
	the Lorenz gauge constraints $Y^\nti{1,2,3}$
	and the RMS-norm of the rescaled Einstein tensor $\rescaledG_{ab}$
	are plotted on the right (logarithmic) scale.
	The time interval (of duration $1253M$) for which
	$\InnerProduct{\UnitVector{\hbarI_\ortho}}{\UnitVector{\hbarI_\hom}}
	\le 0.4$ is shaded.
	}
\label{fig-ppart-ortho-fixed/norms-of-t}
\end{figure}

Figure~\ref{fig-ppart-ortho-fixed/norms-of-t} shows the resulting
fixed-$\lambda$
$\Norm{\hbarI_\ortho}$, $\Norm{Y^\nti{1,2,3}}$, and $\Norm{\rescaledG_{ab}}$
for the case where $\lambda^\fixed$ is set at $t=2000M$
(the \EvolutionName{ppart-ortho-fixed-from-oa-at-t=2000} evolution).
$\Norm{\hbarI_\ortho}$ shows little systematic increase with time
during the time interval (shown shaded) where
$\InnerProduct{\UnitVector{\hbarI_\ortho}}{\UnitVector{\hbarI_\hom}} \le 0.4$,
suggesting that $\hbarI_\ortho$ is \emph{not} dominated by the unstable
mode.
$\Norm{Y^\nti{1,2,3}}$ and $\Norm{\rescaledG_{ab}}$ remain small
at all late times.

In conclusion, using a fixed $\lambda$ seems to provide
an approximately-orthogonalized evolution which is mostly free
of the unstable mode for a period of time which, while finite,
can be considerably longer than $1000M$.

\section{Convergence Tests}
\label{app-convergence-tests}

In any numerical solution of differential equations it's essential to
ensure that as the numerical resolution is increased (i.e., as $\Delta r_*$
is decreased), the numerical solution does indeed converge to to a
continuum solution of the differential equations.
A quantitative test of the rate of this convergence (and whether this
rate matches the theoretical expectation) is an excellent test of the
overall numerical computation (\textcite{Choptuik-1991:FD-consistency}).

This convergence can be assessed by comparing numerical evolutions at
varying numerical resolutions.  Here I use the resolutions
\begin{equation}
\Delta r_* \in \{
		\tfrac{M}{2}, \tfrac{M}{3}, \tfrac{M}{4}, \tfrac{M}{6},
		\tfrac{M}{8}, \tfrac{M}{12}, \tfrac{M}{16}, \tfrac{M}{24},
		\tfrac{M}{32}
		\}
									~ .
					   \label{eqn-convergence-Delta-r_*-set}
\end{equation}
I use the following convergence diagnostics:
\begin{itemize}
\item	For each pair of resolutions differing by
	a factor of~2, I compute the inner-product norm
	\begin{subequations}
						   \label{eqn-convergence-norms}
	\begin{equation}
	\Norm{(\hbarI_\ortho)_\lo - (\hbarI_\ortho)_\hi}
									~ ,
	\end{equation}
	where the subscripts ${}_\lo$ and ${}_\hi$ refer to
	the lower resolution
	and to the higher resolution (subsampled to the lower-resolution grid),
	respectively.
\item	For each resolution, I compute the inner-product norm
	\begin{equation}
	\Norm{Y^\nti{1,2,3}}
	\end{equation}
	and the RMS-norm
	\begin{equation}
	\Norm{\rescaledG_{ab}}
									~ ,
	\end{equation}
	\end{subequations}
	in both cases computed using the $\hbarI_\ortho$
	and their independently-computed time derivatives
	as described in section~\ref{sect-Schw/diagnostics/cmpt-time-derivs}.
\end{itemize}

Given my code's 4th-order finite-differencing scheme (described in detail
in appendix~\ref{app-numerical}), all these norms should ideally be
proportional to $(\Delta r_*)^4$, at least for
sufficiently high resolutions.
As is standard in numerical relativity, I estimate the actual
convergence order $p$ by fitting the model
\begin{equation}
\log \Norm{\,\cdot\,} = p \log \Delta r_* + \constant
\end{equation}
to each of the norms~\eqref{eqn-convergence-norms}
across a range of $\Delta r_*$.

However, there is a complication: the numerical fields
($\hbarI$ for a point-particle scheme,
or $\hbarI_\numerical$ for an effective-source scheme)
are non-smooth at the particle position:
\begin{itemize}
\item	For a point-particle scheme, $\hbarI$ is generically
	$C^0$ at the particle.  Adjusted finite differencing
	across the particle (described in detail in
	appendix~\ref{app-numerical/source/adjusted-FD-ppart})
	compensates for the non-smoothness, but because
	the jump series~\eqref{eqn-ppart-jump-series}
	is truncated at a finite order, the compensation
	isn't perfect, i.e., $\hbar^\adjusted$ as defined
	by~\eqref{eqn-adjusted-FD/ppart} isn't $C^\infty$
	within a finite-difference molecule radius of the particle.
\item	For an effective-source scheme,
	because $\hbar_{ab}^\puncture$ only matches a
	finite number of terms (in my case the first 4~terms)
	of the Laurent series expansion of $\hbar_{ab}^\singular$
	near the particle in powers of the distance from
	the particle, $\hbar^\residual$
	as defined by~\eqref{eqn-hbar-ab-residual}
	is only finitely-differentiable at the particle
	($C^2$ for the puncture used here).
\end{itemize}

Applying finite difference operators to these non-smooth fields results
in finite differencing errors which are ``bump functions''
(\textcite[appendix~F]{Thornburg-1998})
\footnote{
	 Although the discussion of
	 \textcite[appendix~F]{Thornburg-1998}
	 is framed in terms of interpolation,
	 it applies equally well to other finite
	 differencing operations.
	 }
{}
of the particle position modulo the grid spacing.
Because each resolution in the convergence-test
set~\eqref{eqn-convergence-Delta-r_*-set}
samples such a bump function at a quasi-random phase,
\footnote{
	 This phase is effectively
	 $((r_*)_p \bmod \Delta r_*) \big/ \Delta r_*$.
	 While not actually random, this phase is not
 	 controlled by my current numerical scheme.
	 }
{}
each resolution's theoretical
$\O\bigl((\Delta r_*)^4\bigr)$ finite-differencing error term
is effectively multiplied by a quasi-random coefficient,
somewhat spoiling smooth convergence.

I assess the convergence by considering the
norms~\eqref{eqn-convergence-norms} both as functions of time,
and as functions of resolution at three sample late times,
chosen to be
immediately after the 2nd-to-last re-orthogonalization,
midway between the 2nd-to-last and last re-orthogonalizations,
and immediately before the last re-orthogonalization.
These sample times are given in table~\ref{tab-convergence/sample-times}.
By comparing convergence orders across the sample times,
I can determine whether the re-orthogonalizations impair convergence.

\begin{table*}
\begin{center}
\begin{tabular}{l@{\hspace*{1.0em}}c@{\hspace*{1.0em}}c}
Sample time		& \EvolutionName{ppart-ortho-50} evolution
			& \EvolutionName{esrc-ortho-P4} evolution	\\
\hline 
Immediately after the 2nd-to-last
late-time re-orthogonalization
			& $t \seq 1951M$	& $t \sapprox \Z{}972.122M$
									\\
Midway between the 2nd-to-last and last
late-time re-orthogonalizations
			& $t \seq 1975M$	& $t \sapprox \Z{}986.284M$
									\\
Immediately before the last
late-time re-orthogonalization
			& $t \seq 1999M$	& $t \sapprox 1000.446M$
									\\
\hline 
\end{tabular}
\end{center}
\caption[Sample Times for Convergence Tests]
	{
	This table gives the sample times for the convergence
	tests shown in figures~\protect\ref{fig-convergence/hbar},
	\protect\ref{fig-convergence/constraints},
	and~\protect\ref{fig-convergence/Einstein}.
	}
\label{tab-convergence/sample-times}
\end{table*}


\subsection{\allbf{Convergence of $\hbarI_\ortho$}}
\label{app-convergence-tests/hbarI}

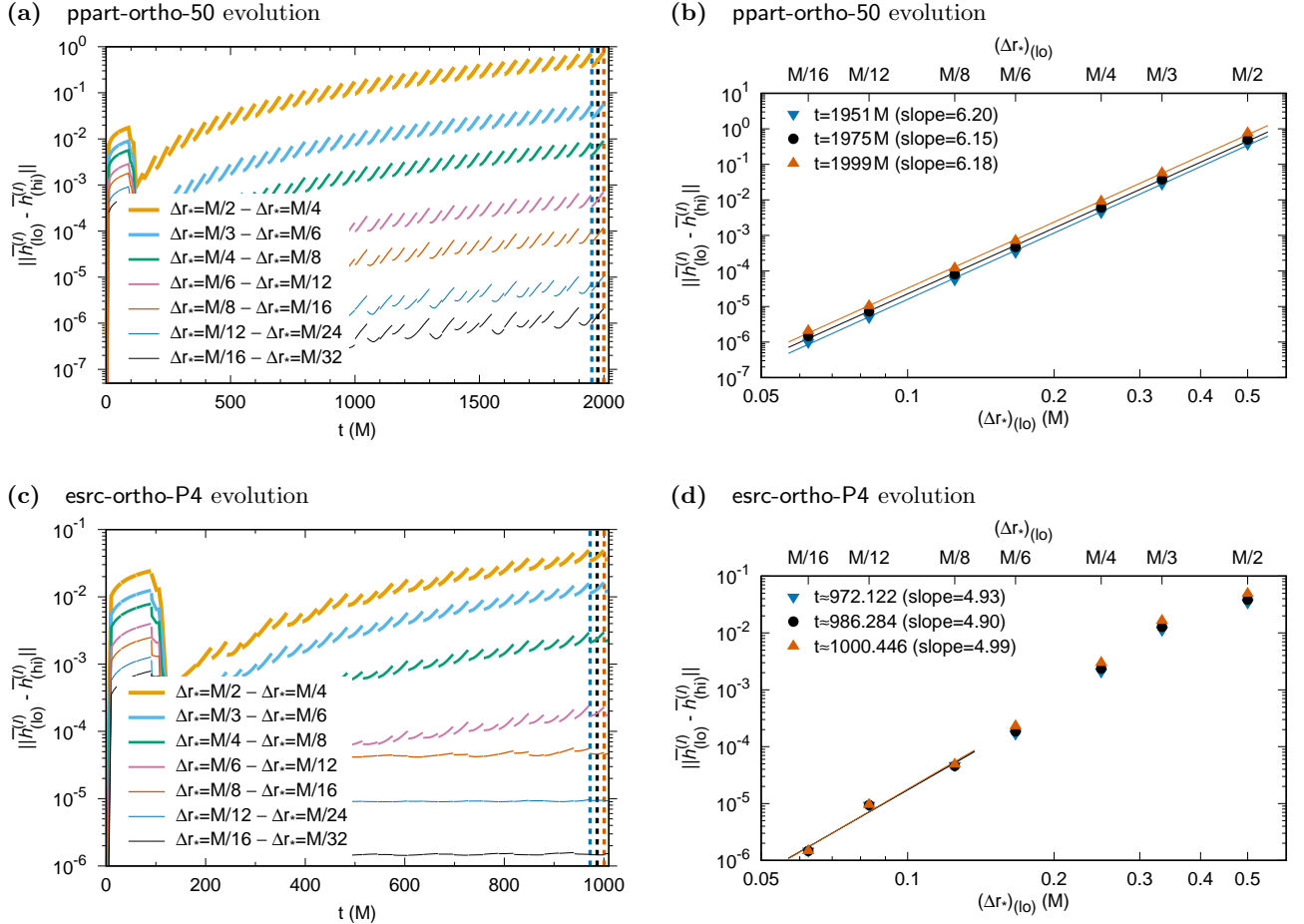
\begin{figure*}
\begin{center}
\begin{picture}(172,121)
%
%
\put(0,64){
  \begin{picture}(0,0)
  \put(0,56){\textbf{(a)}\quad\EvolutionName{ppart-ortho-50} evolution}
  \put(0,0){\includegraphics[scale=1.00]{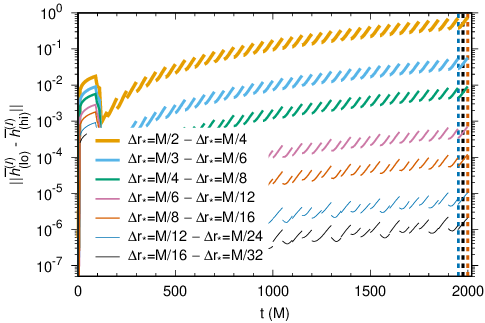}}
  \end{picture}
  }
\put(88,64){
  \begin{picture}(0,0)
  \put(0,56){\textbf{(b)}\quad\EvolutionName{ppart-ortho-50} evolution}
  \put(0,0){\includegraphics[scale=1.00]{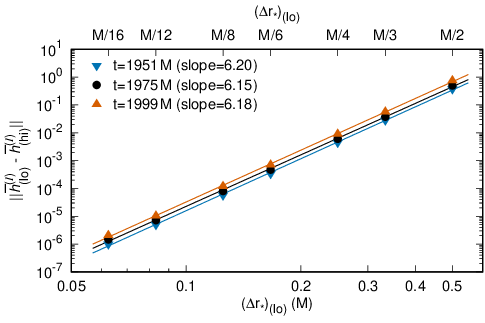}}
  \end{picture}
  }
%
%
\put(0,0){
  \begin{picture}(0,0)
  \put(0,56){\textbf{(c)}\quad\EvolutionName{esrc-ortho-P4} evolution}
  \put(0,0){\includegraphics[scale=1.00]{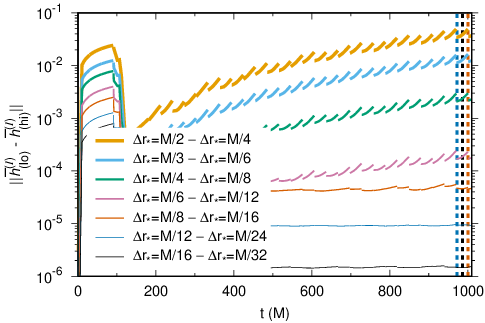}}
  \end{picture}
  }
\put(88,0){
  \begin{picture}(0,0)
  \put(0,56){\textbf{(d)}\quad\EvolutionName{esrc-ortho-P4} evolution}
  \put(0,0){\includegraphics[scale=1.00]{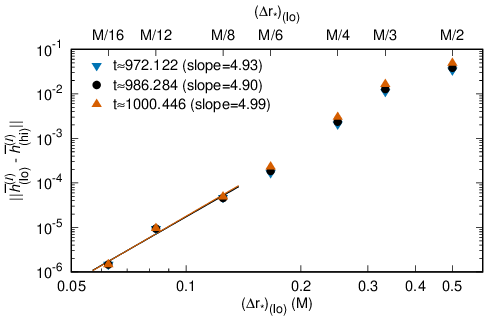}}
  \end{picture}
  }
\end{picture}
\end{center}
\caption[Convergence of $\hbarI$]
	{
	This figure shows the convergence of $\Norm{\hbarI_\lo - \hbarI_\hi}$
	for the \EvolutionName{ppart-ortho-50} 
	and \EvolutionName{esrc-ortho-P4} evolutions.
	The left column (parts~(a) and~(c)) shows
	the norms as a function of time.
	The right column (parts~(b) and~(d) shows
	the norms as a function of the numerical resolution~$\Delta r_*$,
	along with regression lines for fitting the convergence order,
	for the three sample times shown
	as vertical dashed lines in the left column (parts~(a) and~(c)).
	These sample times are 
	given in table~\protect\ref{tab-convergence/sample-times},
	and are chosen to be 
	immediately after a late-time re-orthogonalization,
	midway between late-time re-orthogonalizations,
	and immediately before the next late-time re-orthogonalization,
	respectively.
	See the main text for further discussion.
	}
\label{fig-convergence/hbar}
\end{figure*}

Figure~\ref{fig-convergence/hbar}
shows the convergence of $\Norm{\hbarI_\lo - \hbarI_\hi}$
for the \EvolutionName{ppart-ortho-50} and \EvolutionName{esrc-ortho-P4}
evolutions.  Parts~(a) and~(c) show $\Norm{\hbarI_\lo - \hbarI_\hi}$
as a function of time for each evolution and each resolution pair.
Parts~(b) and~(d) show $\Norm{\hbarI_\lo - \hbarI_\hi}$ as a function
of resolution at three sample times late in each evolution.
The sample times are marked with vertical dashed lines in
parts~(a) and~(c), and are listed in table~\ref{tab-convergence/sample-times}.

For the \EvolutionName{esrc-ortho-P4} evolution,
figure~\ref{fig-convergence/hbar}(c) shows that at late times
$\Norm{\hbarI_\lo - \hbarI_\hi}$ is relatively time-independent
at the three highest resolutions $\big(${}$(\Delta r_*)_\lo \sle M/8${}$\big)$,
but shows substantial secular growth with time at lower resolutions.
Using a combination of $\Norm{\hbarI_\lo - \hbarI_\hi}$
which are relatively time-independent and which are secularly growing with time
would introduce a time-dependent bias into
the convergence-order fit (figure~\ref{fig-convergence/hbar}(d)).
To avoid this time-dependent bias, I use only the three highest resolutions
$\big(${}$(\Delta r_*)_\lo \sle M/8${}$\big)$
in the convergence-order fit (figure~\ref{fig-convergence/hbar}(d)).

For the \EvolutionName{ppart-ortho-50} evolution
$\Norm{\hbarI_\lo - \hbarI_\hi}$ shows comparable secular growth
at all resolutions
(i.e., the different-resolution curves
in figure~\ref{fig-convergence/hbar}(a)
are roughly parallel to each other),
so there's no reason to restrict the convergence-order fits
to a subset of resolutions.

Actually, for both evolutions all the $\Norm{\hbarI_\lo - \hbarI_\hi}$
points are quite close to the fitted convergence-order lines in
figures~\ref{fig-convergence/hbar}(b)
and~\ref{fig-convergence/hbar}(d),
so the choice of which resolutions are included in the fit
makes relatively little difference in the fitted slopes (convergence orders).

For both evolutions, at each sample time the fitted convergence orders
are significantly \emph{greater} than the predicted~$4$.
That is, at late times $\Norm{\hbarI_\lo - \hbarI_\hi}$
is \emph{not} dominated by $\O\bigl((\Delta r_*)^4\bigr)$
finite-differencing errors.
Also, since the norm in $\Norm{\hbarI_\lo - \hbarI_\hi}$
is taken over a set of grid points which is bounded away from the particle
(this is described in detail in
section~\ref{sect-Schw/ortho/choice-of-inner-product}),
the anomalously high convergence orders can't be caused
by the behavior of $\hbarI_\ortho$ near the particle.

The origin of these anomalously high convergence orders remains unclear;
further research is needed to understand this.


\subsection{Convergence of the Lorenz gauge constraints}
\label{app-convergence-tests/constraints}

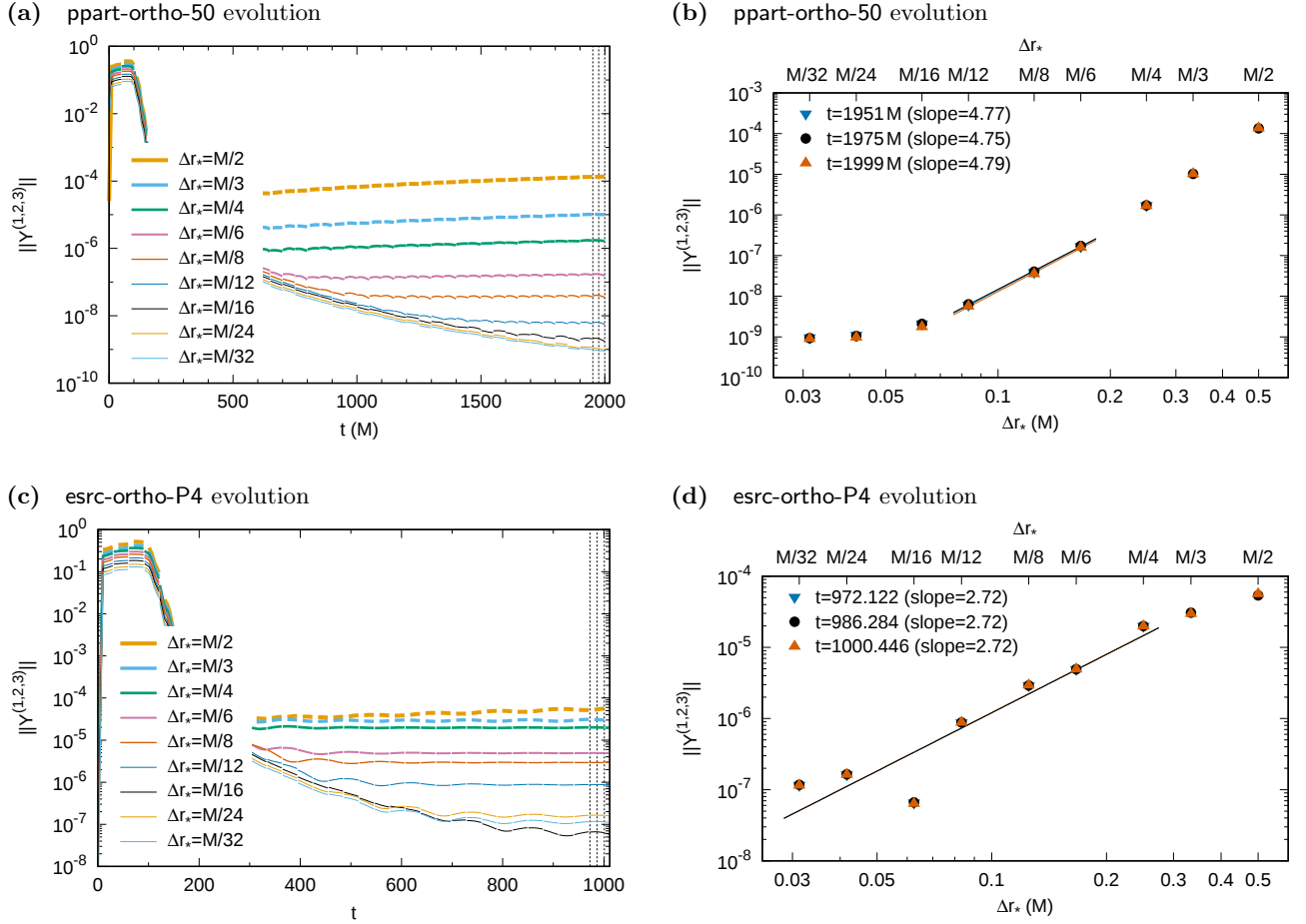
\begin{figure*}
\begin{center}
\begin{picture}(172,122)
%
%
\put(0,64){
  \begin{picture}(0,0)
  \put(0,56){\textbf{(a)}\quad\EvolutionName{ppart-ortho-50} evolution}
  \put(0,0){\includegraphics[scale=1.00]{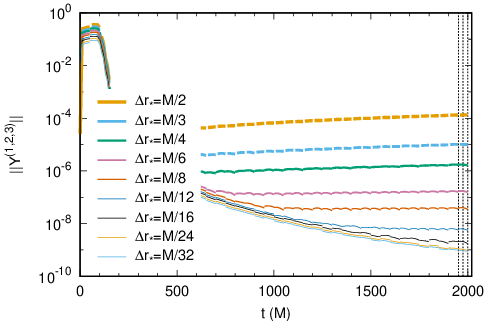}}
  \end{picture}
  }
\put(88,64){
  \begin{picture}(0,0)
  \put(0,56){\textbf{(b)}\quad\EvolutionName{ppart-ortho-50} evolution}
  \put(0,0){\includegraphics[scale=1.00]{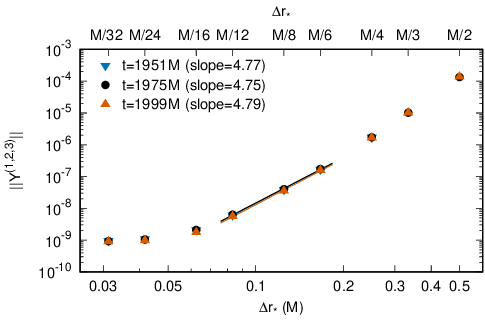}}
  \end{picture}
  }
%
%
\put(0,0){
  \begin{picture}(0,0)
  \put(0,56){\textbf{(c)}\quad\EvolutionName{esrc-ortho-P4} evolution}
  \put(0,0){\includegraphics[scale=1.00]{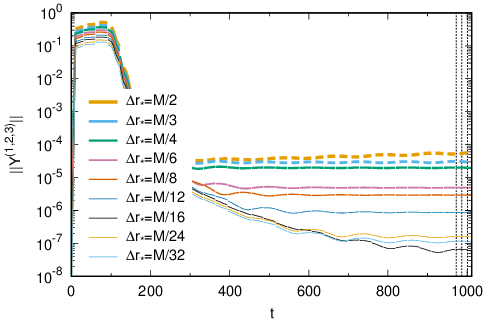}}
  \end{picture}
  }
\put(88,0){
  \begin{picture}(0,0)
  \put(0,56){\textbf{(d)}\quad\EvolutionName{esrc-ortho-P4} evolution}
  \put(0,0){\includegraphics[scale=1.00]{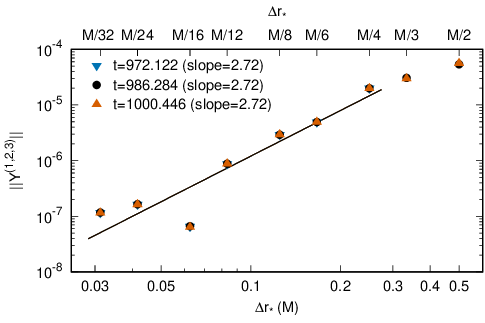}}
  \end{picture}
  }
\end{picture}
\end{center}
\caption[Convergence of Lorenz Gauge Constraints]
	{
	This figure shows the convergence of $Y^\nti{1,2,3}$,
	more precisely the inner-product norm $\Norm{Y^\nti{1,2,3}}$,
	in the \EvolutionName{ppart-ortho-50} 
	and \EvolutionName{esrc-ortho-P4} evolutions.
	The left column (parts~(a) and~(c)) shows
	the norms as a function of time.
	The right column (parts~(b) and~(d) shows
	the norms as a function of the numerical resolution~$\Delta r_*$,
	along with regression lines,
	for the three sample times shown
	as vertical dashed lines in the left column (parts~(a) and~(c)).
	These sample times are 
	given in table~\protect\ref{tab-convergence/sample-times},
	and are chosen to be 
	immediately after a
	late-time re-orthogonalization,
	midway between
	late-time re-orthogonalizations,
	and immediately before the next
	late-time re-orthogonalization, respectively.
	See the main text for further discussion.
	}
\label{fig-convergence/constraints}
\end{figure*}

Figure~\ref{fig-convergence/constraints}
shows the convergence of $\Norm{Y^\nti{1,2,3}}$
for the \EvolutionName{ppart-ortho-50} and \EvolutionName{esrc-ortho-P4}
evolutions.  In each figure, parts~(a) and~(c) show $\Norm{\,\cdot\,}$
as a function of time for each evolution and each resolution.
Parts~(b) and~(d) show $\Norm{\,\cdot\,}$ as a function
of resolution at three sample times late in each evolution.
The sample times are marked with vertical dashed lines in
parts~(a) and~(c), and are listed in table~\ref{tab-convergence/sample-times}.

For the \EvolutionName{ppart-ortho-50} evolution,
figure~\ref{fig-convergence/constraints}(a) shows that
$\Norm{Y^\nti{1,2,3}}$ is still decreasing as a function of time
at the end of the evolution ($t_{\max} \seq 2000M$) for the highest
two resolutions ($\Delta r_* \sle M/24$), and perhaps at the next
lower resolution ($\Delta r_* \seq M/16$) as well.
This suggests that it might be useful to increase
the duration $t_{\max}$ of this evolution.

Given the present data, I have chosen the range of resolutions
$M/12 \sle \Delta r_* \sle M/6$ as having
$\Norm{Y^\nti{1,2,3}}$ being relatively time-independent
for convergence-order fitting (figure~\ref{fig-convergence/constraints}(b)).
At each sample time the fitted convergence orders
are significantly \emph{greater} than the predicted~$4$.
However, at the three highest resolutions the convergence order
is clearly much lower, and is in fact consistent with an
error floor of $\Norm{Y^\nti{1,2,3}} \approx 10^{-9}$.

For the \EvolutionName{esrc-ortho-P4} evolution,
figure~\ref{fig-convergence/constraints}(c) shows that
at late times, $\Norm{Y^\nti{1,2,3}}$
shows slow secular growth with time at the two lowest resolutions
($\Delta r_* \sge M/3$).
At late times, at intermediate resolutions ($M/12 \sle \Delta r_* \sle M/4)$,
$\Norm{Y^\nti{1,2,3}}$ undergoes several particle-orbit-period oscillations
before becoming approximately time-independent,
while at the three highest resolutions ($\Delta r_* \sle M/16$)
these oscillations still have substantial amplitude at
the end of the evolution ($t_{\max} \sapprox 1011.573M$),
suggesting that it might also be useful to increase
the duration $t_{\max}$ of this evolution.

Given the present data, I have chosen the range of resolutions
$\Delta r_* \sle M/4$ for convergence-order fitting
(figure~\ref{fig-convergence/constraints}(d)).
The fitted convergence orders are all much less than the predicted~$4$,
suggesting that $\hbarI_\ortho$ may not be fully smooth
in this evolution (so that the spatial and time derivatives
in~$Y^\nti{1,2,3}$ result in a decrease of convergence order).
Moreover, the $\Delta r_* \seq M/16$ evolution shows a much
smaller $\Norm{Y^\nti{1,2,3}}$ than would be the case for
any fixed convergence order.
This may be due to a fortuitous sampling of a bump-function
finite-differencing error term very close to a zero.

\subsection{Convergence of the rescaled Einstein tensor}
\label{app-convergence-tests/Einstein}

\begin{figure*}
\begin{center}
\begin{picture}(172,122)
%
%
\put(0,64){
  \begin{picture}(0,0)
  \put(0,56){\textbf{(a)}\quad\EvolutionName{ppart-ortho-50} evolution}
  \put(0,0){\includegraphics[scale=1.00]{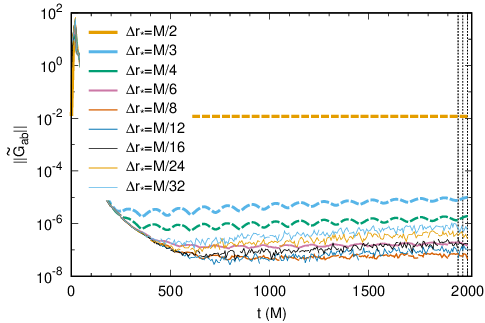}}
  \end{picture}
  }
\put(88,64){
  \begin{picture}(0,0)
  \put(0,56){\textbf{(b)}\quad\EvolutionName{ppart-ortho-50} evolution}
  \put(0,0){\includegraphics[scale=1.00]{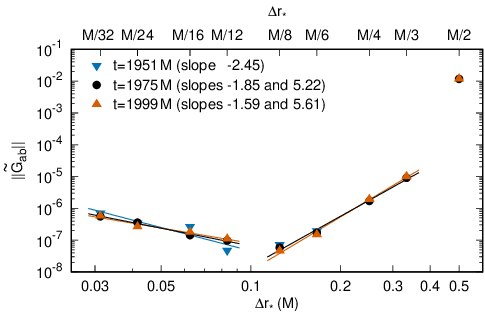}}
  \end{picture}
  }
%
%
\put(0,0){
  \begin{picture}(0,0)
  \put(0,56){\textbf{(c)}\quad\EvolutionName{esrc-ortho-P4} evolution}
  \put(0,0){\includegraphics[scale=1.00]{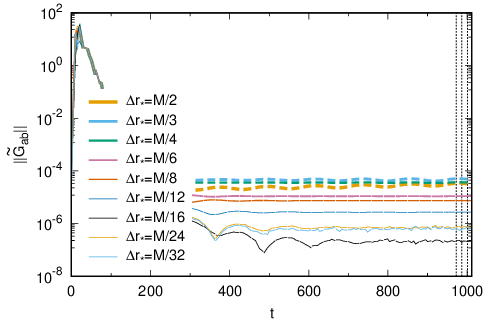}}
  \end{picture}
  }
\put(88,0){
  \begin{picture}(0,0)
  \put(0,56){\textbf{(d)}\quad\EvolutionName{esrc-ortho-P4} evolution}
  \put(0,0){\includegraphics[scale=1.00]{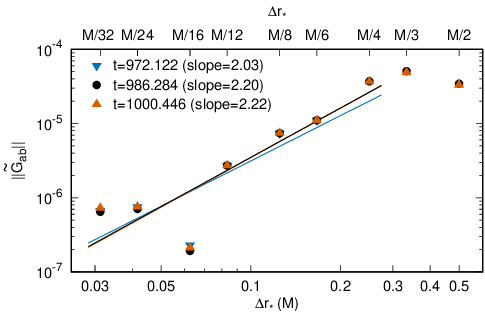}}
  \end{picture}
  }
\end{picture}
\end{center}
\caption[Convergence of rescaled Einstein tensor]
	{
	This figure shows the convergence of $\rescaledG_{ab}$,
	more precisely the RMS-norm $\Norm{\rescaledG_{ab}}$,
	in the \EvolutionName{ppart-ortho-50} 
	and \EvolutionName{esrc-ortho-P4} evolutions.
	The left column (parts~(a) and~(c)) shows
	the norms as a function of time.
	The right column (parts~(b) and~(d) shows
	the norms as a function of the numerical resolution~$\Delta r_*$,
	along with high-resolution and low-resolution regression lines
	for the indicated ranges of points,
	for the three sample times shown
	as vertical dashed lines in the left column (parts~(a) and~(c)).
	(The $t \seq 1951$ low-resolution fit is omitted because
	the lowest~4 resolutions didn't have output at that time.)
	The sample times are 
	given in table~\protect\ref{tab-convergence/sample-times},
	and are chosen to be 
	immediately after a late-time re-orthogonalization,
	midway between late-time re-orthogonalizations,
	and immediately before the next late-time re-orthogonalization,
	respectively.
	See the main text for further discussion.
	}
\label{fig-convergence/Einstein}
\end{figure*}

Figure~\ref{fig-convergence/Einstein}
shows the convergence of $\Norm{\rescaledG_{ab}}$
for the \EvolutionName{ppart-ortho-50} and \EvolutionName{esrc-ortho-P4}
evolutions.  In each figure, parts~(a) and~(c) show $\Norm{\,\cdot\,}$
as a function of time for each evolution and each resolution.
Parts~(b) and~(d) show $\Norm{\,\cdot\,}$ as a function
of resolution at three sample times late in each evolution.
The sample times are marked with vertical dashed lines in
parts~(a) and~(c), and are listed in table~\ref{tab-convergence/sample-times}.

For the \EvolutionName{ppart-ortho-50} evolution,
figure~\ref{fig-convergence/Einstein}(a) shows that
$\Norm{\rescaledG_{ab}}$ has a slow secular increase with time
for almost all resolutions.
Figure~\ref{fig-convergence/Einstein}(b) shows that
$\Norm{\rescaledG_{ab}}$ has two distinct ranges of variation
with $\Delta r_*$:
at lower resolutions ($\Delta r_* \sge M/8$)
$\Norm{\rescaledG_{ab}}$ decreases
with increasing resolution (decreasing $\Delta r_*$)
but at higher resolutions ($\Delta r_* \sle M/12$) $\Norm{\rescaledG_{ab}}$
$\Norm{\rescaledG_{ab}}$ \emph{increases}
with increasing resolution (decreasing $\Delta r_*$).

For the \EvolutionName{esrc-ortho-P4} evolution,
figures~\ref{fig-convergence/Einstein}(c) and~(d)
show very similar behavior
to the corresponding figures for $\Norm{Y^\nti{1,2,3}}$
(figures~\ref{fig-convergence/constraints}(c) and~(d)).
Notably, the $\Delta r_* \seq M/16$ evolution shows a much
smaller $\Norm{\rescaledG_{ab}}$ than would be the case for
any fixed convergence order;
this may again be due to a fortuitous sampling of a bump-function
finite-differencing error term very close to a zero.

\subsection{Convergence Summary}
\label{app-convergence-tests/summary}

Figure~\ref{fig-convergence/hbar} shows that in both the
\EvolutionName{ppart-ortho-50} and \EvolutionName{esrc-ortho-P4} evolutions,
$\Norm{\hbarI_\lo - \hbarI_\hi}$ converges smoothly (to zero)
with increasing resolution $\big($decreasing $(\Delta r_*)_\lo${}$\big)$.

However, figures~\ref{fig-convergence/constraints}
and~\ref{fig-convergence/Einstein}
show significant limitations and non-uniformities in the
convergence of $\Norm{Y^\nti{1,2,3}}$ and $\Norm{\rescaledG_{ab}}$
(which are computed via numerical derivatives of the $\hbarI$),
likely due to low-level non-smoothness in the numerical fields.


\bibliography{jt-new,aei-references,einsteintoolkit}{}


\end{document}